\documentclass[12pt]{article}

\AtBeginDocument{
  \addtocontents{toc}{\normalsize}%\footnotesize
}
\pdfoutput=1

\usepackage{amsmath,amssymb,amsfonts,amsthm,amscd,mathrsfs}
\usepackage{xcolor}
\definecolor{darkblue}{rgb}{0.1,0.1,.7}
\usepackage[colorlinks, linkcolor=darkblue, citecolor=darkblue, urlcolor=darkblue, linktocpage]{hyperref} 
\usepackage[]{version}
\usepackage[]{graphicx}
\usepackage[]{latexsym}
\usepackage{geometry}
\usepackage[all,cmtip]{xy}
\usepackage[margin=10pt,font=small,labelfont=bf]{caption}
\usepackage{ifthen}
\usepackage{tikz}
\usepackage{array,setspace,mathrsfs,amsfonts,yfonts,dsfont,bbm,colonequals}
\usepackage{dsfont}
\usepackage{cite}
\usepackage{xspace}

\numberwithin{equation}{section}

\newcommand{\cO}{\mathcal O}

\newcommand{\reef}[1]{(\ref{#1})}
\newcommand{\be}{\begin{equation}}
\newcommand{\ee}{\end{equation}}
\newcommand{\bea}{\begin{eqnarray}}
\newcommand{\eea}{\end{eqnarray}}
\newcommand{\ba}{\begin{equation}\begin{aligned}}
\newcommand{\ea}{\end{aligned}\end{equation}}

\newcommand{\ud}{\mathrm d}

\newcommand{\Df}{{\Delta_\phi}}
\newcommand{\Dps}{{\Delta_\psi}}

\newcommand{\nnint}{\mathbb{N}}
\newcommand{\DB}{\Delta^{\textrm{B}}}
\newcommand{\DF}{\Delta^{\textrm{F}}}
\newcommand{\Dn}{\Delta_{n}}
\newcommand{\DcO}{\Delta_{\cO }}
\newcommand{\aF}{\alpha^{\textrm{F}}}
\newcommand{\bF}{\beta^{\textrm{F}}}
\newcommand{\aB}{\alpha^{\textrm{B}}}
\newcommand{\bB}{\beta^{\textrm{B}}}

\newcommand{\email}[1]{\vbox{\center\tt#1}\vspace{5mm}}

\begin{document}

\vspace*{-.6in} \thispagestyle{empty}
\begin{flushright}
%CERN PH-TH/2015-200\\
LPTENS/18/17
\end{flushright}
%%
%% Title
%%
\vspace{1cm} {\Large
\begin{center}
{\bf The Analytic Functional Bootstrap II:\\Natural Bases for the Crossing Equation}\\
\end{center}}
\vspace{1cm}
%%%
%% Authors
%%%
\begin{center}
{\bf Dalimil Maz\'a\v{c}$^{\alpha,\beta}$, Miguel F.~Paulos$^{\omega}$}\\[1cm] 
{
\small
$^{\alpha}$ {\em C. N. Yang Institute for Theoretical Physics, Stony Brook University\\Stony Brook, NY 11794, USA}\\
$^{\beta}$ {\em Simons Center for Geometry and Physics, Stony Brook University\\ Stony Brook, NY 11794, USA}\\
$^{\omega}$ {\em Laboratoire de Physique Th\'eorique de l'\'Ecole Normale Sup\'erieure\\ PSL University, CNRS, Sorbonne Universit\'es, UPMC Univ. Paris 06\\ 24 rue Lhomond, 75231 Paris Cedex 05, France
}
\normalsize
}
\\
%\vspace{1cm}\today
\end{center}

\email{dalimil.mazac@stonybrook.edu, miguel.paulos@lpt.ens.fr}

\vspace{4mm}

\begin{abstract}
We clarify the relationships between different approaches to the conformal bootstrap. A central role is played by the so-called extremal functionals. They are linear functionals acting on the crossing equation which are directly responsible for the optimal bounds of the numerical bootstrap. We explain in detail that the extremal functionals probe the Regge limit. We construct two complete sets of extremal functionals for the crossing equation specialized to $z=\bar{z}$, associated to the generalized free boson and fermion theories. These functionals lead to non-perturbative sum rules on the CFT data which automatically incorporate Regge boundedness of physical correlators. The sum rules imply universal properties of the OPE at large $\Delta$ in every unitary solution of $SL(2)$ crossing. In particular, we prove an upper and lower bound on a weighted sum of OPE coefficients present between consecutive generalized free field dimensions. The lower bound implies the $\phi\times\phi$ OPE must contain at least one primary in the interval $[2\Df+2n,2\Df+2n+4]$ for all sufficiently large integer $n$. The functionals directly compute the OPE decomposition of crossing-symmetrized Witten exchange diagrams in $AdS_2$. Therefore, they provide a derivation of the Polyakov bootstrap for $SL(2)$, in particular fixing the so-called contact-term ambiguity. We also use the resulting sum rules to bootstrap several Witten diagrams in $AdS_2$ up to two loops.
\end{abstract}
\vspace{2in}

%\hspace{0.7cm} March 2018

\newpage

{
\setlength{\parskip}{0.05in}
\tableofcontents
\renewcommand{\baselinestretch}{1.0}\normalsize
}

%\newpage

\setlength{\parskip}{0.1in}
\newpage

\section{Introduction}\label{sec:introduction}

The conformal bootstrap constitutes a set of powerful nonperturbative constraints on conformal field theories. Nevertheless, the extraction of concrete physical predictions from the bootstrap equations has proven notoriously hard. As a result, existing analytic approaches typically rely on expanding the equations around a specific point in the space of conformal cross-ratios. Indeed, the subject of modern analytic conformal bootstrap started by studying the double light-cone limit \cite{Komargodski:2012ek,Fitzpatrick:2012yx,Alday:2015eya,Alday:2015ewa,Alday:2016njk}. More recently, progress has been made also by utilizing the Regge limit \cite{Caron-Huot2017b,Li:2017lmh,Costa:2017twz,Kravchuk:2018htv} and the deep Euclidean limit \cite{Qiao:2017xif,Mukhametzhanov:2018zja}.

In this paper, we develop an analytic approach to the conformal bootstrap which does not rely on any kinematical expansion, building on previous work of \cite{Mazac:2016qev,Mazac:2018mdx}. Specifically, we derive useful sum rules satisfied by the CFT data by integrating the crossing equation against appropriate weight-functions in the space of cross-ratios. The integration includes both Euclidean and Lorentzian configurations and combines them in a particularly constraining way.

An important ingredient of our approach is that the crossing equation holds holomorphically as a function of independent \emph{complex} variables $z$ and $\bar{z}$.\footnote{We use the standard convention for the conformal cross-ratios $u=z\bar{z}=\frac{x^2_{12}x^2_{34}}{x^2_{13}x^2_{24}}$, $v=(1-z)(1-\bar{z})=\frac{x^2_{14}x^2_{23}}{x^2_{13}x^2_{24}}$, where $x_{ij}=x_i-x_j$.} The constraints in the Lorentzian and Euclidean signatures, which each correspond to two-dimensional real subsections, thus combine to more powerful constraints in two complex dimensions. The crossing equation expressing the equality of the s- and t-channel expansions holds in this two-complex dimensional space as long as we do not cross branch cuts where a pair of operators becomes null-separated. We will refer to this region of validity of the s=t crossing equation as the \emph{crossing region}.

Analytic control is available at various boundary points of the crossing region, including the double light-cone limit $z\rightarrow0$, $\bar{z}\rightarrow 1$ and the u-channel Regge limit $z,\bar{z}\rightarrow i\infty$. On the other hand, the numerical bootstrap is based on an expansion of the crossing equations around the center of the crossing region $z=\bar{z}=1/2$, going back to the seminal work of \cite{Rattazzi:2008pe}.\footnote{See \cite{Poland:2018epd} for a recent review of some of the developments since then.} The numerical bootstrap effectively constructs distinguished linear functionals acting on the space of functions of $z$ and $\bar{z}$. Those functionals which correspond to optimal bounds on the CFT data also encode the spectra of the optimal solutions to crossing, and are known as extremal functionals \cite{ElShowk:2012hu,El-Showk:2016mxr}. The bounds of the numerical bootstrap typically become optimal only when one goes to an arbitrarily high order in the expansion around $z=\bar{z}=1/2$. In this limit we probe the boundary of the crossing region, including the analytic bootstrap limits. We are led to the conclusion that both the light-cone and Regge limits should play an important role in the numerical bounds on the low-lying CFT spectrum.\footnote{Further evidence for the interrelatedness of the light-cone and numerical bootstrap was provided in \cite{Simmons-Duffin:2016wlq}.}

We can think of the expansion around the crossing-symmetric point as providing a specific basis for the space dual to the space of functions holomorphic in the crossing region. Clearly, it would be of great interest to have instead a basis which extracts the complete information contained in the crossing region, in a manner reflecting our analytic understanding of the corners.

In this paper, we achieve this goal for a simplified version of the crossing equation. The simplification comes from setting $z=\bar{z}$ and expanding the s- and t-channel in $SL(2)$ blocks. When $z\in(0,1)$, our kinematics corresponds to restricting the four operators to lie on a line and using only the conformal group fixing the line. We stress however that the resulting equation holds for complex $z$. In particular, the (u-channel) Regge limit lies within our restricted kinematics and corresponds to $z\rightarrow i\infty$. Indeed, boundedness of physical correlators in this limit plays a crucial role in our analysis. On the other hand, the double light-cone limit lies outside our kinematics so we will not make contact with  the large-spin analytic bootstrap. Our conclusions will be valid for general $D$-dimensional CFTs, but may not be optimal unless $D=1$.

\subsection*{Summary of results and outline}
We will introduce two bases for the crossing equation of the four-point function $\langle \phi \phi \phi \phi\rangle$, where $\phi$ are identical $SL(2)$ primaries. In our restricted kinematics, the equation takes the form
\be
\sum_{\cO \in\phi\times\phi}\!\!a_{\cO }\,F_{\Delta_{\cO }}(z)=0\,,
\label{eq:origcross}
\ee
where the sum runs over the $SL(2)$ primaries present in the $\phi\times\phi$ OPE and $a_{\cO}\equiv (c_{\phi \phi \cO})^2$ is the squared OPE coefficient. $F_{\Delta}(z)$ is defined as the difference of the s- and t-channel conformal blocks
\be
F_{\Delta}(z)=z^{-2\Df}G_{\Delta}(z)-(1-z)^{-2\Df} G_{\Delta}(1-z)\,,
\ee
where $G_{\Delta} = z^{\Delta}{}_2F_1(\Delta,\Delta;2\Delta;z)$. The two bases introduced in this paper are associated to the bosonic and fermionic generalized free field solutions of this equation. Besides the identity, these solutions involve only double-trace operators with dimensions $\DB_n=2\Df+2n$ in the bosonic case and $\DF_n=2\Df+2n+1$ in the fermionic case. For concreteness, let us focus on the somewhat simpler fermionic case and drop the superscript on $\DF_{n}$. As we explain in detail in the main text, the function $F_{\Delta}(z)$ for general $\Delta\geq0$ can be written as a linear combination of $F_{\Dn}(z)$ and $\partial_{\Delta}F_{\Dn}(z)$ for $n\in\mathbb{N}_{\geq0}$:
\be
F_{\Delta}(z)=\sum_{n=0}^{\infty}\left[ \alpha_n(\Delta) F_{\Delta_n}(z)+\beta_n(\Delta) \partial F_{\Delta_n}(z)\right]\,.\label{eq:basisdecomp}
\ee
The general form of the expansion coefficients $\alpha_n(\Delta)$ and $\beta_n(\Delta)$ is rather involved.  Here $\partial F_{\Delta_n}(z)$ stands for the derivative with respect to $\Delta$ of $F_{\Delta}(z)$ evaluated at $\Delta=\Dn$. This equation tells us that if we add an operator of dimension $\Delta$ to the generalized free solution with a small OPE coefficient, then crossing symmetry can be preserved at this order by modifying the scaling dimensions and OPE coefficients of the double traces. 

One interpretation of this equation is that it expresses a general $F_{\Delta}$ in a basis spanned by the $F_{\Delta_n}$ and $\partial F_{\Delta_n}$.
The coefficients of the expansion above can then be obtained by acting on \reef{eq:basisdecomp} with the dual basis, which consists of linear functionals $\alpha_n, \beta_n$ acting on functions of complex variable $z$ and satisfying (for $n,m\in\mathbb{N}_{\geq0}$)
\ba
\label{eq:orthonorm}
\alpha_n(\Delta_m)&=\delta_{nm},&\qquad \alpha_n'(\Delta_m)&=0\\
\beta_n(\Delta_m)&=0,&\qquad \beta_n'(\Delta_m)&=\delta_{nm}\,,
\ea
where we use shorthand notation for the action of a functional $\omega$ on the $F_{\Delta}(z)$: $\omega(\Delta)\equiv \omega[F_{\Delta}]$, and $\omega'(\Delta) = \omega[\partial_{\Delta}F_{\Delta}]$ by linearity of $\omega$. We will show how to construct the functionals $\alpha_n$ and $\beta_n$ explicitly as contour integrals in the complex $z$-plane against appropriate holomorphic kernels. The integration contour probes both the Euclidean OPE limit and the u-channel Regge limit. In particular $\beta_0$ is the extremal functional for the gap maximization problem, constructed in \cite{Mazac:2016qev,Mazac:2018mdx}.

If we insert the decomposition \reef{eq:basisdecomp} into the crossing equation \reef{eq:origcross} and formally demand that the coefficient of each $F_{\Delta_n}(z)$ and $\partial_\Delta F_{\Delta_n}(z)$ vanishes, we find the {\em functional bootstrap equations}:
\be
\boxed{
\sum_{\cO\in\phi\times\phi}\!\! a_{\cO} \,\alpha_n(\Delta_{\cO})=0,\qquad \sum_{\cO\in\phi\times\phi}\!\! a_{\cO} \,\beta_n(\Delta_{\cO})=0\qquad \forall n\in \mathbb{N}_{\geq0}.
}\label{eq:fbes}
\ee
These equations can be derived rigorously by acting with the functionals $\alpha_n$ and $\beta_n$ on the crossing equation \reef{eq:origcross} and swapping them with the infinite sum over operators. This swapping property does not hold for a general functional (see \cite{Rychkov:2017tpc}) but it does for $\alpha_n$ and $\beta_n$. In the main text, we explain a close connection between the swapping property and boundedness in the Regge limit. In particular, \reef{eq:fbes} only hold for the OPE decomposition of Regge-bounded correlators (which includes all correlators in unitary theories). We will prove that \reef{eq:fbes} are a completely equivalent reformulation of the constraints contained in the original crossing equation \reef{eq:origcross}. However, they are much better suited for understanding some of its consequences.

There is another way of thinking about the decomposition \reef{eq:basisdecomp}, namely as expressing the crossing symmetry of the {\em Polyakov block}, defined by
\bea
P_{\Delta}(z)=G_{\Delta}(z)-\sum_{n=0}^{\infty}\left[ \alpha_n(\Delta) G_{\Delta_n}(z)+\beta_n(\Delta) \partial_\Delta G_{\Delta_n}(z)\right]\,.
\eea
The Polyakov block is just the usual conformal block $G_{\Delta}(z)$ ``dressed'' by double trace operators in order to obtain a crossing-symmetric object. In fact, $P_{\Delta}(z)$ is a sum of the s-, t- and u-channel Witten exchange diagrams in $AdS_2$. It is possible to see that if the functional bootstrap equations hold, then we can write
\be
\langle \phi(0)\phi(z)\phi(1)\phi(\infty)\rangle=\sum_{\cO\in\phi\times\phi}\!\!a_{\cO}\, \frac{G_{\Delta_{\cO}}(z)}{z^{2\Df}}=\sum_{\cO\in\phi\times\phi}\!\!a_{\cO}\, \frac{P_{\Delta_{\cO}}(z)}{z^{2\Df}}\,.
\ee
In other words, the correlator can be expanded in a way that makes crossing symmetry manifest. This is precisely the idea behind the Polyakov-Mellin bootstrap \cite{Polyakov:1974gs,Sen:2015doa,Gopakumar2017,Gopakumar2017a}. Our functionals $\alpha_n$ and $\beta_n$ thus provide a derivation of the $SL(2)$ version of the Polyakov-Mellin bootstrap from the standard crossing equation. In \cite{Gopakumar2017,Gopakumar2017a}, one needs to fix the contact-term ambiguity of the Witten exchange diagrams, which has not been done in full generality. We see that in our approach this ambiguity is fixed by demanding that the coefficients $\alpha_n$, $\beta_n$ arise from acting with linear functionals on the standard crossing equation.

Thanks to the orthonormality conditions \reef{eq:orthonorm}, the functionals allow us to solve for perturbations around generalized free fields. Starting with the original solution and perturbing by some fixed $\delta S$,
\bea
\sum_{n=0}^{\infty} {a_n} F_{\Delta_n}(z)=-F_0(z) \quad \Rightarrow \qquad \sum_{n=0}^{\infty}\left[ \delta a_n F_{\Delta_n}(z)+ a_n \delta \Delta_n \partial F_{\Delta_n}(z)\right]=\delta S(z),
\eea
we find simply
\bea
\delta a_n=\alpha_n[\delta S],\qquad a_n \delta \Delta_n=\beta_n[\delta S],
\eea
thus providing an analytic realization of the extremal flows introduced in \cite{El-Showk:2016mxr}.\footnote{Indeed, the generalized free fermion is an extremal solution to crossing, as it saturates the bound on the gap to the leading scaling dimension.} This procedure generalizes to higher orders of perturbation theory, so that in principle one can systematically correct the original solution to any desired order. In the main text, we use this idea to find the OPE decomposition of contact diagrams with external bosons and fermions in $AdS_2$. Carrying the procedure to higher orders, we bootstrap the one- and two-loop contribution to the four-point function in the $\phi^4$ theory in $AdS_2$.

Finally, the orthonormality properties \reef{eq:orthonorm} tell us that the action of the functionals on $F_{\Delta}$ will have double zeros at the double-trace $\Delta$. This implies interesting positivity properties of this action. Since $a_{\cO}$ are positive thanks to unitarity, we can use the functional bootstrap equations to derive sum rules which strongly constrain the OPE data. In particular, we will find both upper and lower bounds on weighted sums of OPE coefficients present between consecutive $\Dn$. Although nontrivial bounds exist for general $n$, their form simplifies as $n\rightarrow\infty$:
\begin{subequations}
\begin{align}
&\limsup_{n\rightarrow\infty}\sum_{\cO:\,|\DcO-\Delta_n|\leq 1}\!\!\!\frac{4\sin^2\!\left[\frac \pi 2 \left(\DcO-\Delta_n\right)\right]}{\pi^2(\DcO-\Delta_n)^2}\,\left(\frac{a_{\cO}}{a^{\textrm{free}}_{\DcO}}\right) \leq 1 \\
&\liminf_{n\rightarrow\infty}\sum_{\cO:\,|\DcO-\Delta_n|\leq 2} \frac{16\sin^2\!\left[\frac{\pi}2 (\DcO-\Delta_n)\right]}{\pi^2(\DcO-\Delta_n)^2(\DcO-\Delta_{n-1})(\Delta_{n+1}-\DcO)}\,\left(\frac{a_{\cO}}{a^{\textrm{free}}_{\DcO}}\right)\geq 1
\end{align}
\label{eq:bounds}
\end{subequations}
Here $a^{\textrm{free}}_{\Delta}$ is an exponentially decreasing function which coincides with the squared generalized free OPE coefficients at $\Delta=\Dn$ 
\be
a^{\textrm{free}}_{\Delta} = \frac{2 \Gamma (\Delta )^2 \Gamma (\Delta+2\Df -1)}{\Gamma (2\Df)^2 \Gamma (2 \Delta -1) \Gamma (\Delta-2\Df +1)}\,.
\ee
In the bounds above the ratio $a_{\cO}/a^{\textrm{free}}_{\DcO}$ is weighted by a positive function bounded above by 1. The bounds tell us that every unitary solution to crossing must be similar to the generalized free field in an appropriate sense at sufficiently large $\Delta$. The first bound essentially means that the total OPE coefficient between $\Dn-1$ and $\Dn+1$ is bounded from above by the mean field OPE coefficient at $\Dn$. This implies an upper bound on OPE coefficients of  \emph{individual} primaries. The second bound in particular implies that at sufficiently large $\Delta$, there must be at least one operator between $\Delta_{n-1}$ and $\Delta_{n+1}$! The inequalities are optimal since they are saturated by the generalized free field.

The outline of this paper is as follows. In the next section we make some general observations on the structure of the crossing equation and on the relationship between Regge boundedness and bootstrap functionals. Section \ref{sec:functionalsPolyakov} discusses in detail how the functional bases described above are related to Witten exchange diagrams and how they can be used to derive a rigorous version of the Polyakov-Mellin bootstrap. The actual construction of the functional basis is postponed to section \ref{sec:construction}. In section \ref{sec:bounds} we study the implications of the functional bootstrap equations in detail, using them to derive upper and lower bounds on the OPE data. The question of {\em completeness}, i.e. whether these equations are not only necessary but also sufficient to ensure that a putative set of OPE data solves crossing, is answered positively in section \ref{sec:completeness}. Section \ref{sec:witten} contains an application of the functional sum rules to bootstrapping tree-level, one- and two-loop Witten diagrams in $AdS_2$. We conclude with a short discussion and outlook. The paper is complemented by an appendix containing some technical details.

%%%%%%%%%%%%%%%%%%%%%%%%%%%%%%%%%%%%%%%
\section{The crossing region and bootstrap functionals}\label{sec:kinematics}
\subsection{The crossing region and analytic bootstrap limits}
In this paper, we will study some aspects of the crossing equation of the four-point function
\be
\langle\phi(x_1)\phi(x_2)\phi(x_3)\phi(x_4)\rangle = \frac{1}{|x_{12}|^{2\Df}|x_{34}|^{2\Df}}\mathcal{G}(z,\bar{z})\,.
\ee
We can take $\phi(x)$ to be a scalar primary operator in a unitary CFT. In fact, for most results of this paper, it will be sufficient if $\phi(x)$ is a primary under $SL(2)$ acting along a spacelike line. The four-point function can be expanded using the OPE in one of the three channels, which leads to constraints on the CFT data. For the present case of four identical scalars, the only independent constraint comes from the equality of the s- and t-channel expansions:
\be
(z\bar{z})^{-2\Df}\!\!\sum\limits_{\cO \in\phi\times\phi}\!\!a_{\cO}\,G_{\Delta_{\cO },J_{\cO }}(z,\bar{z}) =(z\leftrightarrow1-z,\bar{z}\leftrightarrow1-\bar{z})
\,.
\label{eq:crossing1}
\ee
Here $a_{\cO}\equiv (c_{\phi \phi \cO})^2$ are squared OPE coefficients and $G_{\Delta,J}(z,\bar{z})$ is the $d$-dimensional conformal block for the exchange of a symmetric traceless representation of dimension $\Delta$ and spin $J$. In the following, it will be convenient to write this equation as
\be
\sum\limits_{\cO \in\phi\times\phi}\!\!a_{\cO}\,F_{\Delta_{\cO },J_{\cO }}(z,\bar{z})=0\,,
\label{eq:crossing2}
\ee
where we defined
\be
F_{\Delta,J}(z,\bar{z}) = (z\bar{z})^{-\Df}G_{\Delta,J}(z,\bar{z}) - (z\leftrightarrow1-z,\bar{z}\leftrightarrow1-\bar{z})\,.
\ee
We will call the functions $F_{\Delta,J}(z,\bar{z})$ \emph{bootstrap vectors}, since the equation \reef{eq:crossing1} lives in a certain infinite-dimensional vector space of functions of $z$ and $\bar{z}$.

Eventually, we are going to specialize \reef{eq:crossing1} to the section $z=\bar{z}$ and expand both channels in the $SL(2)$ blocks, but first it will be useful to review the region in cross-ratio space where the full crossing equation holds. Let us start in the Euclidean signature, where $z$ and $\bar{z}$ are complex conjugate. The s-channel sum then converges whenever $z\notin[1,\infty)$ and the t-channel sum converges whenever $z\notin(-\infty,0]$. Therefore, in the Euclidean signature, the equation holds whenever $z$ stays away from $(-\infty,0]\cup[1,\infty)$. Moreover, the convergence of both the s-channel and t-channel sum is exponentially fast for every point in the interior of this region \cite{Pappadopulo:2012jk}.

Crucially, equation \reef{eq:crossing1} remains valid when we make $z$ and $\bar{z}$ complex and independent of each other. In order to understand the full region of validity, consider first the s-channel sum
\be
\mathcal{G}(z,\bar{z}) = \!\!\sum\limits_{\cO \in\phi\times\phi}\!\! a_{\cO}\, G_{\Delta_{\cO },J_{\cO }}(z,\bar{z})\,.
\label{eq:sExp}
\ee
The conformal blocks on the RHS are defined for $z$ and $\bar{z}$ complex and independent. Let us now switch to the $\rho,\bar{\rho}$ coordinates of \cite{Hogervorst:2013sma} defined by
\be
z = \frac{4\rho}{(1+\rho)^2}\,,\quad \bar{z} =  \frac{4\bar{\rho}}{(1+\bar{\rho})^2}\,.
\ee
The open unit $\rho$-disk maps to $z\in\mathbb{C}\backslash[1,\infty)$. As shown in \cite{Hartman:2015lfa}, conformal blocks have an expansion into monomials of the form $\rho^{h} \bar{\rho}^{\bar{h}}$ with positive coefficients, where $h,\bar{h}$ are generically non-integer powers. This expansion converges (to a multi-valued function!) whenever both $\rho$ and $\bar{\rho}$ are in the open unit disk. This is also the region of convergence of the sum over primary operators in \reef{eq:sExp}. We conclude that \reef{eq:sExp} converges as long as $z$ and $\bar{z}$ start in a Euclidean configuration and are continued from there such that neither $z$ or $\bar{z}$ passes through the cut at $[1,\infty)$. Similarly, the t-channel expansion converges as long as both $z$ and $\bar{z}$ both stay away from $(-\infty,0]$.

The conclusion is that the crossing equation \reef{eq:crossing1} is valid for $z$ and $\bar{z}$ both on the first sheet such that $(z,\bar{z})\in\mathcal{R}\times\mathcal{R}$, where
\be
\mathcal{R} = \mathbb{C}\backslash((-\infty,0]\cup[1,\infty))\,.
\ee
Correspondingly, we will call $\mathcal{R}\times\mathcal{R}$ the \emph{crossing region}. When we leave the first sheet through one of the branch cuts, either the s- or the t-channel OPE stops converging and the equation \reef{eq:crossing1} becomes meaningless. The crossing region includes the Euclidean section $z=\bar{z}^*\in\mathcal{R}$ as well as the Lorentzian section $0<z,\bar{z}<1$ with $z$ and $\bar{z}$ both on the first sheet. This is the region where all four operators stay spacelike separated.

Inside the crossing region, the sum on either side of \reef{eq:crossing1} converges to a function which is holomorphic in both $z$ and $\bar{z}$. This is because the individual conformal blocks are holomorphic in both $z$ and $\bar{z}$ and the sum converges uniformly inside any compact subregion of the crossing region. This means we can use the powerful tools of complex analysis to study the consequences of crossing.

The crossing region contains various interesting limits of the four-point function where analytic control is available. They correspond to $z$ and $\bar{z}$ approaching $0$, $1$ or $\infty$. These limits lie on the boundary of the crossing region, but need to be approached from its interior in order for the bootstrap equation to be valid. In particular, the point at infinity must be approached along a path away from the real axis.

The simplest are the OPE limits. The s- and t-channel OPE limits correspond to $(z,\bar{z})\rightarrow (0,0)$ and $(z,\bar{z})\rightarrow (1,1)$ respectively. The u-channel OPE limit corresponds to $(z,\bar{z})\rightarrow (i\infty,-i\infty)$ or equivalently $(z,\bar{z})\rightarrow (-i\infty,i\infty)$.\footnote{The four-point function is real in the Lorentzian subsection of the crossing region $0<z,\bar{z}<1$. It follows $\mathcal{G}(z^*,\bar{z}^*) = (\mathcal{G}(z,\bar{z}))^*$ inside the crossing region, where the star stands for complex conjugation. Furthermore, we can assume $\mathcal{G}(z,\bar{z}) = \mathcal{G}(\bar{z},z)$.} In this case $z$ and $\bar{z}$ lie in opposite half-planes since the u-channel OPE limit takes place in the Euclidean signature, where $z$ and $\bar{z}$ are complex conjugates. More generally, if $z$ and $\bar{z}$ approach $\infty$ in any direction away from the real axis but in opposite half-planes, the limit is controlled by the u-channel OPE.

Next, we have the standard double light-cone limit of the analytic light-cone bootstrap, where $x_{12}^2,x_{23}^2\rightarrow 0$, corresponding to $(z,\bar{z})\rightarrow (0,1)$. The other double light-cone limits are specified by $(z,\bar{z})\rightarrow (0,\pm i\infty),\,(1,\pm i\infty)$ and transpositions. For the four-point function of identical operators, these other limits do not carry any new information.
\begin{figure}[ht!]%
\begin{center}
\includegraphics[width=16cm]{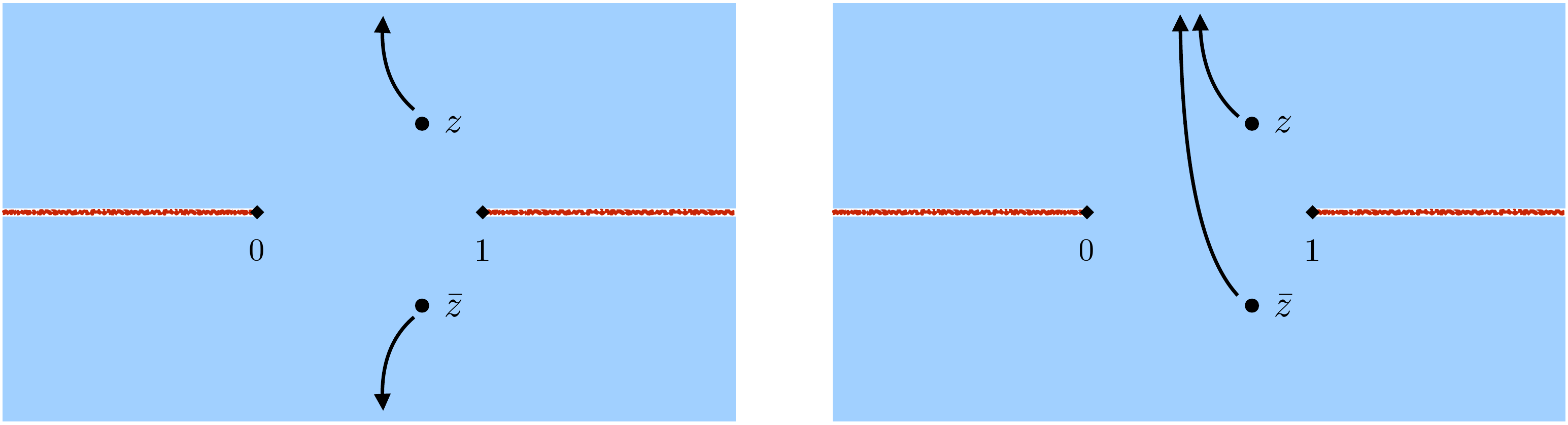}%
\caption{Left: The limit when $z$ and $\bar{z}$ go to infinity in opposite half-planes is controlled by the u-channel OPE. Right: When $z$ and $\bar{z}$ approach infinity in the same half-plane with $z/\bar{z}$ fixed, we get the Regge limit of the u-channel.}
\label{fig:ReggeU}%
\end{center}
\end{figure}

There is precisely one remaining limit where both $z$ and $\bar{z}$ approach $0$, $1$ or $\infty$ from within the crossing region which is not equivalent to one of the above. In this limit, we take $z$ and $\bar{z}$ both to $\infty$ in the same half-plane. This very interesting limit is controlled by the Regge limit of the u-channel. To understand this claim, note that the Regge limit of a given channel is defined in the same way as the OPE limit of that channel, after either $z$ or $\bar{z}$ has been taken around a crossed-channel branch cut. Thus in order to reach the Regge limit of the u-channel, we can start near the u-channel OPE limit where $|z|$ and $|\bar{z}|$ are very large with $z$, $\bar{z}$ in the upper, lower half-plane respectively. Let us keep $z$ fixed and move $\bar{z}$. The u-channel OPE converges as long as $\bar{z}$ stays away from $\bar{z}\in[0,1]$. In order to reach the u-channel Regge limit, we need to pass $\bar{z}$ through $[0,1]$ to the upper half-plane and send $z$ and $\bar{z}$ to $i\infty$.\footnote{Strictly speaking, in the u-channel Regge we take $z,\bar{z}\rightarrow i\infty$ with $z/\bar{z}$ fixed.} See Figure \ref{fig:ReggeU} for an illustration of how the u-channel Regge limit is reached. Both the s- and the t-channel OPE stay convergent during the entire process although they converge very slowly as we approach the u-channel Regge limit. In general, the OPE of a given channel does not converge in the Regge limit of that channel, while the remaining two OPEs converge slowly.

\subsection{Sum rules and functionals}
A fundamental problem in the conformal bootstrap is to use the crossing equation \reef{eq:crossing1} to extract useful constraints on the CFT data. Generally, such constraints take the form of sum rules which arise from applying linear functionals to \reef{eq:crossing2}. Indeed, given a linear functional $\omega$ acting on holomorphic functions of $z$ and $\bar{z}$ in the crossing region, and satisfying certain properties detailed below, we can apply it to the equation \reef{eq:crossing2} to find the sum rule
\be
\sum\limits_{\cO \in\phi\times\phi}\!\!a_{\cO}\,\omega(\Delta_{\cO },J_{\cO })=0\,,
\label{eq:crossingFunctional}
\ee
where we defined $\omega(\Delta,J)$ as the action of $\omega$ on the bootstrap vectors
\be
\omega(\Delta,J) = \omega[F_{\Delta,J}]\,.
\ee
For example, functionals can take the form of finite linear combinations of derivatives with respect to $z$ and $\bar{z}$ evaluated at the crossing-symmetric point $z=\bar{z}=1/2$, which is the usual strategy for the numerical bootstrap. However, more general functionals are possible. In particular, there are functionals which directly probe the interesting limits on the boundary of the crossing region described in the previous subsection.

Of particular interest are the so-called \emph{extremal functionals}, introduced in \cite{ElShowk:2012hu}. The extremal functionals give rise to optimal bounds on the CFT data. Since such bounds are often saturated by interesting strongly-coupled CFTs, the extremal functionals encode the precise manner in which the crossing equation may eventually lead to a nonperturbative solution of such theories.

The extremal functionals should generally be expected to probe the boundary of the crossing region, in particular the analytic bootstrap limits. This is the reason why numerical bootstrap bounds typically become optimal only when derivatives of arbitrarily high order are included. In the limit of infinitely many derivatives, the numerical bootstrap functionals can effectively see all the way to the boundary of the crossing region and probe the analytic bootstrap limits.

This expectation was confirmed in \cite{Mazac:2016qev} and \cite{Mazac:2018mdx}, where examples of extremal functionals were constructed analytically. The functionals take the form of holomorphic contour integrals stretching between the crossing-symmetric point and the boundary of the crossing region. In the present paper, we will generalize the construction and explain the crucial role of the Regge limit.

We will now give a precise definition of a general bootstrap functional. Since the functionals of the greatest interest probe the boundary of the crossing region, where the OPE sums stop converging, we need to be especially careful about which functionals lead to valid sum rules, as emphasized in \cite{Rychkov:2017tpc}. To set up the definition, it will be convenient to include the $(z\bar{z})^{-\Df}$ prefactor in the four-point function and write
\be
\widetilde{\mathcal{G}}(z,\bar{z}) = (z\bar{z})^{-2\Df}\mathcal{G}(z,\bar{z})
\label{eq:gTilde}
\ee
so that crossing symmetry reads
\be
\widetilde{\mathcal{G}}(z,\bar{z}) = \widetilde{\mathcal{G}}(1-z,1-\bar{z})\,.
\ee
Similarly, we define the normalized conformal blocks in the two channels
\ba
\widetilde{G}^{(s)}_{\Delta,J}(z,\bar{z}) &= (z\bar{z})^{-2\Df}G_{\Delta,J}(z,\bar{z})\\
\widetilde{G}^{(t)}_{\Delta,J}(z,\bar{z}) &= [(1-z)(1-\bar{z})]^{-2\Df}G_{\Delta,J}(1-z,1-\bar{z})\,.
\label{eq:gst}
\ea
We can now define a \emph{bootstrap functional} $\omega$ to be a linear functional acting on functions of $z$ and $\bar{z}$ which are holomorphic in both variables in the crossing region, where $\omega$ is subject to the following constraints
\begin{enumerate}
\item Finiteness on conformal blocks: $\omega[\widetilde{G}^{(s)}_{\Delta_,J}]<\infty$, $\omega[\widetilde{G}^{(t)}_{\Delta_,J}]<\infty$ for all unitary representations $(\Delta,J)$.
\item Finiteness on four-point functions: $\omega[\widetilde{\mathcal{G}}]<\infty$, where $\widetilde{\mathcal{G}}(z,\bar{z})$ is any crossing-symmetric four-point function in a unitary theory.
\item Swapping condition: Suppose $\widetilde{\mathcal{G}}(z,\bar{z})$ is a four-point function with conformal block expansions
\be
\widetilde{\mathcal{G}}(z,\bar{z}) = \!\!\sum\limits_{\cO \in\phi\times\phi}\!\!a_{\cO}\,\widetilde{G}^{(s)}_{\Delta_{\cO },J_{\cO }}(z,\bar{z})=\!\!\sum\limits_{\cO \in\phi\times\phi}\!\!a_{\cO}\,\widetilde{G}^{(t)}_{\Delta_{\cO },J_{\cO }}(z,\bar{z})\,,
\ee
where $a_{\cO}>0$. Then the sums
\be
\sum\limits_{\cO \in\phi\times\phi}\!\!a_{\cO}\,\omega[\widetilde{G}^{(s)}_{\Delta_{\cO },J_{\cO }}]\quad\textrm{and}\quad
\sum\limits_{\cO \in\phi\times\phi}\!\!a_{\cO}\,\omega[\widetilde{G}^{(t)}_{\Delta_{\cO },J_{\cO }}]
\ee
are absolutely convergent and equal to $\omega[\widetilde{\mathcal{G}}]$.
\end{enumerate}

When these three conditions are satisfied, the OPE decomposition of any unitary solution to crossing must satisfy \reef{eq:crossingFunctional}. While conditions 2 and 3 may appear at first sight like mere technicalities, we will see that they make contact with some interesting physics. The reason is that the functionals constructed in \cite{Mazac:2016qev,Mazac:2018mdx} and in this paper probe the u-channel Regge limit. As we soon review, four-point functions in unitary theories satisfy a boundedness property in this limit. This allows us to consider functionals whose action on physical four-point functions is finite but which diverge on more general functions that are unbounded in the Regge limit. Such functionals lead to valid sum rules for the CFT data which directly incorporate Regge boundedness.

\subsection{Functionals as contour integrals and Regge boundedness}

In the rest of this paper, we are going to specialize the crossing equation to the section $z=\bar{z}$. $z$ is still allowed to be complex and lie in $\mathcal{R}$. When $z$ is real and in the interval $(0,1)$, this corresponds to restricting the four operators to lie on a spacelike line. This means we have access to the s- and t-channel OPE limits, where $z\rightarrow0,1$. Furthermore, we also have access to the u-channel Regge limit, where $z\rightarrow i\infty$.\footnote{More precisely, we have access to the special case of the Regge limit where $z/\bar{z} = 1$.} It is a wonderful consequence of complex analyticity of the correlator that imposing the crossing equation for operators on a Euclidean line still gives us access to the (very Lorentzian) Regge limit.

Since we are effectively restricting the operators to lie in a line, we will simplify our analysis further by expanding the four-point function in the conformal blocks of the 1D conformal group $SL(2)$ acting along this line. The 1D conformal blocks take the form
\be
G_{\Delta}(z) = z^{\Delta}{}_2F_1(\Delta,\Delta;2\Delta;z)\,.
\ee
The equation that we will study in the rest of this paper is then
\be
\sum\limits_{\Delta}a_{\Delta}F_{\Delta}(z)=0\,,
\label{eq:crossing1D}
\ee
where the sum runs over the scaling dimensions of $SL(2)$ primaries in the OPE, $a_{\Delta}$ is the OPE coefficient squared of the primary and
\be
F_{\Delta}(z) = z^{-2\Df}G_{\Delta}(z) - (1-z)^{-2\Df}G_{\Delta}(1-z)\,.
\label{eq:fVector2}
\ee
Equation \reef{eq:crossing1D} is valid for any four-point function of identical primary operators in general spacetime dimension. For intrinsically 1D conformally-invariant systems, it encodes all information contained in crossing of a given four-point function.

The main technical result of this paper is the construction of an interesting basis for the space of functionals for the crossing equation \reef{eq:crossing1D}. We will work with the same general form of functionals introduced in \cite{Mazac:2016qev,Mazac:2018mdx}. The functionals are specified by a pair of locally holomorphic functions $f(z)$, $g(z)$ and take the form
\be
\omega[\mathcal{F}] = \frac{1}{2}\!\!\!\int\limits_{\frac{1}{2}}^{\frac{1}{2}+i\infty}\!\!\!\!\! dz f(z)\mathcal{F}(z) + 
\int\limits_{\frac{1}{2}}^{1}\!\!dz\,g(z)\mathcal{F}(z)\,,
\label{eq:ffg}
\ee
where $\mathcal{F}(z)$ is a general test function. This form explicitly connects various interesting corners of the crossing region. The first contour integral connects the numerical bootstrap limit $z=1/2$ with the Regge limit $z,\bar{z}\rightarrow i\infty$. The second contour connects the numerical bootstrap limit with the deep Euclidean limit $z,\bar{z}\rightarrow 1$. The sum rule following from the existence of $\omega$ takes the form
\be
\sum\limits_{\Delta} a_{\Delta}\,\omega(\Delta)=0\,,
\label{eq:sumRule1D}
\ee
where we use the shorthand notation
\be
\omega(\Delta)\equiv \omega[F_{\Delta}]\,.
\ee

As a side comment, the first part of the contour corresponds to kinematics which can be used to diagnose chaotic behaviour in out-of-time-order thermal correlators \cite{Maldacena:2015waa,Perlmutter:2016pkf}. Vacuum correlation functions of a CFT in flat space are related by a conformal transformation to thermal correlation functions of the CFT quantized on the hyperbolic space $H_{d-1}$ (see \cite{Roberts:2014ifa} or the above references). One can now compute a thermal correlation function, where the operators $\phi(x_1),\phi(x_2),\phi(x_3),\phi(x_4)$ are inserted in this order with equal distances around the thermal circle at zero Rindler time. This corresponds to the point $z=\bar{z}=1/2$. If we evolve operators $\phi(x_2)$ and $\phi(x_4)$ by the same Rindler time $t$, we get an out-of-time-order thermal correlator. This is equal to the flat-space vacuum correlator at cross-ratios
\be
z=\bar{z} =\frac{1+i \sinh(t)}{2}\,,
\ee
thus precisely tracing out the first contour in \reef{eq:ffg}. The $t\rightarrow\infty$ limit is the Regge limit $z\rightarrow i\infty$.

We would now like to explain more precisely in what sense our functionals probe the Regge limit of physical correlators. The essential fact is that four-point functions of unitary theories are bounded in the Regge limit. Specifically, with $\widetilde{\mathcal{G}}(z,\bar{z})$ normalized as in \reef{eq:gTilde}, we have
\be
\left|\widetilde{\mathcal{G}}\left(\mbox{$\frac{1}{2}+i t,\frac{1}{2}+i t$}\right)\right|\quad\textrm{is bounded as } t\rightarrow\infty\,.
\ee
The boundedness condition can not be improved as there are correlators which approach a constant in the Regge limit, for example the generalized free field. This means that for the functional \reef{eq:ffg} to take a finite value on physical four-point functions, we must have
\be
|f(z)|=O(z^{-1-\epsilon})\quad\textrm{as }z\rightarrow\infty
\label{eq:fCondition}
\ee
with $\epsilon>0$.\footnote{On the other hand, individual s- and t-channel conformal blocks decay in this limit as follows
\be
\widetilde{G}^{(s)}_{\Delta}(z) = O(\log(z)z^{-2\Df})\,,\widetilde{G}^{(t)}_{\Delta}(z) = O(\log(z)z^{-2\Df})\,.
\ee
It follows that assuming \reef{eq:fCondition} holds, the finiteness on conformal blocks and the swapping conditions are satisfied automatically, at least for the first contour integral in \reef{eq:ffg}.} In the examples of extremal functionals constructed in \cite{Mazac:2016qev,Mazac:2018mdx} and here, we have $f(z)=O(z^{-2})$ as $z\rightarrow\infty$ and thus the condition \reef{eq:fCondition} is satisfied with $\epsilon = 1$. Therefore the sum rule \reef{eq:sumRule1D} following from one of these functionals will restrict the correlator $\widetilde{\mathcal{G}}(z)$ to grow at most like $z^{1-\eta}$ for $\eta>0$ in the Regge limit. This behaviour is very different from the numerical bootstrap functionals which do not restrict the Regge behaviour of the correlators unless infinitely many derivatives are included.

In other words, our extremal functionals allow one to bootstrap crossing-symmetric correlators while making Regge boundedness manifest from the start. We will soon illustrate this on the example of contact and Witten exchange diagrams in $AdS_2$.

\section{Extremal functionals and the Polyakov bootstrap}\label{sec:functionalsPolyakov}
\subsection{The fermionic basis}\label{ssec:fermionicBasis}
We are now in a position to explain the main results of the present paper. In section \ref{sec:construction} we will construct two distinguished bases for the space of functionals for the crossing equation \reef{eq:crossing1D}. The two bases are associated to the theory of the generalized free boson and generalized free fermion respectively. The elements of the bases are generalizations of the analytic extremal functionals constructed in \cite{Mazac:2016qev,Mazac:2018mdx}. In particular, all of the basis functionals probe the Regge limit in the sense described in the previous subsection. It will turn out that expressing crossing in the bosonic basis provides a derivation of the $SL(2)$ version of the Polyakov's approach to the conformal bootstrap \cite{Polyakov:1974gs}, recently revisited in \cite{Sen:2015doa,Gopakumar2017,Gopakumar2017a,Dey:2016mcs,Dey:2017fab,Gopakumar:2018xqi}.

Let us focus on the fermionic case first. We claim that there exists a complete basis of functionals for the crossing equation \reef{eq:crossing1D}, which is the dual basis of the vectors $F_{2\Df+2n+1}(z)$ and their $\Delta$-derivatives $\partial F_{2\Df+2n+1}(z)$ with $n$ a non-negative integer. We refer to this as the fermionic basis since $\DF_n\equiv2\Df+2n+1$ is the spectrum of nonidentity primary operators in the $\phi\times\phi$ OPE, where $\phi$ is the generalized free fermion field.
In other words, we claim that for each $\Df\geq0$, there exists a set of bootstrap functionals that we denote as $\aF_n$ and $\bF_n$ (F in the superscript stands for fermion) such that 
\ba
&\aF_n[F_{\DF_m}]=\delta_{mn}\qquad \aF_n[\partial F_{\DF_m}]=0\\
&\bF_n[F_{\DF_m}]=0\qquad\quad\, \bF_n[\partial F_{\DF_m}]=\delta_{mn}
\label{eq:fBasisDef}
\ea
for $m,n\in\nnint$, where we simplified notation by writing $\partial F_{\Delta}(z)\equiv \partial_{\Delta} F_{\Delta}(z)$.\footnote{Here and in the rest of this paper, $\mathbb{N}$ stands for the set of \emph{non-negative} integers.} Thus $\aF_n$ is the functional dual to the vector $F_{\DF_n}(z)$ and $\bF_n$ is dual to $\partial F_{\DF_n}(z)$. Here and in the following, we will frequently denote the action of functionals on bootstrap vectors by the same symbol as the functionals themselves
\ba
\aF_n(\Delta)&\equiv\aF_n[F_{\Delta}],\\
\bF_n(\Delta)&\equiv\bF_n[F_{\Delta}]\,.
\ea
The duality conditions \reef{eq:fBasisDef} imply that $\aF_n(\Delta)$ has double zeros at the locations $\DF_m$ except for $\Delta=\DF_n$, where it is non-vanishing and has a vanishing first derivative. Similarly, $\bF_n(\Delta)$ has double zeros at the same locations except for $\Delta=\DF_n$, where it has a simple zero. In particular, $\bF_0$ is the extremal functional constructed in \cite{Mazac:2016qev,Mazac:2018mdx}, proving that the generalized free fermion four-point function maximizes the gap above identity among $SL(2)$-invariant unitary solutions to crossing. 

Functionals $\aF_n$ and $\bF_n$ are in fact uniquely fixed by the conditions \reef{eq:fBasisDef}. In particular, there is no bootstrap functional that has double zeros on the entire generalized free fermion spectrum. We will construct $\aF_n$ and $\bF_n$ explicitly in Section \ref{sec:construction} in the form \reef{eq:ffg}. We will find that for all basis functionals $f(z)=O(z^{-2})$ as $z\rightarrow\infty$.

Most of the rest of this paper will be devoted to exploring the consequences of the sum rules arising from applying $\aF_n$ and $\bF_n$ to the crossing equation
\be
\boxed{
\sum_{\Delta} a_{\Delta}\,\aF_n(\Delta)=0,\qquad \sum_{\Delta} a_{\Delta}\,\bF_n(\Delta)=0\qquad \forall n\in \mathbb{N}.
}\label{eq:fbes2}
\ee
These equations hold for every unitary solution to crossing. Furthermore, we claim not only that these equations are a consequence of the standard crossing equation \eqref{eq:crossing1D}, but also that they imply \eqref{eq:crossing1D}. In other words, if a putative set of OPE data consisting of $(\Delta_i,a_{\Delta_i})_{i=0,1,\ldots}$ with all $a_{\Delta_i}$ positive satisfies \eqref{eq:fbes2} for all $n$, then it leads to a crossing-symmetric four-point function. The proof of this claim is deferred until Section \ref{sec:completeness}.

We can get some intuition about the meaning of $\aF_n$ and $\bF_n$ by applying the sum rules in perturbation theory around the generalized free fermion. Let us assume that we have a continuous family of crossing-symmetric four-point functions $\mathcal{G}_g(z)$ parametrized by coupling $g$ such that it becomes the generalized free fermion for $g=0$
\be
\mathcal{G}_0(z) = 1+\left(\mbox{$\frac{z}{1-z}$}\right)^{2\Df}-z^{2\Df} =1+\sum\limits_{n=0}^{\infty}a_n^{(0)}G_{\DF_n}(z)\,,
\ee
where $a_n^{(0)} = a^{\textrm{free}}_{\DF_n}$ and
\be
a^{\textrm{free}}_{\Delta} = \frac{2 \Gamma (\Delta )^2 \Gamma (\Delta+2\Df -1)}{\Gamma (2\Df)^2 \Gamma (2 \Delta -1) \Gamma (\Delta-2\Df +1)}\,.
\ee
Let us further assume that making $g$ nonzero has only the following two effects at the leading order. First, a new vector $F_{\Delta_{\mathcal{O}}}$ with general $\DcO$ appears in the $\phi\times\phi$ OPE with coefficient $g$. Second, the double-trace operators acquire anomalous dimensions and anomalous OPE coefficients. These should be of order $g^2$ to match the effect of $\mathcal{O}$. In other words, $\mathcal{G}_g(z)$ admits the following OPE decomposition valid to $O(g^2)$:
\be
\mathcal{G}_g(z) = 1 + g^2\,G_{\DcO}(z)+ \sum\limits_{n=0}^{\infty}a_n(g) G_{\Delta_n(g)}(z)\,,
\ee
where the deformed OPE data can be expanded in $g$:
\ba
\Delta_n(g) &= \DF_n + \gamma_n^{(1)}g^2 + O(g^4)\\
a_n(g) &= a_n^{(0)} + a_n^{(1)} g^2+ O(g^4)\,.
\ea 
This leads to the following perturbative expansion of $\mathcal{G}_g(z)$
\be
\mathcal{G}_g(z) = \mathcal{G}^{(0)}(z) + \mathcal{G}^{(1)}(z)g^2+O(g^4)\,,
\ee
where
\be
\mathcal{G}^{(1)}(z) =G_{\DcO}(z)+ \sum\limits_{n=0}^{\infty}\left[a^{(1)}_n G_{\DF_n}(z) + a^{(0)}_n\gamma_n^{(1)} \partial G_{\DF_n}(z)\right]\,.
\ee
The coefficients $a_n^{(1)}$, $\gamma_n^{(1)}$ are constrained by crossing symmetry
\be
F_{\DcO}(z)+\sum\limits_{n=0}^{\infty}\left[a^{(1)}_n F_{\DF_n}(z) + a^{(0)}_n\gamma_n^{(1)}\partial F_{\DF_n}(z)\right] = 0\,.
\ee
We can solve for $a_n^{(1)}$ and $\gamma_n^{(1)}$ by applying the functionals $\aF_n$ and $\bF_n$. Strictly speaking, we can only use these functionals if $z^{-2\Df}\mathcal{G}^{(1)}(z)$ is bounded in the Regge limit. We know that $z^{-2\Df}\mathcal{G}(g;z)$ is bounded in the Regge limit for any $g$ since it is a unitary solution to crossing by assumption. However, it can happen that Regge boundedness is spoiled at individual orders in perturbation theory and is recovered only in the full finite-coupling answer. Here we will assume that this does not happen at $O(g^2)$, i.e. that $z^{-2\Df}\mathcal{G}^{(1)}(z)$ is bounded in the Regge limit. Functionals $\alpha^{\textrm{F}}_n$ and $\beta^{\textrm{F}}_n$ allow us to pick individual terms from the infinite sum and we find
\ba
a^{(1)}_n &= -\aF_n(\DcO)\\
a^{(0)}_n\gamma^{(1)}_n &= -\bF_n(\DcO)\,.
\ea
Hence the deformation is uniquely fixed at $O(g^2)$ in terms of $\Df$ and $\DcO$. There is a clear interpretation of this claim in terms of field theory in $AdS_2$. The generalized free fermion on the boundary is described by a free massive Majorana fermion $\Psi$ in $AdS_2$. Introducing $\mathcal{O}$ into the OPE is the same as turning on a three-point coupling $\Psi^2\mathcal{O}$ in the bulk with coupling proportional to $g$. $\mathcal{G}^{(1)}(z)$ is therefore equal to the sum of the s-, t- and u-channel Witten exchange diagrams in $AdS_2$, with fermionic bulk-to-boundary propagators of dimension $\Df$ and scalar bulk-to-bulk propagators of dimension $\DcO$. More details on the OPE decomposition of Witten exchange diagrams will be given when we discuss the bosonic case.

Leading-order deformations of the boundary four-point function which only deform the double traces correspond to four-point contact vertices in $AdS_2$. Since the deformation we found above was uniquely fixed, we just discovered using conformal bootstrap that the Majorana fermion in 2D admits no renormalizable four-point interactions. Indeed, $\Psi^4$ vanishes because of fermionic statistics and the simplest interaction is the irrelevant operator $(\Psi\partial\Psi)^2$. Irrelevant interactions lead to $z^{-2\Df}\mathcal{G}^{(1)}(z)$ which are not bounded in the Regge limit and which therefore do not solve the sum rules following from $\aF_n$ and $\bF_n$. In Section \ref{sec:witten}, we will explain how to use these functionals to bootstrap irrelevant interactions as well.

\subsection{The bosonic basis}
\label{ssec:bosonicBasis}

The other complete set of functionals that we construct is associated to the spectrum of the generalized free boson $\DB_n\equiv2\Df+2n$, $n\in\nnint$. An important subtlety arises in this case. By analogy with the fermionic case, we can attempt to construct $\aB_n$, $\bB_n$ satisfying
\ba
&\aB_n[F_{\DB_m}]=\delta_{mn}\qquad \aB_n[\partial F_{\DB_m}]=0\\
&\bB_n[F_{\DB_m}]=0\qquad\quad\, \bB_n[\partial F_{\DB_m}]=\delta_{mn}\,
\label{eq:bBasisDef}
\ea
in the form \reef{eq:ffg}. As we show in Section \ref{sec:construction}, this is possible but only if we relax the constraint on the Regge behaviour of $f(z)$. In fact, $f(z)$ approaches nonzero constants rather than being $O(z^{-2})$ as $z\rightarrow \infty$ for the functionals satisfying \reef{eq:bBasisDef}. This means $\aB_n$ and $\bB_n$ do not satisfy the finiteness on four-point functions and swapping criteria. However, we can cure this problem simply by taking linear combinations which improve the Regge behaviour of $f(z)$ so that $f(z) = O(z^{-2})$. In fact, $f(z)$ is meromorphic at infinity and satisfies $f(z)=f(1-z)$, so that a single subtraction is enough. For example, we can single out $\bB_0$ to improve the Regge behaviour of the remaining functionals and define
\ba
\alpha_n &= \aB_n-c_n \bB_0\\
\beta_n &= \bB_n-d_n\bB_0\,,
\label{eq:fbTilde}
\ea
where $c_n$ and $d_n$ are fixed by the requirement that the $f(z)$ kernels defining $\alpha_n$ and $\beta_n$ satisfy $f(z) = O(z^{-2})$. In practice, we find
\ba
d_n &= \frac{\left(\Delta _{\phi }\right)^4_n \left(4 \Delta _{\phi }-1\right)_{2 n}}{(n!)^2\left(2 \Delta _{\phi }\right)^2_n \left(4 \Delta _{\phi }+2 n-1\right)_{2 n}}\\
c_n &= \frac{1}{2} \partial_n d_n\,.
\label{eq:cdFormula}
\ea
With this definition, $\alpha_n$ for $n=0,1,\ldots$ and $\beta_n$ for $n=1,2,\ldots$ together form a complete set of consistent bootstrap functionals ($\beta_0$ vanishes identically since $d_0=1$). Of course, our choice to improve the Regge behaviour using $\bB_0$ is not canonical -- we could have used any $\aB_n$ or $\bB_n$ instead. However, it is clear that no matter how we choose to perform the improvement, we will never find a consistent functional which vanishes at all double traces $2\Df+2n$ and has a double zero at all but one double trace, unlike what happens in the fermionic case. To emphasize that $\aB_n$ and $\bB_n$ are not full-fledged bootstrap functionals, we refer to them as \emph{pre-functionals}.

There is a simple interpretation of these statements in terms of the crossing-symmetric deformations of the generalized free boson four-point function. Similarly to the fermionic case, we want to classify the deformations of $\mathcal{G}^{(0)}(z)$ which are bounded in the Regge limit, this time assuming that no new operators appear in the OPE at the leading order along the deformation.
\be
\mathcal{G}^{(0)}(z) = 1 + \left(\mbox{$\frac{z}{1-z}$}\right)^{2\Df} + z^{2\Df}
=1+\sum\limits_{n=0}^{\infty}a_n^{(0)}G_{\DB_n}(z)\,,
\label{eq:gfb}
\ee
where $a_n^{(0)}=a^{\textrm{free}}_{\DB_n}$. We parametrize the scaling dimensions and OPE coefficients of deformed double-trace operators as follows
\ba
\Delta_n(g) &= \DB_n + \gamma_n^{(1)}g^2 + O(g^4)\\
a_n(g) &= a_n^{(0)} + a_n^{(1)} g^2+ O(g^4)\,,
\ea
so that the crossing equation at order $g^2$ becomes
\be
\sum\limits_{n=0}^{\infty}\left[a^{(1)}_n F_{\DB_n}(z) + a^{(0)}_n\gamma_n^{(1)}\partial F_{\DB_n}(z)\right] = 0\,.
\label{eq:bDefCrossing}
\ee
Had $\aB_n$ and $\bB_n$ been consistent bootstrap functionals, we could apply them to this equation and conclude the generalized free boson admits no deformations of the kind we are interested in. However, we know that only $\alpha_n$, $\beta_n$ are consistent. Applying them to \reef{eq:bDefCrossing}, we can solve for $\gamma_n^{(1)}$ and $a_n^{(1)}$ up to an overall constant
\ba
a_n^{(0)}\gamma_n^{(1)} &= d_n a_0^{(0)}\gamma_0^{(1)}\\
a_n^{(1)} &= c_n a_0^{(0)}\gamma_0^{(1)}\,,
\label{eq:contact2D}
\ea
with $c_n$ and $d_n$ given in \reef{eq:cdFormula}. The overall constant can be absorbed into the coupling $g^2$ and therefore we find precisely one Regge-bounded deformation. Its OPE decomposition takes the form
\be
A(z) = \sum\limits_{n=0}^{\infty}\left[c_n G_{\DB_n}(z)+d_n \partial G_{\DB_n}(z)\right]\,.
\label{eq:contactOPE}
\ee

These results exactly agree with the expectation from field theory in $AdS_2$. The generalized free boson \reef{eq:gfb} is described by the free real massive scalar field $\Phi$ in $AdS_2$. Crossing-symmetric deformations which involve only corrections to the double traces correspond to quartic vertices. Boundedness of the deformation in the $z\rightarrow i\infty$ limit restricts us to relevant interactions. The only relevant quartic interaction that we can write down is $\Phi^4$. Indeed, one can check that \reef{eq:contactOPE} exactly agrees with the OPE decomposition of the corresponding tree-level Witten contact diagram.

We can generalize this logic to bootstrap higher-derivative contact diagrams in $AdS_2$. More derivatives in the vertex translate into faster polynomial growth of the diagram in the Regge limit. Therefore, functionals of the form \reef{eq:ffg} which are consistent with higher-derivative diagrams need to have $f(z)$ polynomially suppressed by higher inverse powers of $z$ in this limit. We can construct such functionals by taking further linear combinations of the elementary pre-functionals $\aB_n$ and $\bB_n$. Roughly speaking, improving $f(z)$ by a factor $z^{-2}$ costs us one dimension of the space of functionals. This reduction introduces one new dimension in the space of allowed solutions to crossing, corresponding to a new contact diagram. In this way, we can bootstrap all higher-derivative contact interactions, order-by-order in the number of derivatives. We give more details in Section \ref{sec:witten}, where we also explain how to generalize the above procedure to higher orders in $g$ to compute higher-loop diagrams in $AdS_2$.

\subsection{Bootstrapping Witten exchange diagrams}
The functionals $\alpha_n$ and $\beta_n$ defined in \reef{eq:fbTilde} can be given a nice physical interpretation in terms of Witten exchange diagrams in $AdS_2$. This connection also provides a derivation of the Polyakov approach to the conformal bootstrap directly from the position-space crossing equation.

Let us consider Witten exchange diagrams in $AdS_2$. We will denote the diagram for the s-channel exchange of a field with dimension $\Delta$ as
\be
\mathcal{W}^{(s)}_{\Delta}(x_1,x_2,x_3,x_4) = \frac{1}{|x_{12}|^{2\Df}|x_{34}|^{2\Df}}W^{(s)}_{\Delta}(z)\,.
\ee
A priori, $W^{(s)}_{\Delta}(z)$ is defined for $z\in\mathbb{R}$, and has non-analyticities at $z=0,1,\infty$ corresponding to collisions of pairs of boundary operators. Thus $W^{(s)}_{\Delta}(z)$ gives rise to three complex-analytic functions of $z$ -- the analytic continuations of $W^{(s)}_{\Delta}(z)$ from the regions $(-\infty,0)$, $(0,1)$ and $(1,\infty)$. The t- and u-channel exchanges can be obtained from the s-channel exchange by transposing the external points
\ba
&\mathcal{W}^{(t)}_{\Delta}(x_1,x_2,x_3,x_4) = \mathcal{W}^{(s)}_{\Delta}(x_1,x_4,x_3,x_2)\\
&\mathcal{W}^{(u)}_{\Delta}(x_1,x_2,x_3,x_4) = \mathcal{W}^{(s)}_{\Delta}(x_1,x_3,x_2,x_4)\,,
\ea
which leads to the relations
\ba
&W^{(t)}_{\Delta}(z) = \left|\mbox{$\frac{z}{1-z}$}\right|^{2\Df} W^{(s)}_{\Delta}(1-z)\\
&W^{(u)}_{\Delta}(z) = |z|^{2\Df} W^{(s)}_{\Delta}(1/z)\,.
\ea
We will be interested in $W^{(s,t,u)}_{\Delta}(z)$ for $z\in(0,1)$ and their analytic continuation from there to the whole crossing region $\mathcal{R}$. Let us expand these functions in the s-channel OPE. $W^{(s)}_{\Delta}(z)$ contains the ``single-trace'' conformal block of dimension $\Delta$ and double traces of dimensions $\DB_n$ and their $\Delta$-derivatives
\be
W^{(s)}_{\Delta}(z) = G_{\Delta}(z) + \sum\limits_{n=0}^{\infty}\left[\mu^{(s)}_{n}(\Delta)G_{\DB_n}(z) + \nu^{(s)}_{n}(\Delta)\partial G_{\DB_n}(z)\right]\,,
\ee
where we normalized the diagram so that the single trace block appears with a unit coefficient. The s-channel OPE of the t- and u-channel exchange diagrams contains double traces of dimensions $2\Df+j$ with $j\in\nnint$ and their $\Delta$ derivatives
\ba
W^{(t)}_{\Delta}(z) &= \sum\limits_{j=0}^{\infty}\left[\mu^{(t)}_{j}(\Delta)G_{2\Df+j}(z) + \nu^{(t)}_{j}(\Delta)\partial G_{2\Df+j}(z)\right]\\
W^{(u)}_{\Delta}(z) &= \sum\limits_{j=0}^{\infty}(-1)^j\left[\mu^{(t)}_{j}(\Delta)G_{2\Df+j}(z) + \nu^{(t)}_{j}(\Delta)\partial G_{2\Df+j}(z)\right]\,,
\ea
where the equality of their OPE decomposition up to the $(-1)^j$ factor is a consequence of a symmetry exchanging $x_1$ and $x_2$. We will be interested in the crossing-symmetric sum of the exchange diagrams
\be
W^{\textrm{all}}_{\Delta}(z) = W^{(s)}_{\Delta}(z)+W^{(t)}_{\Delta}(z)+W^{(u)}_{\Delta}(z)\,.
\ee
This function satisfies
\be
W^{\textrm{all}}_{\Delta}(z) = \left(\mbox{$\frac{z}{1-z}$}\right)^{2\Df} W^{\textrm{all}}_{\Delta}(1-z)
\ee
in the crossing region $\mathcal{R}$. Its OPE decomposition takes the form
\be
W^{\textrm{all}}_{\Delta}(z) = G_{\Delta}(z) + \sum\limits_{n=0}^{\infty}\left[\mu^{\textrm{all}}_n(\Delta)G_{\DB_n}(z) + \nu^{\textrm{all}}_{n}(\Delta)\partial G_{\DB_n}(z)\right]\,.
\ee
Note in particular that the double traces with dimensions $2\Df$ plus odd integers cancelled between the t- and u-channel exchange.

We would like to bootstrap the coefficients $\mu^{\textrm{all}}_n(\Delta)$, $\nu^{\textrm{all}}_n(\Delta)$ using the complete set of bosonic functionals from the previous subsection. For that to work, we need to make sure $W^{\textrm{all}}_{\Delta}(z)$ is bounded in the Regge limit $z\rightarrow i\infty$. It is possible to show, for example using the Mellin representation, that $W^{\textrm{all}}_{\Delta}(z)$ satisfies
\be
W^{\textrm{all}}_{\Delta}(z) \sim \frac{w(\Df,\Delta)}{z}\quad\textrm{as } z \rightarrow i\infty
\ee
for some constant $w(\Df,\Delta)$. The $\Phi^4$ contact diagram in $AdS_2$ has the same asymptotic behaviour $A(z)\sim a(\Df) z^{-1}$ in this limit. Therefore, there is a linear combination of $W^{\textrm{all}}_{\Delta}(z)$ and $A(z)$ which decays even faster in the Regge limit
\be
\widehat{W}^{\textrm{all}}_{\Delta}(z) = W^{\textrm{all}}_{\Delta}(z) - \frac{w(\Df,\Delta)}{a(\Df)}A(z)\,,
\ee
so that $\widehat{W}^{\textrm{all}}_{\Delta}(z) = O(z^{-2})$ as $z\rightarrow i\infty$. Since the contact diagram $A(z)$ contains only double traces, the structure of the OPE decomposition is unchanged
\be
\widehat{W}^{\textrm{all}}_{\Delta}(z) = G_{\Delta}(z) + \sum\limits_{n=0}^{\infty}\left[\widehat{\mu}^{\textrm{all}}_n(\Delta)G_{2\Df+2n}(z) + \widehat{\nu}^{\textrm{all}}_{n}(\Delta)\partial G_{2\Df+2n}(z)\right]\,.
\ee
The crossing symmetry of $\widehat{W}^{\textrm{all}}_{\Delta}(z)$ translates into
\be
F_{\Delta}(z) + \sum\limits_{n=0}^{\infty}\left[\widehat{\mu}^{\textrm{all}}_n(\Delta)F_{2\Df+2n}(z) + \widehat{\nu}^{\textrm{all}}_{n}(\Delta)\partial F_{2\Df+2n}(z)\right] = 0\,.
\ee
Because of the $O(z^{-2})$ decay of $\widehat{W}^{\textrm{all}}_{\Delta}(z)$ in the Regge limit, we can apply the pre-functionals $\aB_n$ and $\bB_n$ to this equation and swap them with the infinite sum over the double traces. Thanks to the duality of pre-functionals and double-traces \reef{eq:bBasisDef}, only the action on the first term and a single term of the infinite sum survives. We find
\ba
\widehat{\mu}^{\textrm{all}}_n(\Delta) &= - \aB_n[F_{\Delta}]\\
\widehat{\nu}^{\textrm{all}}_n(\Delta) &= - \bB_n[F_{\Delta}]\,.
\ea
The OPE coefficients of double traces in the Regge-improved crossing-symmetric sum of Witten exchange diagrams $\widehat{W}^{\textrm{all}}_{\Delta}(z)$ are precisely given by the pre-functional actions on the bootstrap vectors!

\subsection{Derivation of the Polyakov bootstrap}
We are now one step away from deriving the Polyakov-Mellin approach to the conformal bootstrap for $SL(2)$. Let us first review its basic idea. One starts by postulating the existence of distinguished functions $P_{\Delta}(z)$, which we will call Polyakov blocks. These blocks are required to be crossing-symmetric
\be
P_{\Delta}(z) =  \left(\mbox{$\frac{z}{1-z}$}\right)^{2\Df} P_{\Delta}(1-z)\,.
\ee
Furthermore, the OPE decomposition of these blocks is required to contain the single trace conformal block $G_{\Delta}(z)$ (with coefficient one), as well as double trace conformal blocks $G_{\DB_n}(z)$ and their $\Delta$-derivatives $\partial G_{\DB_n}(z)$ (with coefficients that may depend on $\Df$ and $\Delta$). Finally, and most nontrivially, it is required that for any crossing-symmetric four-point function $\mathcal{G}(z)$ in a unitary theory, we can replace the conformal blocks in its OPE decomposition with the Polyakov blocks without changing the result
\be
\mathcal{G}(z) =\sum\limits_{\Delta}a_{\Delta} G_{\Delta}(z)
= \sum\limits_{\Delta} a_{\Delta} P_{\Delta}(z)\,.
\label{eq:pBootstrap}
\ee
Since the latter expansion is manifestly crossing-symmetric, conformal bootstrap is transformed into the statement that all the double-trace contributions to the individual Polyakov blocks drop out after performing the sum over the physical spectrum on the RHS of the above equation.

A priori, it is not at all clear that objects $P_{\Delta}(z)$ with the above highly-constraining properties should exist. However, we can construct them straightforwardly using the complete set of bosonic functionals. Indeed, we can take
\be
P_{\Delta}(z) = G_{\Delta}(z) - \sum\limits_{n=0}^{\infty}\left[\alpha_n(\Delta)G_{\DB_n}(z) + \beta_n(\Delta)\partial G_{\DB_n}(z)\right]\,.\label{eq:pblock}
\ee
Since $\alpha_n$ and $\beta_n$ are related to the prefunctionals $\aB_n$ and $\bB_n$ via \reef{eq:fbTilde}, this expression is related to the Regge-improved sum of Witten exchange diagrams as follows
\be
P_{\Delta}(z) = \widehat{W}^{\textrm{all}}_{\Delta}(z) + \bB_0(\Delta) A(z)\,,
\label{eq:pSum}
\ee
where $A(z)$ is the $\phi^4$ contact diagram with OPE decomposition \reef{eq:contactOPE}. Hence, $P_\Delta(z)$ is indeed crossing-symmetric since both summands in \reef{eq:pSum} are. Moreover, the equations which express the cancellation of unphysical double-trace contributions on the RHS of \reef{eq:pBootstrap} take the form
\be
\boxed{
\sum_{\Delta}a_{\Delta} \alpha_n(\Delta)=0,\qquad \sum_{\Delta} a_{\Delta} \beta_n(\Delta)=0\qquad \forall n\in \mathbb{N}.
}
\ee
These equations are satisfied in every unitary solution to crossing by construction since $\alpha_n$ and $\beta_n$ are consistent bootstrap functionals. Recall that $\beta_{0} = 0$ identically so that the $n=0$ equation for $\beta$ is satisfied trivially.

In summary, the Polyakov bootstrap equations are the usual bootstrap equations expressed in the basis of functionals $\alpha_n$ and $\beta_n$.

In practice, the quickest way to find $P_{\Delta}(z)$ is to start with $W^{\textrm{all}}_{\Delta}(z)$ without any contact term improvements, and add such multiple of $A(z)$ that precisely cancels $\partial G_{2\Df}(z)$ in the OPE decomposition. Note that just like there is no canonical choice of a basis for the bosonic functionals, there is no canonical choice of the Polyakov blocks. Indeed, one gets equally valid Polyakov blocks from a linear combination of $W^{\textrm{all}}_{\Delta}(z)$ and $A(z)$ where any fixed double trace term is absent, not necessarily $\partial G_{2\Df}(z)$. In general, two consistent choices of the Polyakov blocks will differ by the contact diagram times $\omega[F_{\Delta}]$, where $\omega$ is a bootstrap functional.

The situation in the fermionic case is even simpler. Since $\alpha^{\textrm{F}}_n$ and $\beta^{\textrm{F}}_n$ are consistent bootstrap functionals without any subtractions, we can define the fermionic Polyakov block
\be
P^{\textrm{F}}_{\Delta}(z) = G_{\Delta}(z) - \sum\limits_{n=0}^{\infty}\left[\alpha^{\textrm{F}}_n(\Delta)G_{\DF_n}(z) + \beta^{\textrm{F}}_n(\Delta)\partial G_{\DF_n}(z)\right]\,,
\ee
and the four-point function then satisfies
\be
\mathcal{G}(z) =\sum\limits_{\Delta}a_{\Delta} G_{\Delta}(z)
=\sum\limits_{\Delta} a_{\Delta} P^{\textrm{F}}_{\Delta}(z)\,.
\ee
As discussed above, $P^{\textrm{F}}_{\Delta}(z)$ is the sum of Witten exchange diagrams in the s-, t- and u-channel, where the bulk-to-boundary propagators are fermionic of dimension $\Df$ and the bulk-to-bulk propagators are bosonic of dimension $\Delta$.

The crossing equation in $D>1$ has been used to bootstrap the crossed-channel OPE decomposition of Witten exchange diagrams in $AdS_{D+1}$ in some cases in \cite{Alday:2017gde,Li:2017lmh}. Rather general formulas for the OPE decompositions were found using Mellin-space techniques in \cite{Sleight:2018ryu,Gopakumar:2018xqi}. For $D=2,4$ a closed formula for the coefficient function of the crossed-channel Witten exchange diagrams was found in \cite{Liu:2018jhs} by applying the Lorentzian OPE inversion formula of Caron-Huot \cite{Caron-Huot2017b} to a single crossed-channel conformal block. The direct-channel exchange contributes a part which is not analytic in spin, and is therefore not captured by the standard inversion formula. Functionals of this note automatically incorporate also the direct channel exchange. There is in fact a Lorentzian inversion formula for the principal series of $SL(2)$ which reproduces the full crossing-symmetric sum of Witten exchange diagrams when applied to a single crossed-channel conformal block \cite{Mazac:2018qmi}. We checked our results for the OPE coefficients in Witten exchange diagrams with explicit computations whenever possible.\footnote{We thank Xinan Zhou for providing us with a draft of \cite{Zhou:2018sfz} to facilitate some of these checks.}

Before we conclude this section, we should note that there is an an important subtlety which we skimmed in the above. This is the fact that the equivalence between the Polyakov approach and the functional bootstrap equations is only guaranteed if we are allowed to commute the series running over the functional label $n$ with that over $\Delta$ upon inserting \reef{eq:pblock} into \reef{eq:pBootstrap}. That this is indeed true follows from an upper bound on the OPE data derived in section \ref{sec:bounds}, and will be proven in section \ref{sec:completeness}.

%%%%%%%%%%%%%%%%%%%%%%%%%%%%%%%%%%%%%%%

\section{Construction of the Dual Basis}\label{sec:construction}

\subsection{Functionals for generalized free theory}
We will now find the functional bases with all the properties that were described in previous sections. We will use the construction \cite{Mazac:2018mdx} (which itself builds on \cite{Mazac:2016qev}), to which we refer the reader for further details, and which we now shortly review.\footnote{A different perspective on this construction will be given in \cite{Mazac:2018qmi} in terms of a Lorentzian OPE inversion formula for the principal series of $SL(2)$.}

We begin with the general class of functionals given in equation \eqref{eq:ffg} and consider their action on the $F_{\Delta}$:
\be
\omega[F_{\Delta}] = \frac{1}{2}\!\!\!\int\limits_{\frac{1}{2}}^{\frac{1}{2}+i\infty}\!\!\!\!\! dz f(z)F_{\Delta}(z) + 
\int\limits_{\frac{1}{2}}^{1}\!\!dz\,g(z)F_{\Delta}(z)\,,
\label{eq:ffg2}
\ee
The pair of kernels $f,g$ are assumed to be holomorphic in the upper-half plane, and real along their respective contours of integration in the above definition. It is useful to extend $f(z)$ into the lower half-plane by setting $f(z)=f(1-z)$. The pair $f,g$ should satisfy the so-called gluing condition,
\bea
\mbox{Re} f(z)=-g(z)-g(1-z) \label{eq:gluing}
\eea
for $z\in(0,1)$, which essentially tells us that $f(z),g(z)$ arise as discontinuities of a single kernel $h(z)$. We would like for the functional action $\omega(\Delta)$ to be an oscillating function of $\Delta$, and in particular that it should have double zeros for $\Delta=\Delta_n^{\textrm{B}}, \Delta_n^{\textrm{F}}$ depending on whether we want functionals dual to the generalized free boson or fermion respectively. The trick is to use that
\bea
\lim_{\epsilon\to 0^+}\,\frac{G_\Delta(z+i\epsilon)}{(z+i\epsilon)^{2\Df}}=e^{i\pi (\Delta-2\Df)} \frac{G_\Delta\left(\frac{z}{z-1}\right)}{(-z)^{2\Df}}, \qquad \text{for}\, z<0,
\eea
to obtain the desired oscillations. Let us set
\bea
g(z)=\eta\, (1-z)^{2\Df-2} f\left(\mbox{$\frac{1}{1-z}$}\right),\qquad 0<z<1\label{eq:gfromf},
\eea
with in particular $f(z)$ real for $z\in \mathbb R\backslash (0,1)$, and set $\eta=1,-1$ for the bosonic, fermionic cases respectively. Following \cite{Mazac:2018mdx}, a simple contour-deformation argument gives us
\bea
\omega(\Delta)\equiv\omega[F_\Delta]=\left[1-\eta \cos \pi(\Delta-2\Df)\right] \mathfrak g(\Delta), \qquad \mathfrak g(\Delta)= \int_0^{1} \ud z\,g(z)\,\frac{G_{\Delta}(z)}{z^{2\Df}},
\label{eq:funcact}
\eea
which has the desired double zero structure. For $z\in(0,1)$ equation \reef{eq:gluing} together with \reef{eq:gfromf} implies the {\em fundamental free equation}:
\bea
\boxed{
\eta \,\mbox{Re}\, f(z)
=- (1-z)^{2\Df-2} f\left(\mbox{$\frac{1}{1-z}$}\right)-z^{2\Df-2} f\left(\mbox{$\frac{1}{z}$}\right).
}
\eea
This is the equation that the functionals dual to the generalized free solutions must obey.

\subsection{General solution}
We will now find a full set of solutions to the fundamental free equation subject to appropriate boundary conditions, which we will discuss in detail. As it turns out the actual construction of the solutions will then be fairly simple thanks to a nice property of the equation which allows us to start with one particular solution and derive from it an infinite set of other solutions.

\subsubsection{Boundary conditions}

We begin by discussing constraints on the behaviour of $f(z)$ as $z\to \infty$. One condition arises from demanding that the functional action \reef{eq:ffg2} on the $F_{\Delta}$ should be finite for $\Delta\geq 0$. This imposes in particular that $f(z)$ should decay sufficiently fast as $z$ approaches infinity. For $\Delta \geq 0$, the asymptotics of the $F_\Delta$ imply that we need $f(z)=O(z^{2\Df-\epsilon})$ for some positive $\epsilon$. However, the swapping condition \cite{Rychkov:2017tpc}, which is the requirement that the functional action should commute with (crossing-symmetric) infinite sums of $F_{\Delta}$, actually requires the stronger falloff $z^{-1-\epsilon}$ for some $\epsilon>0$. Since $f(z)$ is analytic for $z$ away from $(0,1)$, we must have that $f(z)$ falls off like an integer power of $z^{-1}$ greater or equal than two. Note that using \reef{eq:gfromf} and the assumed falloff of $f(z)$ we find $g(z)=O[(1-z)^{2\Df}]$ as $z\to 1^-$ and this is sufficient to guarantee that the contribution of $g(z)$ to the functional action is also finite. The swapping condition adds no further constraints on $g(z)$. 
 
Next we discuss the behaviour of $f(z)$ as $z\to 1^+$. A generic solution of the fundamental free equation which is sufficiently bounded at infinity will be divergent as $z\to 1^+$. There are two classes of behaviour, labelled by the presence or absence of a leading logarithmic divergence, along with a specific power law divergence. We can fix this divergent structure by a simple argument. Note that the contour deformation leading to the functional action \reef{eq:funcact} is generically not allowed, since the integral computing $\mathfrak g(\Delta)$ might be divergent. However, since the original functional action was definitely finite, this divergence must cancel against a double zero. The allowed behaviours are hence
\begin{subequations}
\label{eq:possibledivs}
\begin{align}
f(z) &\stackrel{z\to 1^+}{\sim} \frac{a_0 \log(z-1)+b_0}{(z-1)^{2+2m}},\qquad &\eta&=-1& (F)\label{eq:fermiondiv}\\
f(z) &\stackrel{z\to 1^+}{\sim} \frac{a_0 \log(z-1)+b_0}{(z-1)^{1+2m}},\qquad &\eta&=1& (B)\label{eq:bosondiv}
\end{align}
\end{subequations}
with $m \in \mathbb N$.\footnote{Actually, this argument only shows the weaker $m\in \mathbb Z$. For $a_0=0$ positivity of $f(z)$ together with the right falloff at infinity set $m\geq 0$. We do not however have a general proof of this statement.} If $a_0=0$, we say the functionals are of type $\beta$, otherwise of type $\alpha$. Note that a functional with a given $m$ is only defined up to addition of lower $m$ ones. Furthermore, $\alpha$-type functionals are also ambiguous to the addition of a $\beta$ functional with the same $m$. The functionals are characterized by the fact that both have double zeros for $\Delta_p$ with $p>m$, while for $\Delta_m$ the $\beta_m$ functionals have simple zeros and the $\alpha_m$ functionals are non-zero.

\subsubsection{General solution}
Let us denote functionals by $\alpha_m^\eta, \beta_m^{\eta}$  and write $f(z)\to f^{\Df,\eta}_{\alpha_m}, f^{\Df,\eta}_{\beta_m}$ for the corresponding $f(z)$ kernels, keeping in mind that $\eta=1,-1$ corresponds to bosons, fermions respectively. Previously we constructed those solutions with $\eta=-1$ for all $\Delta_\phi$  \cite{Mazac:2018mdx}. We recall their form:
\begin{subequations}
\ba
f_{\beta_0}^{\Df,-}(z)=
-\kappa(\Df)
&\frac{2z-1}{w^{3/2}}\left[\, _3\widetilde{F}_2\left(-\frac{1}{2},\frac{3}{2},2
   \Df+\frac{3}{2};\Df+1,\Df+2;-\frac{1}{4 w}\right)+\right.\\
	&\,\;\;\left.+\frac{9}{16 w} \,
   _3\widetilde{F}_2\left(\frac{1}{2},\frac{5}{2},2 \Df+\frac{5}{2};\Df+2,\Df+3;-\frac{1}{4w}\right)\right],
\label{eq:allaf}
\ea
\ba
f_{\alpha_0}^{\Df,-}(z)=
\kappa(\Df)\frac{2(z-2)(z+1)}{(2z-1)w^{3/2}}
&\left[
{}_3\widetilde{F}_2\left(-\frac{1}{2},-\frac{1}{2},2\Df +\frac{3}{2};\Df +2,\Df +2;-\frac{1}{4 w}\right)+\right.\\
+\frac{(2 \Df +3) (2 \Df +5)}{16 w} &{}_3\widetilde{F}_2\left(\frac{1}{2},\frac{1}{2},2 \Df +\frac{5}{2};\Df +3,\Df +3;-\frac{1}{4 w}\right)-\\
-\frac{3 (4 \Df +5)}{256 w^2}&\left.{}_3\widetilde{F}_2\left(\frac{3}{2},\frac{3}{2},2 \Df +\frac{7}{2};\Df +4,\Df +4;-\frac{1}{4 w}\right)\right].
\label{eq:fAlphaTilde}
\ea
\end{subequations}
Here ${}_3\widetilde F_2$ stands for the regularized hypergeometric function, $w=z(z-1)$ and the normalization factor reads
\be
\kappa(\Df) = \frac{\Gamma(4\Df+4)}{2^{8\Df+5}\Gamma(\Df+1)^2}\,.
\ee
These solutions falloff as $z^{-2}$ as required, and satisfy \reef{eq:fermiondiv} with $m=0$. The ambiguity in $\alpha_0$ has been fixed here by demanding $\alpha_0'(\Delta_0^{\textrm{F}})=0$. 

In order to obtain other solutions, we will resort to the following very useful property of the fundamental free equation. Consider some solution $f_{\omega}^{\Df,\eta}(z)$, where $\omega=\alpha,\beta$, and define a new function
\bea
f^{\widehat \Delta_\phi,\widehat \eta}_{\widehat \omega}(z)\equiv \frac{[1+z(z-1)]^p}{[z(1-z)]^k} f^{\Df,\eta}_{\omega}(z),
\eea
for integer $k,p$. Then it is easy to check that this will also satisfy the fundamental free equation, as long as we change
\bea
\eta\to \widehat \eta=(-1)^k \eta, \qquad \Df \to \widehat \Delta_\phi=\Df-\frac{3}2 k+p.
\eea
Multiplication by the prefactor modifies the behaviour near $z=1$ and $z=\infty$, and it can also give us a bosonic functional from a fermionic one and vice-versa (i.e. change $\eta$). Starting from the elementary $m=0$ fermionic functionals written above we can use these shifts to obtain all other solutions of interest. 

First let us obtain obtain all fermionic solutions, i.e. those with $\eta=-1$ but arbitrary $m\geq 0$. To do this we simply define
\bea
f^{\Df,-}_{\omega_{m}}(z)=\left[\frac{1+z(z-1)}{z(z-1)}\right]^{2m} f^{\Df+m,-}_{\omega_0}(z).
\eea
The resulting functionals still have the required falloff at infinity and the right divergent structure as $z\to 1^+$. Since the prefactor is positive, they also preserve any positivity properties of the original $m=0$ kernels. Hence these provide good functionals with all the right properties, and we remind the reader that each of them can be redefined by adding lower $m$ solutions. We can use this freedom to make certain nice choices detailed further below. To obtain bosonic functionals we can define:
\bea
f^{\Df,+}_{\omega_{m+1}}(z)=\frac{1}{z(1-z)} f^{\Df+\frac 32,-}_{\omega_m}(z),
\eea
since we have just obtained all solutions on the righthand side. The minus sign is included so that positivity of the fermionic functional action translates into positivity of the bosonic one, cf. equations \reef{eq:gfromf} and \reef{eq:funcact}. 

This leaves a priori two other sets of solutions to be constructed, namely those bosonic solutions which satisfy \reef{eq:bosondiv} with $m=0$. Note that multiplication of a fermionic functional by $z(1-z)$ could do the trick, but this ruins the falloff at infinity; we cannot correct this by dividing by $1+z(z-1)$ since this would destroy analyticity of $f(z)$ away from the real axis. The only way out is to first take a specific linear combination of $m=0$ fermionic functionals which falls off as $z^{-4}$ and only then multiply by $z(1-z)$, to obtain a bosonic $\alpha$ functional with $m=0$. In detail this is achieved by setting:
\bea
f^{\Df,+}_{\alpha_0}(z)=z(1-z)\left[f^{\Df-\frac 32,-}_{\alpha_0}(z)+\frac{1}{2\Df-1} f^{\Df-\frac 32,-}_{\beta_0}(z)\right].
\eea
This still leaves the $\beta_0^+$ functional to be constructed. However, this is fine since such a functional actually does not exist for general $\Df$.\footnote{An exception is the degenerate case $\Df=0$ where $f(z)=\frac{1}{z(z-1)}$ does the trick.}. Indeed, as we have discussed in section \ref{sec:functionalsPolyakov}, it is precisely this missing functional which allows the deformation of the generalized free boson solution by a contact term in AdS$_2$. Another argument is that such a functional would rule out the generalized free fermion solution to crossing, since the associated functional action would be non-negative for all $\Df\geq 2\Df$ and zero on the identity. Finally, this result can also be seen directly from the fundamental equation and boundary conditions. Suppose such a functional did exist. Then we could form a new fermionic functional with $m=0$ by dividing it by $z(z-1)$. But then this would mean that there exists a fermionic functional with $m=0$ that falloff as $z^{-4}$, which is false: the unique $m=0$ functional is the one in equation \reef{eq:allaf}, which decays as $z^{-2}$ \cite{Mazac:2018mdx}. We conclude the only way to recover the missing functional is to relax the required asymptotics to $O(1)$ instead of $O(z^{-2})$. This leads precisely to the prefunctionals discussed in section~\ref{ssec:bosonicBasis}.

\subsection{Orthonormal bases and special cases}
\label{sec:constructortho}
The construction described above provides us with a full set of functionals with prescribed boundary conditions. They have the property that the associated functional actions satisfy
\ba
\alpha_m(\Delta_n)&=\delta_{nm},&\qquad \alpha_m'(\Delta_n)&=0\\
\beta_m(\Delta_n)&=0,&\qquad \beta_m'(\Delta_n)&=\delta_{nm}
\ea
for $n\geq m$, for both bosons and fermions. In order to extend these orthogonality conditions to $0\leq n<m$ we can use the freedom to redefine a given $m$ functional by lower $m$ ones, and that of shifting an $\alpha$ functional by the corresponding $\beta$ functional with the same $m$. We were able to find closed form expressions for such orthogonal functionals only for special values of $\Df$. For instance, for $\Df=1/2$ the set of orthogonal fermionic ($\eta=-1$) functionals is given by
\bea
f_{\alpha_m}(z)&=&\frac 12\partial_m f_{\beta_m}(z)+\frac{2}{\pi^2}\frac{\Gamma(2+2m)^4}{\Gamma(3+4m)\Gamma(4+4m)}\, G_{2m+2}(1/z)\\
f_{\beta_m}(z)&=&\frac{2}{\pi^2}\frac{\Gamma(2+2m)^2}{\Gamma(3+4m)}\left[ \frac{P_{2m+1}\left(\frac{z-2}z\right)}z+\frac{P_{2m+1}\left(\frac{1+z}{z-1}\right)}{1-z}\right]
\eea
with $P_m(z)$ the Legendre polynomials. We have checked (numerically) that the functionals so defined satisfy the orthogonality properties above for all $n,m \in \mathbb N$.

%\begin{figure}%
%\begin{center}
%\includegraphics[width=10cm]{plots/betans}
%\caption{The fermionic $\beta_n$ functionals for $n=0,\ldots, 3$. The functionals have been rescaled %by a common positive function of $\Delta$ for clarity and plotted from $\Delta=0$}%
%\label{fig:beta0gff}%
%\end{center}
%\end{figure}

 Another case which will be useful to us later are the bosonic functionals with $\Df=1$ and $\eta=1$:
\begin{equation}
\begin{split}
f_{\alpha_m}(z)&=\frac 12\partial_m f_{\beta_m}(z)-\frac{2}{\pi^2}\frac{\Gamma(2+2m)^4}{\Gamma(3+4m)\Gamma(4+4m)}\, G_{2m+2}(1/z)\\
f_{\beta_{m}}(z)&=\frac{2}{\pi^2}\frac{\Gamma(2+2m)^2}{\Gamma(3+4m)}\left[P_{2m+1}\left(\frac{z-2}z\right)+P_{2m+1}\left(\frac{1+z}{z-1}\right)\right.\\&\left.-P_{1}\left(\frac{z-2}z\right)-P_{1}\left(\frac{1+z}{z-1}\right)\right].
\end{split}
\end{equation}
It can be checked that the corresponding functionals satisfy the relations:
\ba
\alpha_m(\Delta_n)&=\delta_{nm},& \alpha'_m(\Delta_n)&=-c_m \delta_{n0},& n,m\geq 0\\
\beta_m(\Delta_n)&=0,& \beta_m'(\Delta_n)&=\delta_{nm}-d_m\delta_{n0},& n,m\geq 0
\ea
where the constants $c_m,d_m$ are related to a contact diagram in $AdS_2$, cf. \reef{eq:cdFormula}. Again, we should set $\beta_0\equiv 0$. 

\begin{figure}[ht!]
\begin{center}
\includegraphics[width=\textwidth]{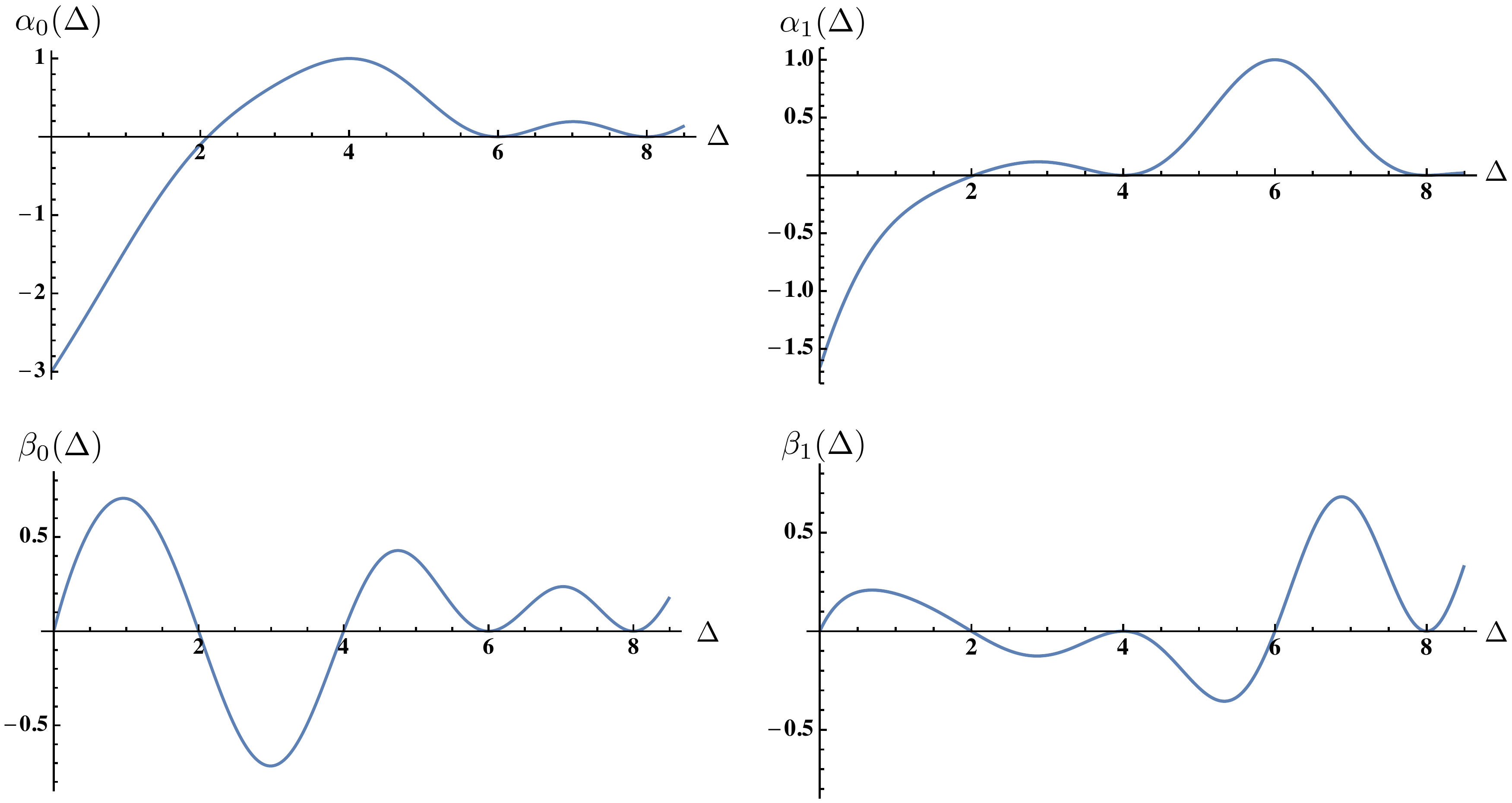}
\caption{The action of the fermionic $\alpha_n$ and $\beta_n$ functionals for $\Df=\frac{3}{2}$ and $n=0,1$. The functionals have double zeros at $\Delta=2\Df+2m+1$ for $m\neq n$ and $m,n\in\mathbb{N}_{\geq 0}$. The action of $\alpha_n$ on the identity ($\Delta=0$) is equal to $-a^{\textrm{free}}_{2\Df+2n+1}$, while $\beta_n$ vanishes there. $\alpha_n(\Delta)$ and $\beta_n(\Delta)$ also describe the OPE decomposition of the crossing-symmetric sum of Witten exchange diagrams with exchanged dimension $\Delta$. }%
\label{fig:FunctionalsAB}%
\end{center}
\end{figure}

These examples illuminate a more general pattern of orthonormal functionals. The basic building blocks 
\bea
f^{\Df,\eta}_{\beta,m}(z)&=&\left( \frac{P_{2m+1}\left(\frac{z-2}z\right)}{z^{2-2\Df}}+\eta \frac{P_{2m+1}\left(\frac{1+z}{z-1}\right)}{(z-1)^{2-2\Df}}\right)\\
f^{\Df,\eta}_{\alpha,m}(z)&=& \frac 12\partial_m f^{\Df,\eta}_{\beta,m}(z)-\eta \,\frac{\Gamma(2+2m)^2}{\Gamma(4+4m)}\, G_{2m+2}(1/z)
\eea
satisfy the free fundamental equation for all $m$, for $\eta=1$ and integer $\Df$ or $\eta=-1$ and half-integer $\Df$. Note the asymptotic behaviour near $z=1$ takes the form
\ba
f^{+}_{\beta,m}(z)&\stackrel{z\to 1^+}{=}O[(z-1)^{-1-2m}],& m&\geq \Df-1, & \Df&\in \mathbb N\\
f^{-}_{\beta,m}(z)&\stackrel{z\to 1^+}{=}O[(z-1)^{-2-2m}],& m&\geq \Df-\frac 12,&  \Df&\in \mathbb N+\frac 12
\ea
and similarly for the $\alpha$-type functionals with extra logarithmic factors. This means that as we increase $\Df$ we have the freedom to include building blocks with lower $m$, which we must use to obtain a sufficiently fast falloff (i.e. $O(z^{-2})$) near $z=\infty$. As we increase $\Df$ by one unit the behaviour near infinity of the building blocks gets multiplied by $z(z-1)$, but at the same time we gain two new lower $m$ blocks ($\alpha$ and $\beta$), so it may seem the system is under-constrained. However, experimentally we find that there are always identities among these lower $m$ building blocks which reduces the number of degrees of freedom in precisely the right way. Furthermore, the net result is always that after properly orthonormalized $\beta_m$ functionals have been constructed, the $\alpha_m$ functionals are always given by
\begin{subequations}
\bea
f_{\alpha_m}^{\Df,-}(z)&=&\frac 12 \partial_m f_{\beta_m}^{\Df,\eta}(z)\nonumber\\
&&+\frac{2}{\pi^2} \frac{\Gamma(1+2\Df+2m)^4}{\Gamma(2+4\Df+4m)\Gamma(1+4\Df+4m)} G_{2m+1+2\Df}(1/z),\\
f_{\alpha_m}^{\Df,+}(z)&=&\frac 12 \partial_m f_{\beta_m}^{\Df,\eta}(z)\nonumber \\
&&-\frac{2}{\pi^2} \frac{\Gamma(2\Df+2m)^4}{\Gamma(4\Df+4m)\Gamma(4\Df-1+4m)} G_{2m+2\Df}(1/z),
\eea
\end{subequations}
for half-integer and integer $\Df$ respectively.

As a simple example, the fermionic basis for $\Df=\frac 32$ is given by
\bea
f_{\beta_m}^{\frac 32,-}(z)=\frac{2}{\pi^2}\frac{\Gamma(4+2m)^2}{\Gamma(5+4m)\Gamma(6+4m)}\left( f_{\beta,m+1}^{\frac 32,-}(z)- \frac 13 c_m f_{\beta,0}^{\frac 32,-}(z)\right)
\eea
with $c_m=23+28 m+8 m^2$, and $\alpha_m$ kernels given as above with $\Df=\frac 32$. The identity among lower $m$ kernels in this case is:
\bea
f_{\beta,0}^{\frac 32,-}(z)=f_{\alpha,0}^{\frac 32,-}(z)=-3.
\eea

As a concrete illustration, the actions of the first few fermionic functionals at $\Df=\frac{3}{2}$ are plotted in Figure \ref{fig:FunctionalsAB}. 

%%%%%%%%%%%%%%%%%%%%%%%%%%%%%%%%%%%%%%%
\section{Functional bootstrap equations and their implications}\label{sec:bounds}
\subsection{General idea}
The functional bases constructed in the previous section provide us with an infinite but countable set of constraints on the CFT data. These constraints are obtained by acting with the functionals on the crossing equation and using the swapping property to find the {\em functional bootstrap equations} \reef{eq:fbes} which we repeat here:
\be
\sum_{\Delta} a_\Delta \alpha_n(\Delta)=0, \qquad \sum_{\Delta} a_\Delta \beta_n(\Delta)=0, \qquad  n\in \mathbb{N}
\ee
where one may choose to use the bosonic or fermionic basis.
In either case, these equations provide necessary, and as we will argue in the next section, sufficient conditions for the OPE data to satisfy the crossing equation. In other words, these equations contain the same information as the original crossing equation, but in a form that is much more amenable to analytic (and numeric \cite{pauloszan}) studies. In this section we will use these equations to extract universal properties that must hold for any solution to crossing.

Previous work \cite{Pappadopulo:2012jk,Qiao:2017xif,Mukhametzhanov:2018zja} appealing to Tauberian theory has been able to derive powerful constraints on moments of OPE density at large $\Delta$. These results may be summarized as establishing that~1) the OPE decomposition converges exponentially fast in the crossing region $\mathcal R$ away from its boundary and 2) the integrated or average OPE density of any solution to crossing must behave asymptotically like that of a generalized free field (with calculable corrections). Here we will be able to make more refined statements: we will prove an upper bound on individual OPE coefficients, as well as the total OPE density present inside intervals of finite size. We can establish these bounds for any value of $\Delta$, but they take a particularly simple form for $\Delta \gg 1$. Moreover, we will obtain a {\em lower} bound on a suitable average of the OPE density. The result is that both upper and lower bounds strongly constrain variations of the OPE density away from the generalized free answer.

In both cases the derivation of the bounds follows in a straightforward manner from the functional bootstrap equations, and in particular from the $\alpha_n$ sum rules. The $\alpha_n$ functionals satisfy $\alpha_n(\Delta_m)=\delta_{nm}$ and also know about the generalized free OPE density since $\alpha_n(0)=-a^{\text{free}}_{\Delta_n}$, which follows from the existence of the generalized free solutions to the functional equations. For both upper and lower bounds this will essentially imply that contributions to the sum rule from the OPE density in the vicinity of $\Delta_n$ will have to cancel the contribution of the identity, which is controlled by the generalized free result. For the upper (lower) bound we will take $\alpha_n$ ($-\alpha_n$) and modify it appropriately by $\beta$ functionals so as to obtain an object with suitable positivity properties. The positive contributions can be ignored to obtain inequalities, and potential undesired negative contributions are always sub-leading for large $n$, which leads to the desired bounds.

Throughout this section, it will be useful to keep in mind the definition:
\bea
a_{\Delta}^{\text{free}}=2\frac{\Gamma(\Delta)^2}{\Gamma(2\Delta-1)}\,\frac{\Gamma(\Delta+2\Df-1)}{\Gamma(2\Df)^2\,\Gamma(\Delta-2\Df+1)},
\eea
which yields the correct GFF OPE coefficients when evaluated for $\Delta=\Delta^{\textrm{F}}_n,\Delta^{\textrm{B}}_n$.

\begin{figure}[ht!]%
\begin{center}
\includegraphics[width=0.7\textwidth]{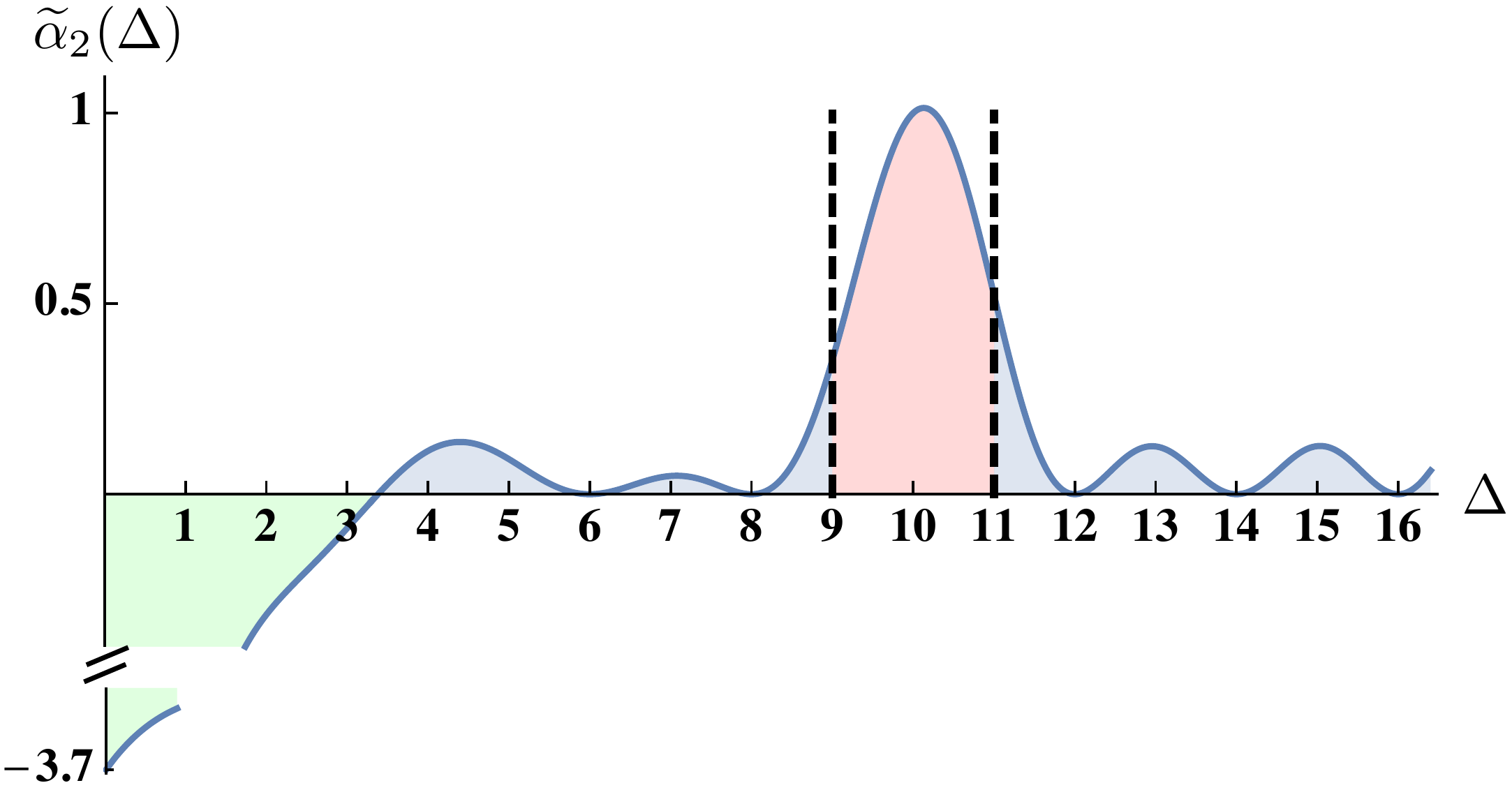}
\caption{Functional $\widetilde{\alpha}_n$ used to prove the upper bound on the OPE coefficients \eqref{eq:upperbound}, here shown in the fermionic case for $\Df=\frac{5}{2}$, $n=2$. The contribution to the sum rule from operators in $[\Delta_n-1,\Delta_n+1]$ (shown in red) must be bounded above by minus the contribution from where $\widetilde{\alpha}_n(\Delta)<0$ (shown in green). The negative region stays finite in extent as $n\rightarrow\infty$ and is dominated by the identity in this limit, giving rise to the RHS of \eqref{eq:upperbound}.}%
\label{fig:aTilde}%
\end{center}
\end{figure}

\subsection{Upper bound on the OPE data}

We will work with the fermionic functional bootstrap equations and at the end we will comment on how similar results may be derived by resorting to the bosonic ones. We want to find a functional for which contributions to the corresponding sum rule from values of $\Delta$ in a neighbourhood of $\Delta_n$ must cancel that of the identity. Unfortunately, the basic $\alpha_n$ functional will not do, but a simple modification will suffice. 

Let us begin from the $\alpha_n$ equations.  Ignoring various positive contributions to the corresponding sum rules, they imply the following bound: 
\bea
\sum_{\Delta\in B_n} a_\Delta \alpha_n(\Delta)\leq a_{\Delta_n}^{\text{free}} +\sum_{\Delta\in S_n^-} a_\Delta[-\alpha_n(\Delta)]
\eea
Here we have defined the bins $B_n\equiv [\Delta_n-1,\Delta_n+1]$ centered around the generalized free fermion values $\Delta_n^{\textrm{F}}=1+2\Df+2n$, and the set $S_n^-$ where the functional is negative, formally $S_n^{-}=\{\Delta>0: ( \Delta\notin B_n) \cap (\alpha_n(\Delta)\leq 0)\}$. We have also used $\alpha_n(0)=-a_{\Delta_n}^{\text{free}}$. The equation above establishes a bound on a certain average OPE density inside any given bin, but it is not very useful as one can check that the set $S_n^-$ contains operators with arbitrarily large~$\Delta$. Our strategy is to introduce new functionals:
\bea
\widetilde \alpha_n\equiv \alpha_n+b_n \beta_n.
\eea
The coefficient $b_n$ is chosen so that the associated functional kernel has a stronger falloff behaviour,\footnote{To justify this prescription, we point out that for $f_{\alpha_n}(z)$ we have $\lim_{z\to \infty} z^2 f_{\alpha_n}(z)>0$, which implies the functional action \reef{eq:funcact} will be necessarily negative for sufficiently large $\Delta$. On the other hand sub-leading terms turn out to have positive coefficients, so it is natural to remove the offending piece.}
\bea
f_{\widetilde \alpha_n}(z) \overset{z\to \infty}{\sim} O(z^{-4}).
\eea
The shape of the resulting functional is illustrated in Figure \ref{fig:aTilde}. The point of this improved definition is that the functional action $\widetilde \alpha_n(\Delta)$ turns out to have much nicer positivity properties, namely
\bea
\widetilde \alpha_n(\Delta)\geq 0 \qquad \text{for}\quad \Delta>\Delta^{\text{pos}}_n,
\eea
with $\Delta^{\text{pos}}_n$ roughly $2\Delta_\phi$ for all $n$. We have checked this numerically in several cases. It can also be proven rigorously at large $n$ for specific, half-integer values of $\Df$ using the asymptotic expressions for the functional actions derived in appendix \ref{app:funcs}, as we will see momentarily. With the $\widetilde \alpha$ functionals our new improved bound is
\bea
\sum_{\Delta\in B_n} a_\Delta \widetilde \alpha_n(\Delta)\leq a_{\Delta_n}^{\text{free}}+\sum_{0<\Delta\leq \Delta_n^{\text{pos}}} a_\Delta[-\widetilde \alpha_n(\Delta)]. \label{eq:bound}
\eea
This is an exact bound valid for any $n$, constraining the average OPE density in the bin $B_n$. The bound is non-trivial, as the $\widetilde \alpha_n(\Delta)$ are bounded from below in $B_n$ by an order one number. Although valid, this result is perhaps not so useful as expressions for the $\widetilde \alpha_n(\Delta)$ are generically quite complicated. However we can obtain a cleaner version of the bound by considering large $n$. For a generic functional $\omega_n$ let us define:
\bea
\omega_n(\Delta)=\frac{4\sin^2\left[\frac{\pi}2(\Delta-\Delta_n)\right]}{\pi^2}\left(\frac{a_{\Delta_n}^{\text{free}}}{a_\Delta^{\text{free}}}\right)\, R_{\omega}(\Delta,\Delta_n|\Df).
\eea
The goal of this rewriting of $\omega_n(\Delta)$ is to factor out the fast, exponential dependence on $\Delta$ and $n$. Using the results of appendix \ref{app:funcs} we can obtain
\begin{subequations}
\bea
R_\beta(\Delta, \Delta_n|\Df)&\underset{\Delta,\Delta_n\to \infty}{\sim} & \frac{4 \Delta \Delta_n^2}{\Delta^4-\Delta_n^4}\,,\label{eq:rbeta1}\\
R_{\widetilde \alpha}(\Delta,\Delta_n|\Df)&
\underset{\Delta,\Delta_n\to \infty}{\sim}
&\frac{16 \Delta  \Delta _n^5 }{\left(\Delta ^4-\Delta
   _n^4\right){}^2}\,,
\eea
\end{subequations}
where the limits are taken holding the ratio $\Delta/\Delta_n$ fixed. To be precise, we have checked these expressions hold for several half-integer $\Df$ where the functionals take a simpler form.\footnote{We believe that the expressions hold for general $\Df$. For instance, assuming this is indeed the case it is easy to show that the generalized free boson OPE density will satisfy the (fermionic) $\beta_n$ sum rule at large $n$ for any $\Df$, which follows from $$\text{P}\int_0^{\infty} \ud x\,  \frac{x}{x^4-1}=0$$.
}

 If we make $\Delta_n$ large but keep $\Delta$ fixed, we again find simplifications, but the result now depends on $\Df$. In the simplest case $\Df=1/2$ we find:
\begin{subequations}
\bea
\frac{R_\beta(\Delta,\Delta_n|\frac 12)}{(2\Delta-1)}&\overset{\Delta_n\to \infty}{\sim}&-\frac{2}{\Delta_n^2}-\frac{\Gamma(\Delta)^4}{\Gamma(2\Delta)}\,\frac{\sin[\pi(\Delta-\Delta_n)]}{\pi\,\Delta_n^{2\Delta}}, \\
\frac{R_{\widetilde \alpha}(\Delta,\Delta_n|
\frac 12)}{(2\Delta-1)}\!\!\!&\overset{\Delta_n\to \infty}{\sim}&\frac{8}{\Delta_n^3}-\frac{a_\Delta^{\text{free}}}{2\Delta-1}\,\frac{\Gamma(\Delta)^2}{\Delta_n^{2\Delta}}\,\cos^2\left(\frac{\pi \Delta}2\right)
\eea
\end{subequations}
We have written the last expression in a funny way because the ratio $a_\Delta^{\text{free}}/(2\Delta-1)$ is positive for $\Delta>0$. From the expression above we see explicitly that for large $n$ the functional action $\widetilde \alpha_n(\Delta)$ is positive beyond $\Delta>3/2$. Furthermore for $\Delta<3/2$ we obtain
\bea
\widetilde \alpha_n(\Delta)=-\frac{a_{\Delta_n}^{\text{free}}}{\Delta_n^{2\Delta}\Gamma(1-\Delta)^2}.
\eea
In particular, in the large $n$ limit all values of $\widetilde \alpha_n(\Delta)$ for $\Delta<3/2$ are suppressed relative to the value at the identity $\widetilde \alpha_n(0)=-a^{\text{free}}_{\Delta_n}$. This means that in equation \reef{eq:bound} we can safely neglect the sum on the righthand side for sufficiently large $n$. For other values of $\Df$ we find the same pattern: the $R_{\widetilde \alpha}$ functions contains two pieces, one analytic and one non-analytic, with the latter dominating for small enough $\Delta$, and for which the identity contribution is always leading in the limit $n\to \infty$.

Overall, we can now write the bound in following simple form:
\bea
\boxed{\limsup_{n\to \infty}\sum_{|\Delta-\Delta_n|\leq 1} \frac{4\sin^2\left[\frac \pi 2 \left(\Delta-\Delta_n\right)\right]}{\pi^2(\Delta-\Delta_n)^2}\left(\frac{a_{\Delta}}{a_\Delta^{\text{free}}}\right)\,\leq 1} \label{eq:upperbound}
\eea
Although we have only derived this bound for a few specific values of $\Df$, we believe this to be actually true for all $\Df$. A few notes are in order. Firstly, we have found the very same bound by working with the bosonic functional basis, upon suitably reinterpreting the meaning of the $\Delta_n$ above (i.e. changing them from $\Delta_n^{\textrm{F}}$ to $\Delta_n^{\textrm{B}}$). Secondly, by using the fact that the bound above holds for all large $n$ together with the specific form of the averaging kernel, we are free to extend the sum over a wider range in both directions, as long as $\Delta$ stays large. Finally, it is clear that an equally good bound can be obtained by shrinking the region over which we are averaging as much as we wish, so that the result above implies also a bound on individual OPE coefficients, namely:
\bea
\limsup_{\Delta \to \infty} \frac{a_{\Delta}}{a_\Delta^{\text{free}}}\leq C, \qquad 1\leq C\leq \pi^2/4.
\eea
The bound is strongest when $\Delta$ is at the centre of the bin, where $C=1$, weakening as we move close to the edge by up to a factor of $\pi^2/4$. However, we can combine this bound with the one obtained from the bosonic basis (where the bins are shifted by one unit) to improve the upper range of $C$ to $\pi^2/8\sim 1.2$. While we cannot rigorously prove it, it is tempting to conjecture that the actual bound is actually
\bea
\limsup_{\Delta \to \infty} \frac{a_{\Delta}}{a_\Delta^{\text{free}}}\leq 1.
\eea

\begin{figure}[ht!]%
\begin{center}
\includegraphics[width=0.7\textwidth]{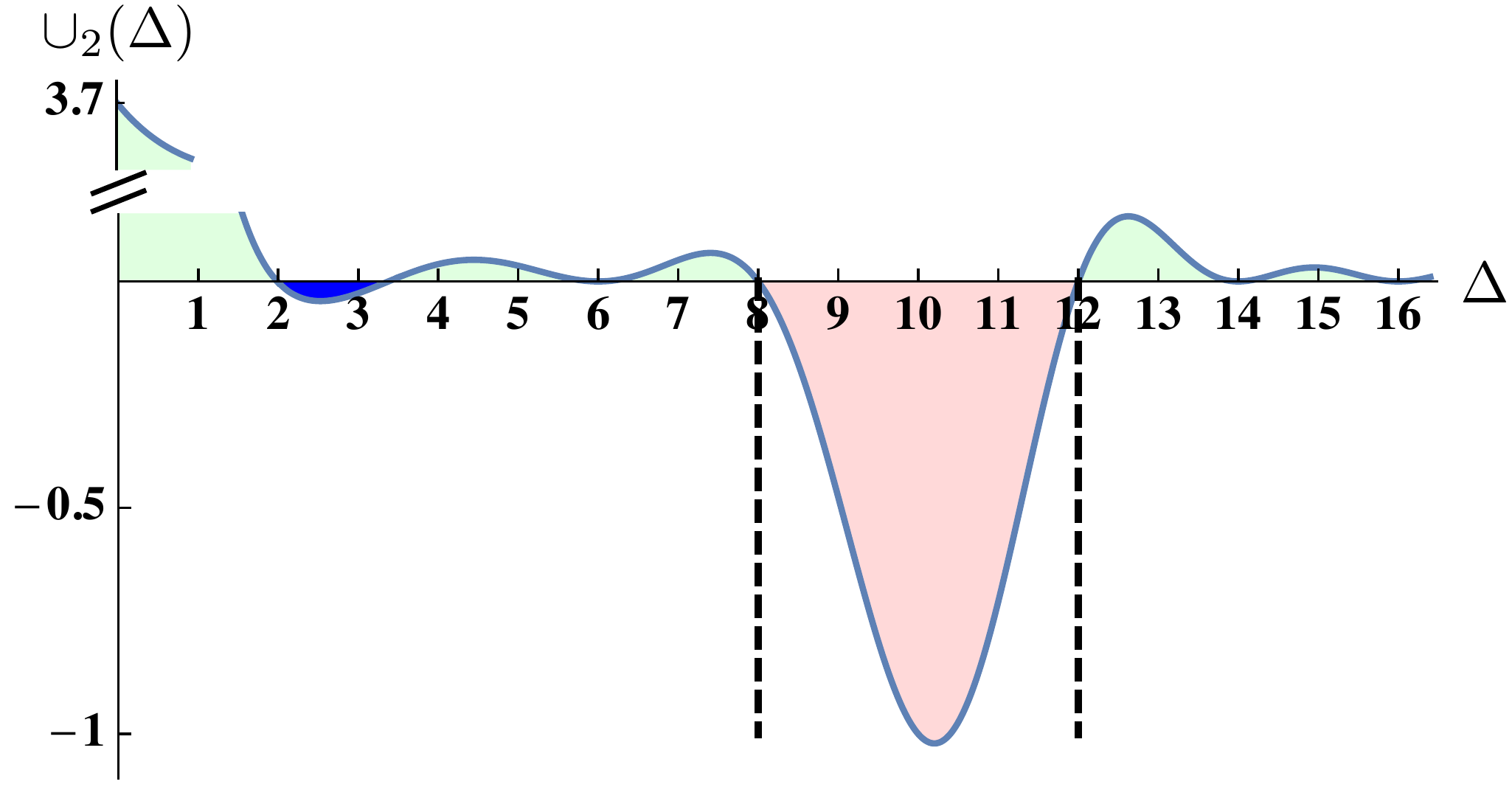}
\caption{Functional $\cup_n$ used to prove the lower bound on the OPE coefficients \eqref{eq:lowerbound}, here shown in the fermionic case for $\Df=\frac{5}{2}$, $n=2$. The contribution to the sum rule from operators in $[\Delta_n-2,\Delta_n+2]$ (shown in red) must compensate the contribution from where $\cup_n(\Delta)>0$ (shown in green). The latter always contains the identity, which gives rise to the RHS of \eqref{eq:lowerbound}. There could be an additional negative region (shown in blue), but its extent stays finite as $n\rightarrow\infty$ and its contribution sub-leading compared to that of the identity in this limit.}%
\label{fig:dBin}%
\end{center}
\end{figure}

\subsection{Lower bound on the OPE data}

We will now establish a lower bound on the OPE density. The strategy to obtain such a bound is to cook up a functional $\cup_n$, such that its action looks essentially as illustrated in Figure \ref{fig:dBin}. The functional action is negative for $[\Delta_{n-2},\Delta_{n+2}]$, with all other negative regions being suppressed for large $n$. The functional is constructed such that its action on the identity is positive and given by the generalized free OPE coefficient $a_{\Delta_n}^{\text{free}}$. This implies then that in the sum rule for $\cup_n$, the OPE in the aforementioned negative region must be large enough to at least cancel the identity contribution, establishing a lower bound.

Consider then the following combination which we call the {\em double bin} functional:
\bea
\cup_n\equiv \frac {a_{\Delta_n}^{\textrm{free}}}4\left[\frac{\beta_{n+1}}{a^{\textrm{free}}_{\Delta_{n+1}}}-\frac{\beta_{n-1}}{a^{\textrm{free}}_{\Delta_{n-1}}}\right]-\left [\alpha_n-\frac{\partial_{n} a^{\text{free}}_{\Delta_n}}{2a_{\Delta_n}^{\text{free}}}\,\beta_n\right].
\eea
The reason for this name is that the functional action dips to negative values when $\Delta\in[\Delta_{n-1},\Delta_{n+1}]$. This particular combination of functionals is chosen so that the functional action looks like in Figure \ref{fig:dBin}, in particular it has first order zeros at $\Delta_{n\pm 1}$, $\cup_n(\Delta_n)=-1$ and $\cup_n(0)=a_{\Delta_n}^{\text{free}}$.

Let us denote the second square bracket above by $\widehat \alpha_n$. For large $\Delta, \Delta_n$ we have $R_{\widehat \alpha}=\partial_{\Delta_n} R_{\beta}$ with exponentially small corrections. 
We find then
\bea
\cup_n(\Delta)\underset{\Delta,\Delta_n,|\Delta-\Delta_n|\to \infty}\sim \frac{4\sin^2\left[\frac{\pi}2(\Delta-\Delta_n)\right]}{\pi^2} \left(\frac{a_{\Delta_n}^{\text{free}}}{a_\Delta^{\text{free}}}\right)\,\left[\frac{2}{3} \partial^3_{\Delta_n} R_{\beta}(\Delta,\Delta_n|\Df)\right]\,,
\eea
which is positive, as can be checked using \reef{eq:rbeta1}. In the limit of large $\Delta_n$ with fixed $\Delta$ we again lose universality in $\Df$ and must check case by case. For $\Df=1/2$, we get:
\bea
\cup_n(\Delta)&\overset{\Delta_n\to \infty}{\sim}& \frac{4\sin^2\left[\frac{\pi}2(\Delta-\Delta_n)\right]}{\pi^2}\,a_{\Delta_n}^{\text{free}} \frac{\Gamma(2\Delta)}{\Gamma(\Delta)^2} \left[\,\frac{\Gamma(\Delta)^4}{\Gamma(2\Delta)}\frac{\cos^2\left(\frac{\pi \Delta}2\right)}{\Delta_n^{2\Delta}}+\frac{16}{\Delta_n^5}\right],
\eea
where we have shown the leading analytic and non-analytic pieces in $\Delta_n$. In particular, this shows that the functional is positive in this limit, and positive contributions to the sum rule can always be ignored when we want to obtain a bound. For more general $\Df$, we find a similar story. One always finds that for $\Delta$ larger than $2\Df$, the functional action is always non-negative, apart from the double bin region $(\Delta_{n-1},\Delta_{n+1})$. On the identity it is clearly positive, since it must be equal to the free OPE value. For intermediate values of $\Delta$ we find that there can be negative regions of $\cup_n(\Delta)$, but these turn out to be always suppressed in the large $\Delta_n$ limit, much as we saw in the previous subsection in the determination of the upper bound on the OPE density.

Ignoring subleading and positive contributions, the sum rule arising from $\cup_n$ leads to a simple bound: 
\bea
\boxed{
\liminf_{n\to \infty} \sum_{|\Delta-\Delta_n|\leq 2}\frac{16\sin^2\!\left[\frac{\pi}2 (\Delta-\Delta_n)\right]}{\pi^2(\Delta-\Delta_n)^2(\Delta-\Delta_{n-1})(\Delta_{n+1}-\Delta)}\,\left(\frac{a_{\Delta}}{a^{\textrm{free}}_{\Delta}}\right)\geq 1.
}
\label{eq:lowerbound}
\eea
Here the averaging kernel can be obtained by considering the functional action in the limit of large $\Delta, \Delta_n$ with fixed difference. This result establishes a lower bound on the OPE density, which is valid in the limit of large $n$. Just like in the previous subsection, we have explicitly derived this bound for several half-integer $\Df$, and have also done the same with the bosonic basis of functionals (for integer $\Df$), finding the exact same result upon  reinterpreting the meaning of $\Delta_n$. We conjecture the bound holds for all $\Df$ and for either basis. 

A few comments are in order. It is possible to obtain a lower bound for finite $n$, just like we did for the upper bound in the previous subsection, but we will not write it down explicitly here. The bound above is optimal, in the sense that it is saturated by the generalized free solutions. Moreover, it is possible to check that the fermionic bound is satisfied by the bosonic solution and vice versa. A more non-trivial test is that the $\langle \sigma(0) \sigma(1) \sigma(z) \sigma(\infty)\rangle$ correlator for the 2d Ising model satisfies the above bound non-trivially, with the left-hand side evaluating to $\approx 1.06$. 

A straightforward consequence of the bound is that one cannot have large gaps in the OPE, since there must always be at least one state in between $\Delta_n$ and $\Delta_{n+2}$. Combining the bosonic and fermionic bounds we find the remarkable result that the spacing between consecutive primaries can be no larger than five at sufficiently large $\Delta$.

\section{Completeness}
\label{sec:completeness}
In this section we will demonstrate that the full set of functional bootstrap equations are equivalent to the crossing equation, assuming unitarity. That is, we will prove
\bea
\sum_{\Delta} a_{\Delta} F_{\Delta}(z)=0\qquad \Leftrightarrow \qquad
\sum_{\Delta} a_\Delta \alpha_n(\Delta)=0, \qquad \sum_{\Delta} a_\Delta \beta_n(\Delta)=0, \qquad \forall n\in \mathbb N,
\eea
where the sums range over $\Delta\geq 0$, and $a_{\Delta}\geq 0$. The rough intuition for why this should be the case is that equation \reef{eq:decomp} below allows us to express $F_\Delta$ in a basis formed by the $F_{\Delta_n}, \partial F_{\Delta_n}$, with uniqueness of the decomposition guaranteed by demanding that the coefficients are functional actions. The functional bootstrap equations are then the coefficients of this basis decomposition of the crossing equation.

For thoroughness, let us first briefly review why the functional equations follow from crossing, i.e. why they are necessary. Necessity follows from
\bea
\sum_\Delta a_{\Delta} F_{\Delta}=0\quad \overset{?}{\Rightarrow}\quad \omega_n\left[\sum_\Delta a_{\Delta} F_{\Delta}\right]=\sum_{\Delta} a_{\Delta} \omega_n(\Delta),
\eea
that is to say, if the swapping condition holds, with $\omega=\alpha,\beta$. By linearity of the functionals this is proven if
\bea
\lim_{\Delta^*\to \infty} \omega_n\left[\sum_{\Delta>\Delta^*} a_\Delta F_{\Delta}\right]=0.\label{eq:swaptail}
\eea
Our functionals were defined by integrals with kernels $f,g$ as in our basic definition \reef{eq:ffg}. Then, as long as $\omega_n(\Delta)$ is finite for all $\Delta\geq 0$, the only danger comes from the region of integration close to $z=\infty$ \cite{Rychkov:2017tpc}. In that region we bound
\bea
\left|\sum_{\Delta>\Delta^*} a_{\Delta} F_{\Delta}\right|&\leq &\sum_{\Delta>\Delta^*} a_{\Delta}\left[\frac{G_{\Delta}\left(\left|\frac{z}{z-1}\right|\right)}{|z|^{\Df}}+\frac{G_{\Delta}\left(\left|\frac{z-1}{z}\right|\right)}{|z-1|^{\Df}}\right]\nonumber\\
&\underset{|z|\to \infty} \lesssim& G_{\Delta_0}\left(1/|z|\right)\underset{|z|\to \infty} \sim O(|z|^{-\Delta_0})
\eea
where in the second step we have assumed crossing holds and that $\Delta_0$ is the smallest dimension for which $a_{\Delta}>0$. Equation \reef{eq:swaptail} will hold if the $f(z)$ kernel behaves near $z=\infty$ as $z^{\Delta_0-1-\epsilon}$ for some $\epsilon>0$. Depending on $\Delta_0$ we have different possibilities, which leads to different possible choices of functional bases. Here we are interested in the case $\Delta_0=0$, and our functionals were constructed so as to satisfy this precise requirement. This proves necessity.

To show that the functional equations are also sufficient, we will use the crucial relation:
\bea
F_{\Delta}(z)=\sum_{n=0}^{\infty}\left[ \alpha_n(\Delta) F_{\Delta_n}(z)+\beta_n(\Delta) \partial F_{\Delta_n}(z)\right]\label{eq:decomp}.
\eea
Establishing that \reef{eq:decomp} actually holds is a priori no easy feat. Luckily we already know that such expressions do exist, as they follow from the conformal block decomposition of (crossing-symmetric sums of) Witten exchange diagrams. More precisely, those objects guarantee that decompositions of $F_{\Delta}$ exist in terms of $F_{\Delta_n}, \partial F_{\Delta_n}$. Once this is established, the coefficients in the decomposition can be chosen as functional actions as we have proven in section \ref{sec:functionalsPolyakov}. 

%The intuition is that this equation expresses $F_\Delta$ in the basis formed by the $F_{\Delta_n}, \partial F_{\Delta_n}$. The uniqueness of the decomposition is then guaranteed by demanding that the coefficients are functional actions. The functional bootstrap equations are then simply the basis decomposition of the crossing equation. Our strategy here will be to plug in this expression into the crossing equation and show that the two series over $\Delta,n$ may be commuted. This will be possible thanks to the upper bound on the OPE density derived in the previous section, which will give us control over the tails of the double series. 

Using the decomposition \reef{eq:decomp}, sufficiency can be rephrased as the commuting of a double series:
\bea
\sum_{\Delta} a_{\Delta} F_{\Delta}&=&\sum_{\Delta}\sum_{n=0}^{\infty} a_{\Delta} \alpha_n(\Delta) F_{\Delta_n}+\sum_{\Delta}\sum_{n=0}^{\infty} a_{\Delta} \beta_n(\Delta) \partial F_{\Delta_n}\nonumber\\
&\overset{?}{=}&\sum_{n=0}^{\infty}\left(\sum_{\Delta} a_{\Delta} \alpha_n(\Delta)\right) F_{\Delta_n}+\sum_{n=0}^{\infty}\left(\sum_{\Delta} a_{\Delta} \beta_n(\Delta)\right) \partial F_{\Delta_n}
\eea
If this is true, the functional bootstrap equations are not only necessary but sufficient. The detailed argument proving this below  is somewhat technical, but the main idea is that to show the double series commutes we need to have sufficient control over the OPE density at large $\Delta$, and this is possible thanks to the upper bound derived in the previous section.

Let us argue why it is indeed possible to commute the series. We will show this for the $\alpha_n$ functionals, the other case being completely analogous. We will start from the top expression and show that the functional equations imply it is zero. First note that
\bea
\sum_\Delta \sum_{n=0}^{\infty} a_{\Delta} \alpha_n(\Delta) F_{\Delta_n}(z)&=&\sum_{\Delta}\left(\sum_{n=0}^{N-1}+\sum_{n=N}^{\infty}\right) a_{\Delta}\alpha_n(\Delta) F_{\Delta_n}(z)\nonumber\\
&=& \sum_{\Delta} \sum_{n=N}^{\infty} a_{\Delta} \alpha_n(\Delta) F_{\Delta_n}(z)
\eea
We have commuted the sum with the series since each individual series is finite by assumption (and zero). Next we do the following manipulation:
\bea
&&\sum_{\Delta} \sum_{n=N}^{\infty} a_{\Delta} \alpha_n(\Delta)F_{\Delta_n}(z)=\lim_{\Delta_M\to \infty} \sum_{\Delta<\Delta_M-1} \sum_{n=N}^{\infty} a_{\Delta} \alpha_n(\Delta) F_{\Delta_n}(z)\nonumber \\
 &=& %\lim_{\Delta_M\to \infty} \sum_{n=N}^{\infty}\sum_{\Delta<\Delta_M}  a_{\Delta} \alpha_n(\Delta) F_{\Delta_n}(z)
-\lim_{\Delta_M\to \infty} \sum_{n=N}^{\infty}\sum_{\Delta\geq \Delta_M-1}  a_{\Delta} \alpha_n(\Delta) F_{\Delta_n}(z)
\eea
where in the second step we commuted a finite sum with an infinite one and used the functional equations once more. The unit shift in $\Delta_M$ is for later convenience. If we can show that this last expression vanishes we are done.

Let us then examine the limit. We begin by pointing out that since for any finite sum in $n$ the limit is indeed zero, it is not hard to see that the desired result can be established by showing that:
\bea
\forall \epsilon>0, \quad \exists N_{\epsilon},M_{\epsilon}:\qquad \left|\sum_{n=N_{\epsilon}}^{\infty}\sum_{\Delta\geq \Delta_{M_\epsilon}-1} a_{\Delta} \alpha_n(\Delta) F_{\Delta_n}(z)\right|<\epsilon \label{eq:epscond}.
\eea
To prove this we will use the upper bound \reef{eq:upperbound} on the OPE density. First, we bound the lefthand side by the double sum of the modulus of the summands, and consider the inner sum, which we rewrite as:
\bea
&&\sum_{\Delta\geq \Delta_{M_\epsilon}-1} \left|a_{\Delta} \alpha_n(\Delta)\right|=\sum_{k=M_{\epsilon}}^{\infty}\sum_{\Delta\in B_{k}} |a_{\Delta} \alpha_n(\Delta)|\nonumber \\
&= & a_{\Delta_n}^{\text{free}}\,\sum_{k=M_{\epsilon}}^{\infty} \sum_{\Delta \in B_{k}} \left|\frac{a_{\Delta}}{a_{\Delta}^{\text{free}}}\, \left\{\frac{2 \sin\left[\frac\pi 2(\Delta-\Delta_n)\right]}{\pi(\Delta-\Delta_k)}\right\}^2\, R_\alpha(\Delta,\Delta_n|\Df) (\Delta-\Delta_k)^2\right|.
\eea
As we show in appendix \ref{app:funcs}, the important property of $R_\alpha(\Delta,\Delta_n|\Df)$ is that for $\Delta,\Delta_n$ both large, it is given by a simple rational function with a double pole for $\Delta=\Delta_n$. For those bins $B_k$ such that $|\Delta_n-\Delta_k|\gg 1$, we can use the OPE upper bound to get:
\bea
\sum_{\Delta \in B_{k}} \left|\frac{a_{\Delta}}{a_{\Delta}^{\text{free}}}\, \frac{4 \sin^2\left[\frac\pi 2(\Delta-\Delta_n)\right]}{\pi^2(\Delta-\Delta_k)^2}\,  R_\alpha(\Delta,\Delta_n|\Df)(\Delta-\Delta_k)^2\right|\leq  R_\alpha(\Delta_k,\Delta_n|\Df)<1
\eea
In particular,
\bea
R_{\alpha}(\Delta_k,\Delta_n|\Df)\sim O(\Delta_k^{-3}),\qquad \Delta_k\gg \Delta_n\gg 1.
\eea
For those bins where $|\Delta_n-\Delta_k|\sim 1$ we have
\bea
R_\alpha(\Delta,\Delta_n)(\Delta-\Delta_k)^2\sim  R_\alpha(\Delta,\Delta_n)(\Delta-\Delta_n)^2\sim 1
\eea
and we can use the OPE upper bound again to obtain
\bea
\sum_{\Delta \in B_{k}} \left|\frac{a_{\Delta}}{a_{\Delta}^{\text{free}}}\, \frac{\left[\frac 2\pi \sin\left(\frac{\pi \Delta}2\right)\right]^2}{(\Delta-\Delta_k)^2}\,  R_\alpha(\Delta,\Delta_n)(\Delta-\Delta_k)^2\right|\lesssim 1
\eea
It is clear that the full sum over bins converges, and we conclude
\bea
\left|\sum_{\Delta\geq \Delta_{M_\epsilon}} a_{\Delta}\alpha_n(\Delta)\right|\leq a_{\Delta_n}^{\text{free}}\, p_n
\eea
where $p_n$ is some $n$-dependent polynomial of fixed degree independent of $M_{\epsilon}$. We are now nearly done since we have shown:
\bea
\sum_{n=N_{\epsilon}}^{\infty}\,\left|\sum_{\Delta\geq \Delta_{M_\epsilon}} a_{\Delta} \alpha_n(\Delta) F_{\Delta_n}(z)\right|\leq \sum_{n=N_\epsilon}^{\infty} |p_n| a_{\Delta_n}^{\text{free}} |F_{\Delta_n}(z)|.
\eea
Since we know that the OPE for the generalized free field solution converges exponentially fast \cite{Pappadopulo:2012jk}, the extra polynomial growth factor is irrelevant and hence by choosing $N_\epsilon$ appropriately we can make this arbitrarily small. We conclude that \reef{eq:epscond} is true, and that the functional bootstrap equations are indeed necessary and sufficient to establish that a set of OPE data satisfy the crossing equation.

%%%%%%%%%%%%%%%%%%%%%%%%%%%%%%%%%%%%%%%

\section{Witten diagrams in $AdS_2$}\label{sec:witten}
\subsection{Continuous families of solutions to crossing}
Let us imagine we are given the full set of CFT data of a conformal-invariant theory, satisfying unitarity and crossing. The theory may or may not include the stress tensor. It is natural to ask whether the theory admits a deformation of the CFT data which preserves unitarity and crossing. We can further restrict to deformations which do not introduce any additional degrees of freedom. Thus the deformation is described by a continuous family of CFT data $\Delta_i(g)$, $c_{ijk}(g)$ where $i,j,k$ run over all primary operators in the theory, and where $g$ is a real deformation parameter such that for $g=0$ we recover the original theory. For simplicity, let us focus on the four-point function $\mathcal{G}_g(z,\bar{z})$ of identical scalar primaries along such deformation. Thanks to crossing and unitarity, $\mathcal{G}_g(z,\bar{z})$ is bounded in the Regge limit for any $g$. If we assume the CFT data admit a series expansion for small $g$, we get a corresponding expansion of $\mathcal{G}_g(z,\bar{z})$
\be
\mathcal{G}_g(z,\bar{z}) = \mathcal{G}^{(0)}(z,\bar{z}) + \mathcal{G}^{(1)}(z,\bar{z})g^2 + \mathcal{G}^{(2)}(z,\bar{z})g^4+\ldots\,,
\ee
where we chose $g^2$ as the small parameter for future convenience. While $\mathcal{G}_g(z,\bar{z})$ has to be bounded in the Regge limit, there is in general no guarantee that the terms in the expansion $\mathcal{G}^{(m)}(z,\bar{z})$ with $m>0$ are themselves bounded in the Regge limit. However, one expects that the Regge boundedness of the finite-coupling correlator constrains the rate of growth of the perturbative terms in the Regge limit.\footnote{Consider the toy example $f_g(z)=\frac{1}{1-g^2 z}$, which is bounded as $z\rightarrow\infty$, but whose terms of the perturbative expansion diverge there. The rate of growth of the perturbative terms is proportional to the order in perturbation theory.}

The situations where we start with a theory with a stress tensor and where the stress tensor is present along the deformation are quite rare. Indeed, the only known examples of local theories with nontrivial conformal manifolds are two-dimensional or supersymmetric and come from deforming the path integral by an exactly marginal local operator. On the other hand, the existence of deformations is presumably more generic in non-local conformal theories. For example, the 3D Ising CFT admits a non-local deformation in the form of the long-range Ising model \cite{Paulos:2015jfa,Behan2017,Behan:2017emf,Behan:2018hfx}. Therefore, the picture of the space of conformal theories that our current understanding suggests is that of a finite-dimensional space of non-local theories, where local theories arise generically as isolated points, or sub-manifolds of nonzero codimension.

Many examples of non-local conformal theories can be obtained as the boundary duals of standard local, UV-complete quantum field theories placed in $AdS$ \cite{Paulos:2016fap,Carmi:2018qzm}. Any deformation of the bulk QFT preserving its UV-completeness and overall consistency will give rise to a family of unitary and crossing-symmetric boundary CFT data. In the simplest case, we can start with the theory of a single free massive scalar field in $AdS_{d+1}$. Its set of boundary correlators is known as the generalized free scalar theory. The theory can be deformed by local interaction vertices in the bulk Lagrangian. If we want to preserve UV-completeness in the bulk and thus get a full-fledged theory on the boundary, the interaction should be renormalizable. The simplest example is the mass term, which simply interpolates between generalized free fields of different scaling dimensions.

Non-renormalizable bulk vertices still give rise to perturbative deformations of the boundary CFT data, but there is no guarantee that such perturbative expansion can be completed to a non-perturbative family of conformal theories. References \cite{Heemskerk:2009pn,Heemskerk:2010ty} found that (in $d\geq 2$) there is in fact a one-to-one correspondence between bulk four-point vertices and leading-order deformations of the scalar four-point functions $\mathcal{G}^{(1)}(z,\bar{z})$ on the boundary, assuming the deformation only modifies the double-trace data at this order. Note that of these infinitely many vertices, only the scalar $\Phi^4$ interaction gives rise to $\mathcal{G}^{(1)}(z,\bar{z})$ which is bounded in the Regge limit.

\subsection{Scalar contact diagrams}\label{ssec:scalarContacts}
We can use the bases of functionals constructed in this paper to repeat this exercise in $AdS_2$, and then carry the procedure to higher orders in the coupling. Consider the theory of a real scalar field $\Phi$ in $AdS_2$ with mass fixed in the units of the AdS scale so that the boundary $\phi$ has dimension $\Df$. The counting of physically distinct bulk four-point vertices is equivalent to the counting of crossing-symmetric polynomial S-matrices in $2D$ Minkowski space. In $2\rightarrow2$ s-channel scattering in $2D$, the Mandelstam variable $u$ vanishes, while $s+t=4m^2$. Since crossing corresponds to $s\mapsto4m^2-s$, the crossing-symmetric S-matrices are linear combinations of $s^{b}(4m^2-s)^{b}$ with $b=0,1,\ldots$. The corresponding complete and independent set of quartic vertices can be written schematically as $(\partial^{b}\Phi)^4$ with $b=0,1,\ldots$.\footnote{The notation simply means that the vertex contains four $\Phi$ fields and $4b$ derivatives.} In a $2\rightarrow2$ scattering process in $2D$, there is only one way to take the high-energy limit, namely $s\rightarrow\infty$. We can think of it as the u-channel Regge limit since we can get it by boosting particles 1 and 3 by a large boost to the right and particles 2 and 4 by the same boost to the left. In other words, $u=0$ stays fixed, while $s$ and $t$ are becoming large. In this limit, the S-matrix of vertex $(\partial^{b}\Phi)^4$ behaves as $s^{2b}$.

In the boundary theory, we consider the four-point function $\langle\phi(x_1)\phi(x_2)\phi(x_3)\phi(x_4)\rangle$ and proceed exactly as in section \ref{ssec:bosonicBasis}. We want to solve for the double-trace data at tree-level $\gamma_n^{(1)}$ and $a_n^{(1)}$, which yield the OPE decomposition of the tree-level four-point function
\be
\mathcal{G}^{(1)}(z) = \sum\limits_{n=0}^{\infty}\left[a^{(1)}_n G_{2\Df+2n}(z) + a^{(0)}_n\gamma_n^{(1)} \partial G_{2\Df+2n}(z)\right]\,.
\label{eq:deformation1}
\ee
What is the expected Regge behaviour of $\mathcal{G}^{(1)}(z)$ corresponding to the bulk vertex $(\partial^{b}\Phi)^4$? In general, if the S-matrix goes as $s^{2j}$ at large $s$ with $u$ fixed, we expect
\be
z^{-2\Df}\mathcal{G}^{(1)}(z)\sim z^{j-1}\quad\textrm{as }z\rightarrow i\infty\,.
\ee
This can be derived for example from the Mellin representation of the contact diagrams. Therefore, for the vertex $(\partial^{b}\Phi)^4$ we expect $z^{-2\Df}\mathcal{G}^{(1)}(z)\sim z^{2b-1}$. In particular, only the interaction with no derivatives is bounded in the Regge limit.

This structure is reproduced in the language of functionals in the following way. As explained in section \ref{ssec:bosonicBasis}, the bosonic prefunctionals $\aB_n$, $\bB_n$ imply there is no consistent deformation of the form \reef{eq:deformation1} which decays at infinity as $z^{-2\Df}\mathcal{G}^{(1)}(z)=O (z^{-1-\epsilon})$ with $\epsilon>0$. This of course agrees with our counting of bulk vertices, since the interaction with the softest Regge behaviour goes as $z^{-1}$. The functionals $\aB_n$, $\bB_n$ are defined using the contour integrals \reef{eq:ffg} with $f(z)=O(1)$ as $z\rightarrow\infty$. In order to construct functionals which are compatible with solutions to crossing with worse Regge behaviour, we need to take linear combinations of $\aB_n$ and $\bB_n$ with improved Regge behaviour of $f(z)$. This can be accomplished from knowing the large-$z$ expansion of $f(z)$ for the prefunctionals. Recall that $f(z) = f(1-z)$ for all functionals. As in section \ref{ssec:bosonicBasis}, we can subtract an appropriate multiple of $\bB_0$ from the remaining prefunctionals to cancel the constant term of $f(z)$ at $z=\infty$. Thanks to the symmetry $f(z)=f(1-z)$, this automatically cancels the coefficient of $1/z$ as well. Therefore, we obtain functionals $\alpha_{0,n}$ for $n\geq 0$ and $\beta_{0,n}$ for $n\geq 1$ which are compatible with any crossing-symmetric deformation such that $z^{-2\Df}\mathcal{G}^{(1)}(z)=O (z^{1-\epsilon})$ with $\epsilon>0$. The first subscript refers to them being obtained by a subtraction of $\bB_0$ from the other elementary prefunctionals. These functionals fix all the double-trace data in terms of $\gamma_0^{(1)}$ and therefore, they allow precisely one linearly independent solution to crossing. This is precisely the contact diagram with no derivatives, which is indeed the unique solution with the stated Regge behaviour.

We can generalize this procedure to encompass solutions with arbitrarily fast Regge growth as follows. Let us single out the functionals $\bB_n$ with $n=0,\ldots,N$ and use them to improve the Regge behaviour of the remaining functionals. Thanks to the symmetry $f(z)=f(1-z)$, we can thus cancel all inverse powers of $z$ up to and including $z^{-2N-1}$. We call the resulting subtracted functionals $\alpha_{N,n}$ (with $n\geq 0$) and $\beta_{N,n}$ (with $n\geq N+1$). These functionals are compatible with any crossing-symmetric deformation such that $z^{-2\Df}\mathcal{G}^{(1)}(z)=O(z^{2N+1-\epsilon})$ with $\epsilon>0$, and they fix the double-trace data in terms of $\gamma_n^{(1)}$ with $n=0,\ldots,N$. Thus, there are precisely $N+1$ linearly independent crossing-symmetric deformations of the form \reef{eq:deformation1} with the stated Regge behaviour. These are exactly the contact interactions $(\partial^{b}\Phi)^4$ with $b=0,\ldots,N$.

We have used the above procedure to find the OPE decomposition of the first few contact interactions.\footnote{In practice, we can perform this procedure fully rigorously only when we have full control over the functional kernels, i.e. $\Df\in\mathbb{N}$ in the bosonic and $\Df\in\mathbb{N}+\frac{1}{2}$ in the fermionic case. The presented results are obtained by analytic continuation from these values.} For the contact diagram with no derivatives, we find
\be
\gamma^{(1)}_n(0) = \frac{(2 n)!\left(\Delta _{\phi }\right)_n^4 \left(4 \Delta _{\phi }-1\right)_{2 n}}{(n!)^2\left(2 \Delta _{\phi }\right)_n^2 \left(2 \Delta _{\phi }\right)_{2 n}^2}\,,
\label{eq:contact0}
\ee
in agreement with the known result. Here the argument in the round bracket refers to the label $b=0$ of the interaction. The normalization is chosen so that $\gamma^{(1)}_0=1$. For the contact diagram with four derivatives (up to a possible addition of the non-derivative diagram), we find
\ba
\gamma^{(1)}_n(1) = &\left(16 \Delta _{\phi }^5-13 \Delta _{\phi }^3-3 \Delta _{\phi }^2+16 n^4 \Delta _{\phi }+8 n^4+64 n^3 \Delta _{\phi }^2+\right.\\
&\left.+16 n^3 \Delta _{\phi }-8 n^3+96 n^2 \Delta _{\phi }^3+8 n^2 \Delta _{\phi }^2-24 n^2 \Delta _{\phi }-2 n^2+\right.\\
&\left.+64 n \Delta _{\phi }^4-28 n \Delta _{\phi }^2-2 n \Delta _{\phi }+2 n\right)\frac{n \left(4 \Delta _{\phi }+2 n-1\right)}{\left(\Delta _{\phi }+n-1\right) \left(2 \Delta _{\phi }+2 n+1\right)}\gamma^{(1)}_n(0)\,.
\ea
We also verified that the corrections to the OPE coefficients in all contact diagrams are given by the standard relation
\be
a_n^{(1)} = \frac{1}{2}\frac{\partial (a_n^{(0)}\gamma^{(1)}_n)}{\partial n}\,.
\label{eq:ope1Formula}
\ee

\subsection{Fermionic contact diagrams}
It is straightforward to repeat the reasoning of the previous section for the theory of a single massive Majorana fermion $\Psi$ in $AdS_2$. The boundary correlators in the free theory are those of the generalized free fermion. In this case we can not write down any renormalizable bulk interactions. There is no $\Psi^4$ vertex because it vanishes thanks to the fermionic statistics. Similar counting as in the bosonic case leads to the conclusion that the independent quartic vertices have $4b+2$ derivatives with $b\geq 0$ and there is a unique vertex for each $b$.

In order to bootstrap these contact diagrams in the boundary theory, we consider the four-point function $\langle\psi(x_1)\psi(x_2)\psi(x_3)\psi(x_4)\rangle$, where $\psi$ is a 1D Majorana fermion operator, dual to the one-particle state in the bulk. The set-up is the same as in section \ref{ssec:fermionicBasis}. Power-counting suggests that the bulk four-point vertex with $4b+2$ derivatives will give a correction to a four-point function such that $z^{-2\Dps}\mathcal{G}^{(1)}(z)\sim z^{2b+1}$ as $z\rightarrow i\infty$. The fermionic functionals $\aF_n$ and $\bF_n$ show that there are indeed no deformations of the kind we are interested in such that $z^{-2\Dps}\mathcal{G}^{(1)}(z)=O(z^{1-\epsilon})$ as $z\rightarrow i\infty$ with $\epsilon>0$. Similarly to the previous section, we can subtract appropriate linear combinations of $\bF_n$ with $n=0,\ldots,N$ from the remaining functionals to obtain functionals which are compatible with any crossing-symmetric four-point function such that $z^{-2\Dps}\mathcal{G}^{(1)}(z)= O(z^{2N+3-\epsilon})$ with $\epsilon>0$. The subtracted functionals fix the double-trace data in terms of $\gamma_n^{(1)}$ with $n=0,\ldots,N$. There are thus precisely $N+1$ independent deformations involving only corrections to the fermionic double-trace data such that $z^{-2\Dps}\mathcal{G}^{(1)}(z)= O(z^{2N+3-\epsilon})$ with $\epsilon>0$ in the Regge limit. These are precisely the fermionic four-point interactions with $4b+2$ derivatives and $b=0,\ldots,N$.

We used the large-$z$ expansion of the $f(z)$ kernel of the functionals $\aF_n$, $\bF_n$ to find the OPE data for the first few values of $b$. The expressions quickly get complicated so we include only the leading contact interaction (with $b=0$, i.e. two derivatives)
\ba
\gamma_n=&\left(64 \Delta _{\phi }^4+64 \Delta _{\phi }^3-4 \Delta _{\phi }^2-16 \Delta _{\phi }+64 n^4+256 n^3 \Delta _{\phi }+64 n^3+384 n^2 \Delta _{\phi }^2+\right.\\
&\left.+192 n^2 \Delta _{\phi }-8 n^2+256 n \Delta _{\phi }^3+192 n \Delta _{\phi }^2-16 n \Delta _{\phi }-12 n-3\right)\times\\
&\times\frac{\Gamma \left(n+\frac{3}{2}\right)\Gamma \left(n+\Delta _{\phi }-\frac{1}{2}\right) \Gamma \left(n+\Delta _{\phi }+\frac{1}{2}\right) \Gamma \left(n+2 \Delta _{\phi }+\frac{1}{2}\right)}{\Gamma (n+1) \Gamma \left(n+\Delta _{\phi }+1\right) \Gamma \left(n+\Delta _{\phi }+2\right) \Gamma \left(n+2 \Delta _{\phi }\right)}\,.
\ea
The corrections to the OPE coefficients are again given by the formula \reef{eq:ope1Formula}.

\subsection{Higher orders: universality up to two loops}
Intuition from flat space indicates that loop-level diagrams are not independent of tree-level diagrams. Indeed, if one knew all tree-level scattering amplitudes in a perturbative QFT, one could fix all the vertices in the Lagrangian, which could then be used to find the amplitudes at arbitrarily high order in perturbation theory, at least in principle. Therefore, tree-level determines the theory to all orders. At its core, this principle is just unitarity of the underlying theory. Quantitatively, it can be expressed as cutting rules for Feynman diagrams \cite{Cutkosky:1960sp}. We expect a similar principle to apply for weakly-coupled field theory in $AdS$, or equivalently for the perturbation theory in CFTs around the generalized free field. Indeed, the authors of \cite{Aharony:2016dwx,Liu:2018jhs} found that the conformal bootstrap equations fix certain one-loop diagrams in $AdS_{d+1}$ for $d\geq 2$. Here we would like to explain how one can use bootstrap functionals to find some one- and two-loop Witten diagrams in $AdS_2$.

Before we compute the diagrams, we will make general comments about what kinds of contributions can be expected at increasing orders in perturbation theory. We consider the most general renormalizable Lagrangian for a real scalar field in 2D which preserves the global $\mathbb{Z}_2$ symmetry
\be
\mathcal{L} = \frac{1}{2}(\partial\Phi)^2 - \frac{1}{2}m^2\Phi^2 - \sum\limits_{n=2}^{\infty}\frac{\lambda_{2n}}{(2n)!}\Phi^{2n}\,.
\label{eq:lagrangian}
\ee
All UV divergences of Feynman graphs coming from this Lagrangian can be removed by normal-ordering the vertices. We want to perform the standard perturbation theory in the number of loops. This is equivalent to writing
\be
\lambda_{2n}=\mu_{2n} g^{2n-2}\,,
\ee
and performing perturbation theory in $g$ while keeping $\mu_n$ fixed. Consider the Feynman graphs contributing to the flat space $2\rightarrow 2j$ scattering amplitude in this theory. The leading contribution comes from the tree-level diagram produced by vertex $\Phi^{2j+2}$ and the amplitude is thus proportional to $g^{2j}$. Thanks to unitarity, the $2\rightarrow2$ scattering amplitude will have branch cuts starting at $s=(2 j M)^2$ corresponding to the production of the $2j$-particle states, where $M$ is the physical mass of the single-particle states. The discontinuity across the branch cut is proportional to the square of the $2\rightarrow 2j$ amplitude, and hence to $g^{4j}$. Since any $L$-loop graph for the $2\rightarrow2$ scattering goes as $g^{2L+2}$, we see that the $2j$-particle cut appears first at $2j-1$ loops. In particular, diagrams up to and including two loops only contain two-particle cuts.

Let us place the theory with Lagrangian \reef{eq:lagrangian} inside $AdS_2$ in a way respecting the isometries. We adjust $m^2$ in such a way that $\Df$ is independent of $g$. For $g=0$, the spectrum of the primary operators on the boundary consists of the identity, the single-particle state $\phi$ of dimension $\Df$, two-particle states $[\phi^2]_n$ of dimensions $\Delta_n = 2\Df+2n$, as well as $k$-particle states $[\phi^k]_I$ of dimensions $k\Df+\textrm{integers}$ for all $k\geq 3$. Consider the $\phi\times\phi$ OPE. For $g=0$, only the identity and two-particle states appear in the OPE. For $g>0$, higher-particle states can appear in this OPE. By the global $\mathbb{Z}_2$ symmetry, only states with even $k$ can appear. The set of Witten graphs contributing to the boundary three-point function of two $\phi$s and one $2j$-particle state is the same as the set of Feynman graphs contributing to the $2\rightarrow 2j$ amplitude. Therefore, this three-point function goes as $g^{2j}$ as $g\rightarrow 0$. If we expand the four-point function $\langle\phi\phi\phi\phi\rangle$ in the OPE, we find that the leading contribution of the exchange of $2j$-particle states goes as $g^{4j}$ for $j>1$ and therefore these states first appear at $2j-1$ loops.\footnote{We can include the $j=1$ case by saying that the double discontinuity of the four-point function first receives contribution from the exchange of $2j$-particle states at $2j-1$ loops. In this way, the statement is completely analogous to the flat-space discussion of the previous paragraph if we replace double discontinuity of the four-point function with the discontinuity of the scattering amplitude.} In other words, the four-point function up to and including two loops comes entirely from perturbative corrections to the two-particle states.

\subsection{Using functionals to calculate loop diagrams}
\begin{figure}[ht!]%
\begin{center}
\includegraphics[width=0.9\textwidth]{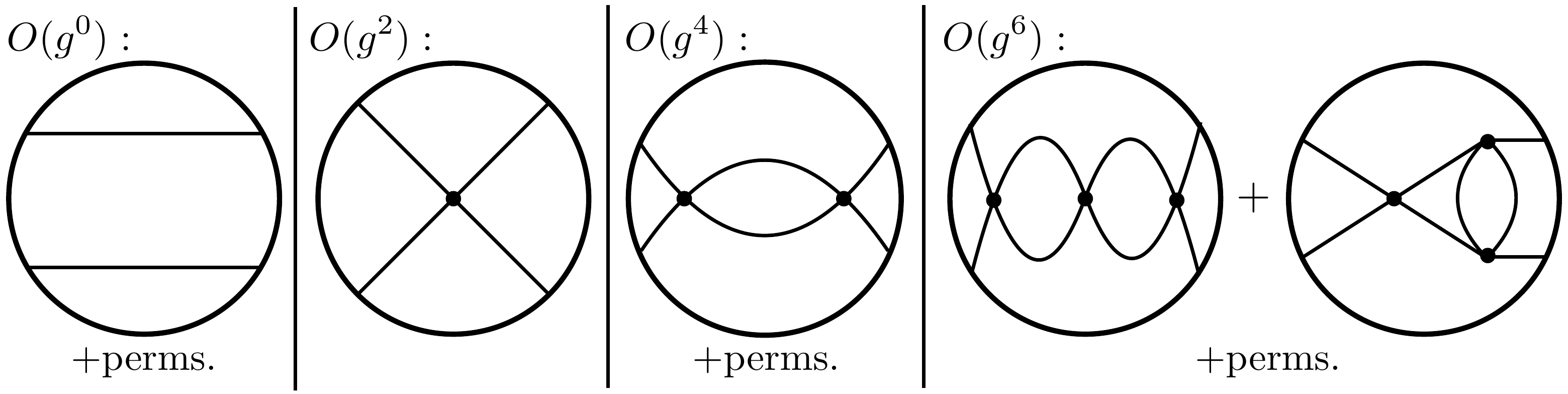}%
\caption{The Witten diagrams contributing to the four-point function at increasing orders in perturbation theory. ``$+\textrm{perms.}$'' means that diagrams obtained by permuting the external legs should be included.}
\label{fig:Witten}%
\end{center}
\end{figure}

\begin{figure}[ht!]%
\begin{center}
\includegraphics[width=0.17\textwidth]{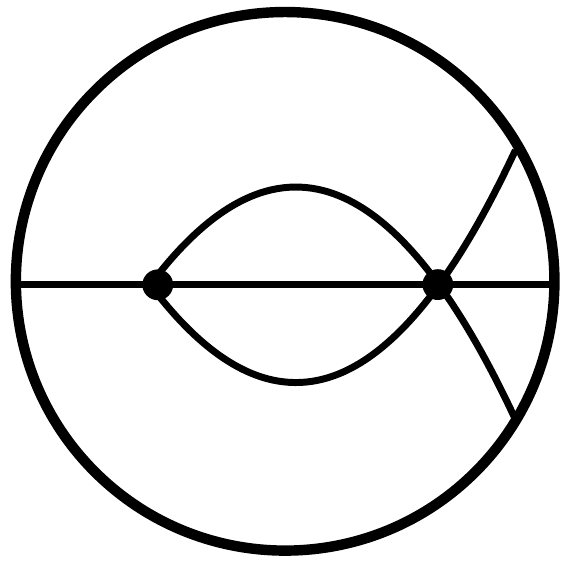}%
\caption{This $O(g^6)$ diagram does not need to be included because it is proportional to the four-point contact diagram, and therefore only renormalizes the four-point coupling.}
\label{fig:Witten46}%
\end{center}
\end{figure}

We will now explain how to use bootstrap functionals to compute the one- and two-loop contributions to the four-point function $\langle\phi\phi\phi\phi\rangle$. The basic idea is that once we have determined all double-trace data at a given order in perturbation theory, we can use the functionals $\alpha_n$ and $\beta_n$ to compute the OPE coefficient and anomalous dimension of $[\phi^2]_n$ at the next order. Let us write the perturbative expansion of the four-point function up to two loops as follows
\be
\mathcal{G}(z) = \mathcal{G}^{(0)}(z)+\mathcal{G}^{(1)}(z)g^2+\mathcal{G}^{(2)}(z)g^4+\mathcal{G}^{(3)}(z)g^6+O(g^8)\,.
\ee
Figure \ref{fig:Witten} shows the Witten diagrams contributing at the various orders of perturbation theory. Except for $O(g^0)$, we only include connected diagrams. The only role of higher-order disconnected diagrams is to renormalize $m^2$ of the boundary-to-boundary propagators, which we are fixing by keeping the external dimension fixed at $\Df$. Similarly, only amputated diagrams are included, since we are keeping $\Df$ fixed in the bulk-to-boundary propagators. Finally, one might think that the four-point function at $O(g^6)$ can depend on the $\Phi^6$ coupling thanks to the diagram including one four-point and one six-point vertex, shown in Figure \ref{fig:Witten46}. However, this diagram is in fact proportional to the four-point contact diagram already included at $O(g^2)$, and therefore its only effect is to renormalize the four-point coupling.

In summary, one can think of our set-up as a kind of on-shell renormalization scheme, where the renormalization conditions are to keep $\Df$ and $\Delta_{\phi^2}$ fixed, the perturbative parameter being $\Delta_{\phi^2}-2\Df\sim g^2$. The four-point function up to $O(g^6)$ is uniquely fixed in terms of $\Df$ and $\Delta_{\phi^2}$.

We will assume that $z^{-2\Df}\mathcal{G}^{(i)}(z)$ for $i=1,2,3$ are bounded as $z\rightarrow i\infty$. This assumption is presumably possible to prove from the fact that all interaction vertices are relevant, and is ultimately justified by the self-consistency of the answers we find. Let us write the perturbative expansion of the double-trace data up to two loops as follows
\ba
\Delta_n(g) &= 2\Df+2n + \gamma_n^{(1)}g^2+ \gamma_n^{(2)}g^4+ \gamma_n^{(3)}g^6+O(g^8)\\
a_n(g) &= a_n^{(0)} + a_n^{(1)}g^2+ a_n^{(2)}g^4+ a_n^{(3)}g^6+O(g^8)\,.
\ea
We would like to fix the anomalous dimensions and OPE coefficients from crossing symmetry. Before we do that, note that the bulk coupling $g^2$ does not map to any natural boundary observable. Thus instead of parametrizing the deformation by $g^2$, we will parametrize it by the anomalous dimension of $\phi^2$. Thus we define a new coupling
\be
\widetilde{g}^2 = \Delta_0(g) - 2\Df = \gamma_0^{(1)}g^2+ \gamma_0^{(2)}g^4+ \gamma_0^{(3)}g^6+O(g^8)
\ee
and express all OPE data in terms of $\widetilde{g}^2$. This procedure is well-defined since $\gamma^{(1)}_0\neq 0$. In fact, we will drop the tilde and call the new coupling $g^2$ by a small abuse of notation. In other words, we are fixing the gauge of the coupling reparametrization invariance by requiring
\be
\gamma_0^{(1)} = 1,\quad \gamma_0^{(j)}= 0\,\textrm{ for }j\geq 2\,.
\label{eq:gaugeFix}
\ee
Note that as a result of this reparametrization, our results for the OPE data at one and two loops will in general contain contributions from Witten diagrams of lower loop orders.

The OPE data yields the following OPE sums for the perturbative correlator, where we introduced the shorthand notation $\Delta_n = 2\Df + 2n$
\begin{align}
\mathcal{G}^{(0)}(z) &= \sum\limits_{n=0}^{\infty}a^{(0)}_nG_{\Delta_n}(z)\\
%%%
\mathcal{G}^{(1)}(z) &= \sum\limits_{n=0}^{\infty}\left[a^{(1)}_nG_{\Delta_n}(z)+a^{(0)}_n\gamma_n^{(1)}\partial G_{\Delta_n}(z)\right]\\
%%%
\mathcal{G}^{(2)}(z) &= \sum\limits_{n=0}^{\infty}\left\{
a^{(2)}_nG_{\Delta_n}(z)+
\left[a^{(0)}_n\gamma^{(2)}_n + a^{(1)}_n\gamma_n^{(1)}\right]\partial G_{\Delta_n}(z)+\frac{1}{2}a^{(0)}_n(\gamma^{(1)}_n)^2\partial^2 G_{\Delta_n}(z) \right\}\\
%%%
\mathcal{G}^{(3)}(z) &= \sum\limits_{n=0}^{\infty}\left\{
a^{(3)}_nG_{\Delta_n}(z)+
\left[a^{(0)}_n\gamma^{(3)}_n + a^{(1)}_n\gamma_n^{(2)}+ a^{(2)}_n\gamma_n^{(1)}\right]\partial G_{\Delta_n}(z) +\phantom{\frac{1}{2}}\right.\\\nonumber
&\left. \qquad\qquad +\left[a^{(0)}_n\gamma^{(1)}_n\gamma^{(2)}_n+\frac{1}{2}a^{(1)}_n(\gamma^{(1)}_n)^2\right]\partial^2 G_{\Delta_n}(z) + \frac{1}{6}a^{(0)}_n(\gamma^{(1)}_n)^3\partial^3 G_{\Delta_n}(z)\right\}\,,
\end{align}
where $\partial^j G_{\Delta_n}(z)$ denotes the $j$th derivative of the conformal block with respect to $\Delta$, evaluated at $\Delta=\Delta_n$. $\mathcal{G}^{(j)}(z)$ includes $\Delta$-derivatives of the double-trace conformal blocks of maximal order $j$. The OPE data $\gamma^{(j)}_n$ and $a^{(j)}_n$ appear in $\mathcal{G}^{(j)}(z)$ only in coefficients of $\partial G_{\Delta_n}(z)$ and $G_{\Delta_n}(z)$ respectively, but not in any terms with a higher $\Delta$-derivative of conformal blocks.

The last fact implies that we can solve for $\gamma^{(j)}_n$ and $a^{(j)}_n$ recursively in the order of perturbation theory. In section \ref{ssec:scalarContacts}, we have already found the tree-level quantities $\gamma_n^{(1)}$ and $a^{(1)}_n$, see equations \reef{eq:contact0} and \reef{eq:ope1Formula}. We will now use the crossing symmetry and Regge boundedness of $\mathcal{G}^{(2)}(z)$ to solve for $\gamma^{(2)}_n$ and $a^{(2)}_n$. The crossing equation at one loop reads
\be
\sum\limits_{n=0}^{\infty}\left\{
a^{(2)}_nF_{\Delta_n}(z)+
\left[a^{(0)}_n\gamma^{(2)}_n + a^{(1)}_n\gamma_n^{(1)}\right]\partial F_{\Delta_n}(z)+\frac{1}{2}a^{(0)}_n(\gamma^{(1)}_n)^2\partial^2 F_{\Delta_n}(z) \right\} = 0\,.
\label{eq:crossingOneLoop}
\ee
Recall from section \ref{ssec:scalarContacts} the complete basis of bosonic functionals $\alpha_m$ and $\beta_m$ obtained by an appropriate subtraction of $\beta^{\textrm{B}}_0$ from $\alpha^{\textrm{B}}_m$ and $\beta^{\textrm{B}}_m$:
\ba
\alpha_m &= \alpha^{\textrm{B}}_m - \frac{a^{(1)}_m}{a^{(0)}_0\gamma^{(1)}_0}\beta^{\textrm{B}}_0 = \alpha^{\textrm{B}}_m - \frac{a^{(1)}_m}{2}\beta^{\textrm{B}}_0\\
\beta_m &= \beta^{\textrm{B}}_m - \frac{a^{(0)}_m\gamma^{(1)}_m}{a^{(0)}_0\gamma^{(1)}_0}\beta^{\textrm{B}}_0=\beta^{\textrm{B}}_m - \frac{a^{(0)}_m\gamma^{(1)}_m}{2}\beta^{\textrm{B}}_0\,,
\ea
where we used \reef{eq:gaugeFix} to simplify the expressions. Let us apply $\beta_m$ to the one-loop crossing equation \reef{eq:crossingOneLoop}. $\beta_m$ commutes with the infinite sum because of Regge boundedness. Inside the sum, we get zero contribution from the term proportional to $F_{\Delta_n}(z)$ since $\beta_m(\Delta)$ vanishes on all double traces. From the term proportional to $\partial F_{\Delta_n}(z)$, only the summands with $n=m$ and $n=0$ contribute thanks to the defining Kronecker-delta behaviour of $\beta_m$ on derivatives of double traces. Therefore, we find the equations
\be
a^{(0)}_m\gamma^{(2)}_m + a^{(1)}_m\gamma_m^{(1)} + r^{(2)}_m = 
\frac{a^{(0)}_m\gamma^{(1)}_m}{a^{(0)}_0\gamma^{(1)}_0}\left[a^{(0)}_0\gamma^{(2)}_0 + a^{(1)}_0\gamma_0^{(1)} + r^{(2)}_0\right]\,,
\ee
where we defined the infinite sums
\be
r^{(2)}_m = \sum\limits_{n=0}^{\infty}\frac{1}{2}a^{(0)}_n(\gamma^{(1)}_n)^2\partial^2\beta^{\textrm{B}}_m(\Delta_n)\,.
\ee
Note that $r^{(2)}_m$ is fixed by the tree-level OPE data. The $m=0$ equation is satisfied trivially, and the $m\geq1$ equations allow us to solve for the one-loop anomalous dimensions
\ba
\gamma^{(2)}_m &= -\frac{r^{(2)}_m}{a^{(0)}_m} + \left[\frac{\gamma^{(2)}_0}{\gamma^{(1)}_0} + \frac{a^{(1)}_0}{a^{(0)}_0} - \frac{a^{(1)}_m}{a^{(0)}_m} + \frac{r^{(2)}_0}{a^{(0)}_0\gamma^{(1)}_0}\right]\gamma^{(1)}_m =\\
&=-\frac{r^{(2)}_m}{a^{(0)}_m} + \left[\frac{a^{(1)}_0}{2} - \frac{a^{(1)}_m}{a^{(0)}_m} + \frac{r^{(2)}_0}{2}\right]\gamma^{(1)}_m\,.
\label{eq:gammaOneLoop}
\ea
Thus the one-loop anomalous dimensions are determined provided we can evaluate the sums $r^{(2)}_m$. Similarly, in order to find the one-loop OPE coefficients $a^{(2)}_m$, we apply the functional $\alpha_m$ to the crossing equation \reef{eq:crossingOneLoop}. We find
\be
a^{(2)}_m =- q^{(2)}_m+ \frac{a^{(1)}_m}{a^{(0)}_0\gamma^{(1)}_0}\left[a^{(0)}_0\gamma^{(2)}_0 + a^{(1)}_0\gamma_0^{(1)} + r^{(2)}_0\right]=
 -q^{(2)}_m+\frac{a^{(1)}_0 + r^{(2)}_0}{2}a^{(1)}_m
\,,
\label{eq:opeOneLoop}
\ee
where
\be
q^{(2)}_m = \sum\limits_{n=0}^{\infty}\frac{1}{2}a^{(0)}_n(\gamma^{(1)}_n)^2\partial^2\alpha^{\textrm{B}}_m(\Delta_n)\,.
\ee

In the next subsection, we will present a worked example at $\Df=1$, where the infinite sums can be found analytically. Having found the double-trace OPE data at one loop, it is straightforward to carry the argument to two loops. The two-loop crossing equation takes the form
\ba
\sum\limits_{n=0}^{\infty}&\left\{
a^{(3)}_nF_{\Delta_n}(z)+
\left[a^{(0)}_n\gamma^{(3)}_n + a^{(1)}_n\gamma_n^{(2)}+ a^{(2)}_n\gamma_n^{(1)}\right]\partial F_{\Delta_n}(z) +\phantom{\frac{1}{2}}\right.\\\nonumber
&\left.\;\; +\left[a^{(0)}_n\gamma^{(1)}_n\gamma^{(2)}_n+\frac{1}{2}a^{(1)}_n(\gamma^{(1)}_n)^2\right]\partial^2 F_{\Delta_n}(z) + \frac{1}{6}a^{(0)}_n(\gamma^{(1)}_n)^3\partial^3 F_{\Delta_n}(z)\right\} = 0\,.
\ea
If we apply the functional $\beta_m$ to this equation, we can solve for $\gamma^{(3)}_m$ in terms of the OPE data at lower orders. Similarly, we can solve for $a^{(3)}_m$ by applying the functional $\alpha_m$. The result is
\ba
\gamma^{(3)}_m &= -\frac{r^{(3)}_m}{a^{(0)}_m}-\frac{a^{(1)}_m}{a^{(0)}_m}\gamma^{(2)}_m
 + \left[
 \frac{r^{(3)}_0+a^{(2)}_0}{2} - \frac{a^{(2)}_m}{a^{(0)}_m}\right]
 \gamma^{(1)}_m\\
 a^{(3)}_m &=-q^{(3)}_m+\frac{a^{(2)}_0 + r^{(3)}_0}{2}a^{(1)}_m\,,
 \label{eq:twoLoopData}
\ea
where we defined the following infinite sums over the lower-order OPE data
\ba
r^{(3)}_m &= \sum\limits_{n=0}^{\infty}\left\{
\left[a^{(0)}_n\gamma^{(1)}_n\gamma^{(2)}_n+\frac{1}{2}a^{(1)}_n(\gamma^{(1)}_n)^2\right]\partial^2\beta^{\textrm{B}}_m(\Delta_n) + \frac{1}{6}a^{(0)}_n(\gamma^{(1)}_n)^3\partial^3\beta^{\textrm{B}}_m(\Delta_n)\right\}\\
q^{(3)}_m &= \sum\limits_{n=0}^{\infty}\left\{
\left[a^{(0)}_n\gamma^{(1)}_n\gamma^{(2)}_n+\frac{1}{2}a^{(1)}_n(\gamma^{(1)}_n)^2\right]\partial^2\alpha^{\textrm{B}}_m(\Delta_n) + \frac{1}{6}a^{(0)}_n(\gamma^{(1)}_n)^3\partial^3\alpha^{\textrm{B}}_m(\Delta_n)\right\}\,.
\ea

\subsection{Example: $\Df=1$}
We will now carry out the above algorithm in practice for $\Df=1$. In this case, we have
\be
a^{(0)}_n = \frac{(2 n+1)! (2 n+2)!}{(4 n+1)!}\,.
\ee
From \reef{eq:contact0} and \reef{eq:ope1Formula}, we get the OPE data at tree-level
\ba
\gamma^{(1)}_n &= \frac{1}{(n+1)(2n+1)}\\
a^{(1)}_n &=-\left[2 H_{4 n+1}-H_{2 n}-H_{2 n+1}\right] \frac{2 (2 n+1)! (2 n)!}{(4 n+1)!}\,,
\ea
where $H_{x}$ is the harmonic number. The tree-level four-point function can in fact be found in closed form
\be
\mathcal{G}^{(1)}(z) = 2z^2 \left[\frac{\log (1-z)}{z}+\frac{\log (z)}{1-z}\right]\,.
\ee
We can now use formula \reef{eq:gammaOneLoop} to find $\gamma^{(2)}_n$. It is possible to find a closed formula valid for general $n$
\ba
\gamma^{(2)}_n =
&-\frac{2 n (2 n+3)}{(n+1) (2 n+1)} \zeta (3)+\frac{H_{2 n}}{(n+1)^2 (2 n+1)^2}
+\frac{n \left(2 n^2+5 n+4\right)}{2 (n+1)^3 (2 n+1)^2}-\\
&-\frac{1}{(2 n+2) (2 n+1)}
\sum\limits_{j=1}^{2 n} \frac{(-1)^j (2 n-j+1)_{2 j+2}}{j ((j+1)!)^2}\left(H_j^2+H_j^{(2)}\right)\,.
\ea
Here we use the following notation for harmonic numbers of higher rank
\be
H^{(r)}_n = \sum\limits_{j=1}^{n}\frac{1}{j^r}\,,
\ee
so that in particular $H_n=H^{(1)}_n$. In practice, the result was found by evaluating the sums $r^{(2)}_n$ to a high precision, recognizing them as a rational linear combination of $\zeta(3)$ and 1, and finally recognizing the form for general $n$. For the first few values of $n$, the formula gives
\ba
\gamma^{(2)}_{0} &= 0\quad\textrm{(by definition of the coupling)}\\
\gamma^{(2)}_{1} &= -\frac{5}{3} \zeta (3)+ \frac{317}{144}\\
\gamma^{(2)}_{2} &= -\frac{28}{15} \zeta (3)+ \frac{25127}{10800}\,.
\ea
Similarly, we can apply formula \reef{eq:opeOneLoop} to find the corrections to OPE coefficients at one loop. In this case, we were not able to find a formula valid for general $n$. The first few values read
\ba
a^{(2)}_0 &= \frac{\pi ^4}{15}-4 \zeta (3)+\frac{5}{2}\\
a^{(2)}_1 &= \frac{\pi ^4}{25}+\frac{19}{15}\zeta (3)-\frac{612119}{108000}\\
a^{(2)}_2 &=\frac{\pi ^4}{126}+\frac{1177}{2835} \zeta (3)-\frac{3889170127}{3000564000}\,.
\ea
We were able to determine the full four-point function at one loop, from which the coefficients $a^{(2)}_n$ can be read off
\ba
\mathcal{G}^{(2)}&(z) =
\frac{1}{(1-z)^2}\left\{\phantom{\frac{\pi ^2}{3}\!\!\!\!\!\!\!}
4 (z-2) z^3 \text{Li}_4(1-z)+4 (1-z)^2 \left(z^2-1\right) \text{Li}_4(z)+\right.\\
&+4 (2 z-1) \text{Li}_4\left(\mbox{$\frac{z}{z-1}$}\right)
-\frac{ \pi ^4}{90} z^2 \left(z^2-2 z-6\right)-\frac{4z-2}{3}\log (z) \log ^3(1-z)-\\
&-2 (1-z)^2\left[\left(z^2-1\right) \log (1-z)+\left(z^2+2\right)\log (z)\right]\text{Li}_3(1-z)-\\
&-2z^2\left[(z-2) z \log (z)+\left(z^2-2 z+3\right) \log (1-z)\right]\text{Li}_3(z)-(z-1) z^2 \log (z)+\\
&+\left[2 \log (z)-2 \left(2 z^3-3 z^2+4 z-1\right) \log \left(\mbox{$\frac{z}{1-z}$}\right)\right]\zeta (3)
+\frac{2 z-1}{6} \log ^4(1-z)+\\
&+\left[\frac{\pi ^2}{3} (2 z-1)-(z-1)^2 \left(z^2+1\right) \log ^2(z)\right] \log ^2(1-z)+\\
&\left.+(1-z)^2 \left[\frac{\pi ^2}{3} \left(z^2+2\right) \log (z)+z\right] \log (1-z)
\right\}\,.
\ea
One can explicitly check that this expression is crossing symmetric. We also checked that it agrees with the crossing-symmetric combination of one-loop bubble diagrams in $AdS_2$, up to a tree-level contact interaction.

Finally, let us present the results at two loops, coming from formulas \reef{eq:twoLoopData}. The evaluation of the infinite sums over double-trace operators becomes much more demanding. We could find the first few anomalous dimensions and the first OPE coefficient:
\ba
\gamma^{(3)}_0 &= 0\quad\textrm{(by definition of the coupling)}\\
\gamma^{(3)}_1 &= -\frac{20 \zeta (3)^2}{3}-10 \zeta (5)+\frac{\pi ^4}{18}-\frac{329 \zeta (3)}{36}+\frac{1225 \pi ^2}{2592}+\frac{209}{486}\\
\gamma^{(3)}_2 &= -\frac{112 \zeta (3)^2}{15}+\frac{56 \zeta (5)}{5}+\frac{7 \pi ^4}{150}-\frac{29503 \zeta (3)}{2700}+\frac{174979 \pi ^2}{162000}-\frac{45033217}{16200000}
\ea
\be
a^{(3)}_0 =\frac{4 \pi ^4 \zeta (3)}{15} +\frac{22 \pi ^6}{945}-56 \zeta (3)^2-10 \pi ^2 \zeta (3)+136 \zeta (5)+\frac{\pi ^4}{90}-\frac{\pi ^2}{2}+3\,.
\ee
These results are a prediction of the conformal bootstrap for two-loop Witten diagrams in $AdS_2$. It would be interesting to check them against an explicit evaluation of said diagrams.

%%%%%%%%%%%%%%%%%%%%%%%%%%%%%%%%%%%%%%%

\section{Discussion}\label{sec:discussion}

%%%%%%%%%%%%%%%%%%%%%%%%%%%%%%%%%%%%%%%

In this paper, we have constructed two interesting bases for the $SL(2)$ crossing equation. The bases provide a direct bridge between the analytic and numerical bootstrap. Expressing the crossing equation in either basis leads to sum rules satisfied by the OPE data of any crossing-symmetric four-point function in a unitary theory. Regge boundedness of such four-point functions plays a crucial role in deriving the sum rules, and indeed the sum rules will not be satisfied by a general crossing-symmetric four-point unless it is bounded in the Regge limit. The sum rules can be obtained rigorously by applying suitable linear functionals to the standard crossing equation.

The elements of our bases are labelled by the double-trace operators in mean field theory. More precisely, we get two functionals (basis elements) for every double trace, denoted $\alpha_n$ and $\beta_n$, where $n$ labels the double traces. The contribution of a primary operator of dimension $\Delta$ to the sum rules is weighted by universal functions $\alpha_n(\Delta)$ and $\beta_n(\Delta)$. These functions coincide with coefficients of the double-trace conformal blocks in the OPE decomposition of the crossing-symmetric sum of Witten exchange diagrams in $AdS_2$ with exchanged dimension $\Delta$. We call the latter objects the Polyakov blocks. The validity of the sum rules discussed in this paper implies that the four-point function can be expanded not only in conformal blocks but also in Polyakov blocks, with the same coefficients. In this way, our approach gives a derivation of the Polyakov-Mellin approach to the conformal bootstrap for the $SL(2)$ crossing equation.

We have discussed several applications of the sum rules. One of them is the analytic bootstrap of the anomalous dimension and OPE coefficients in perturbative theories in $AdS_2$ up to two loops. Another application is the derivation of upper and lower bounds on OPE coefficients in general unitary solutions to crossing, valid for sufficiently heavy exchanged operators. These results are similar to those of the analytic Euclidean bootstrap of \cite{Mukhametzhanov:2018zja}, which studied the large-$\Delta$ tails of the OPE density. The advantage of our bounds is that they apply to \emph{individual} primary operators at large but \emph{finite} scaling dimension.\footnote{Of course, the advantage of the results of \cite{Mukhametzhanov:2018zja} is that they apply to OPE density of primaries under $SO(2,D)$ for $D>1$ rather than $SO(2,1)$ primaries as is the case in our work.} Our bounds apply to squared OPE coefficients of primary operators $a_{\mathcal{O}}\equiv (c_{\phi\phi\mathcal{O}})^2$ divided by an exponentially decreasing function $a_{\Delta}^{\textrm{free}}$ which interpolates between the squared OPE coefficients in the generalized free field. The upper bound implies that the sum of $a_{\mathcal{O}}/a_{\Delta_{\mathcal{O}}}^{\textrm{free}}$ over all primary operators in the $\phi\times\phi$ OPE with $\Delta_{\mathcal{O}}$ between $2\Df+2n-1$ and $2\Df+2n+1$ is at most $\frac{\pi^2}{4}+\epsilon_n$, where $\epsilon_n\rightarrow 0$ as $n\rightarrow\infty$. This implies the same upper bound on $a_{\mathcal{O}}/a_{\Delta_{\mathcal{O}}}^{\textrm{free}}$ for any individual primary $\mathcal{O}\in\phi\times\phi$ present in the same interval.

The lower bound implies that the sum of $a_{\mathcal{O}}/a_{\Delta_{\mathcal{O}}}^{\textrm{free}}$ for $\Delta_\mathcal{O}$ between $2\Df+2n-2$ and $2\Df+2n+2$ must be at least $1-\epsilon'_n$, where $\epsilon'_n\rightarrow0$ as $n\rightarrow\infty$. In particular, there must be at least one operator in this interval for sufficiently large $n$. Since this result holds also for the fermionic mean field theory, we conclude that there are no gaps larger than 5 in the spectrum of primaries in the OPE, from some finite $\Delta$ onwards.

The existence of the upper bound discussed above allowed us to prove that the functional bootstrap equations are completely equivalent to the standard crossing equation. In other words, if a putative set of OPE data satisfies all the sum rules from our basis, then it gives rise to a crossing-symmetric four-point function.

The essential feature of the sum rules of this article are the prefactors $\sin^2\!\left[\frac{\pi}{2}(\Delta-2\Df)\right]$, which provide the double zeros at double traces while maintaining positivity. This structure is reminscent of the double discontinuity which enters in the Lorentzian OPE inversion formula of Caron-Huot \cite{Caron-Huot2017b}. It turns out that there is indeed a version of the Lorentzian inversion formula which underlies the functionals of the present article, as will be explained in an upcoming work \cite{Mazac:2018qmi}.

There are several natural generalizations and possible future directions stemming from this work. We believe the logic presented here should carry over universally to various problems governed by a version of the crossing equation. Most importantly, these are the crossing equation in $D>1$. The notions of the mean field theory and also of Witten exchange diagrams exist also in that case so that a generalization should be feasible. The case of nonidentical external operators, and systems of multiple correlators should also be addressed.

Another natural avenue is the modular bootstrap, where there exists a direct connection between our approach and the recent solution of the sphere-packing problem in 8 and 24 dimensions \cite{2016arXiv160304246V,2016arXiv160306518C,ModularSphere}. It is likely that adapting the techniques of the present work would be effective for improving the bounds of \cite{Hellerman:2009bu,Friedan:2013cba,Collier:2016cls} at large central charge.

The construction of the functionals $\alpha_n$ and $\beta_n$ central to this paper came directly from an effort to understand the optimal functionals of the numerical bootstrap and how optimal solutions may be smoothly deformed or ``flowed'' into each other \cite{El-Showk:2016mxr}. Indeed, our functionals are the optimal functionals whenever the generalized free field is the optimal solution to crossing. It is therefore natural to expect that $\alpha_n$ and $\beta_n$ are a much better starting point for the numerical bootstrap even in the cases when the optimal solution is not the mean field theory. This will be explored in an upcoming work \cite{pauloszan}.

In some sense, $\alpha_n$ and $\beta_n$ exist precisely because mean field theory saturates appropriate bootstrap bounds. Should we expect that a similarly useful basis of functionals exists for every theory that saturates some bootstrap bound? This seems to be the case perturbatively around mean field theory and probably also more generally in 1D. Further developments in our understanding of the analytic conformal bootstrap will be needed to answer this important question in $D>1$.

\vspace{1 cm}
\section*{Acknowledgments}
We would like to thank Fernando Alday, Connor Behan, Rajesh Gopakumar, Apratim Kaviraj, Shota Komatsu, Slava Rychkov, Aninda Sinha, Bernardo Zan, Sasha Zhiboedov, Xinan Zhou and especially Leonardo Rastelli for useful discussions and/or comments on the draft. DM is grateful to LPT ENS in Paris for hospitality as well as to the OIST March 2018 workshop. DM and MFP also acknowledge the role of the bootstrap collaboration workshops in Azores and Caltech where parts of this work were completed. DM is supported by a Simons Collaboration grant.

\newpage 
\appendix

\section{Functional actions and asymptotics}
\label{app:funcs}
In this appendix we will show how to determine the bosonic (fermionic) functional actions for $\Df$ (half-)integer for the building blocks constructed in section \ref{sec:constructortho}. We will then show how these may be expanded in various regimes.

\subsection{Building block actions}
For convenience let us work with the $g$ kernel:
\bea
g(z)\equiv \eta (1-z)^{2\Df-2} f\left(\frac{1}{1-z}\right).
\eea
For $\Delta$ sufficiently large, the functional actions are computed by:
\bea
\omega(\Delta)=\left[1-\eta \cos \pi(\Delta-2\Df)\right]\int_0^1 \ud z g(z) \frac{G_\Delta(z)}{z^{2\Df}}.
\eea
In the end the result will be valid for generic $\Delta$ by analytic continuation. We will find the actions for the building blocks introduced in section \ref{sec:construction}, which correspond to:
\bea
g_{\beta,h}^{\Df,\eta}&=&z^{2\Df-2} P_{h-1}\left(\frac{2-z}z\right)+\eta P_{h-1}(2z-1)\\
g_{\alpha,h}^{\Df,\eta}&=&\partial_h g_{\beta,h}^{\Df,\eta}(z)-\frac{\Gamma(h)^2}{\Gamma(2h)}\frac{ G_{h}(1-z)}{(1-z)^{2-2\Df}}
\eea
where we have swapped the $m$ label by $h\equiv 2+2m$. We can now compute
\bea
\int_0^1 \frac{\ud z}{z^2}\, P_{h-1}\left(\frac{2-z}z\right) G_{\Delta}(z)&=&\frac{\Gamma(\Delta)^2}{\Gamma(2\Df)}\,\frac{1}{\Delta(\Delta-1)-h(h-1)}\\
\int\limits_{0}^{1}\!\ud z\,P_{h-1}(2z-1)\frac{G_{\Delta}(z)}{z^{2\Df}}&=:&\frac{\Gamma(2\Delta)}{\Gamma(\Delta)^2}\,s( h ;\Delta|\Df)\\
\int\limits_{0}^{1}\!\frac{\ud z}{z^{2}}\,G_ h (1-z)\left(\mbox{$\frac{z}{1-z}$}\right)^{2-2\Df}G_{\Delta}(z)\,.&=:&\frac{\Gamma(2\Delta)}{\Gamma(\Delta)^2}\frac{\Gamma(2h)}{\Gamma(h)^2}\, \widetilde{s}( h ;\Delta|\Df)
\eea
with
\ba
s( h ;\Delta|\Df) = & \frac{\Gamma(\Delta)^2\Gamma(\Delta-2\Df+1)^2}{\Gamma(2\Delta)\Gamma(\Delta- h -2\Df+2)\Gamma( h +\Delta-2\Df+1)}\times\\
&\times{}_4F_3\left( {\begin{array}{*{20}{c}}
{\Delta ,\Delta ,\Delta -2 \Df+1,\Delta -2 \Df+1}\\
{2 \Delta,\Delta- h  -2 \Df+2 , h +\Delta -2 \Df+1}
\end{array};1} \right)
\ea
and
\be
\widetilde{s}( h ;\Delta|\Df)=
\frac{\pi\left[s( h ;\Delta|\Df)-s(\Delta; h |1-\Df)\right]}{\sin[\pi(\Delta- h -2\Df)]}\,.
\ee
With these ingredients we may compute any functional action for fermionic functionals with half-integer $\Df$ or bosonic with integer $\Df$, after the functional kernels are constructed as explained in section \ref{sec:constructortho}. In some cases, $s( h ;\Delta|\Df)$ can be expressed in terms of more elementary functions. See \cite{Hyp} for an illustration of the techniques involved in such simplifications.

\subsection{Asymptotic expansions}
The basic tool for performing asymptotic expansions of the above integrals is the formula
\be
J_t( h ) \equiv \frac{\Gamma( h )^2}{\Gamma(2 h )}
\int\limits_{0}^{1}\!\ud z z^{-2}\,G_ h (z)\left(\mbox{$\frac{1-z}{z}$}\right)^{t} =
\frac{\Gamma (t+1)^2 \Gamma ( h -t -1)}{\Gamma ( h +t+1)}\,.
\ee
It has the following expansion for $ h \gg1$
\be
J_t(h) = \Gamma (t+1)^2 h ^{-2 t-2}\times\left[1+O( h ^{-1})\right]\,.
\ee
This makes it easy to find the asymptotic expansion of $\widetilde{s}( h ;\Delta|\Df)$ for $ h \gg1$. We simply need to expand $\left(\mbox{$\frac{z}{1-z}$}\right)^{2\Df}G_{\Delta}(1-z)$ in powers of $y=\mbox{$\frac{1-z}{z}$}$. The leading terms is just $y^{\Delta-2\Df}$ so that we find
\be
\widetilde{s}( h ;\Delta|\Df) = \frac{\Gamma (\Delta)^2 \Gamma (\Delta-2 \Df+1)^2}{\Gamma (2 \Delta)} h  ^{4\Df-2 \Delta-2}\times\left[1+O( h ^{-1})\right]\,.
\ee
Similarly, we can find the asymptotic expansion of $s( h ;\Delta|\Df)$ for $\Delta\gg 1$ by writing
\be
s( h ;\Delta|\Df)=\frac{\Gamma(\Delta)^2}{\Gamma(2\Delta)}
\int\limits_{0}^{1}\!\frac{\ud z}{z^2}\,G_{\Delta}(z)z^{2-2\Df}P_{h-1}(2z-1)
\ee
and expanding $z^{2-2\Df}P_{h-1} (2z-1)$ in powers of $y=\mbox{$\frac{1-z}{z}$}$. We find
\be
s( h ;\Delta|\Df)=\frac{1}{\Delta ^2}+\frac{1}{\Delta ^3}+\frac{2\Df-h^2+h-1}{\Delta ^4}+O(\Delta^{-5})\,.
\ee
In order to find the asymptotic expansion of $s( h ;\Delta|\Df)$ for $h \gg 1$, we can first use the relation between $s( h ;\Delta|\Df)$, $s(\Delta;h|1-\Df)$ and $\widetilde{s}( h ;\Delta|\Df)$
\be
s( h ;\Delta|\Df) = s(\Delta;h|1-\Df) + \frac{\sin[\pi(\Delta- h -2\Df)]}{\pi}\widetilde{s}( h ;\Delta|\Df)
\ee
and use the expansions derived above to find
\ba
&s( h ;\Delta|\Df)=\frac{1}{h^2}+\frac{1}{h^3}+\frac{1-2\Df-\Delta^2+\Delta}{h^4}+O(h^{-5})-\\
&\;-\frac{\Gamma (\Delta)^2 \Gamma (\Delta-2 \Df+1)^2}{\pi\Gamma (2 \Delta)} \sin[\pi(h-\Delta+2\Df)]h^{4\Df-2 \Delta-2}\times\left[1+O( h ^{-1})\right]\,.
\ea
Which term dominates depends on the value of $\Delta$. Finally, it is useful to find the expansion of $s( h ;\Delta|\Df)$ in the regime where both $h$ and $\Delta$ are becoming large with a fixed ratio. One way to do this is to first perform the expansion of $s( h ;\Delta|\Df)$ in $1/\Delta$ and fixed $h$ and keep only the maximal power of $h$ at each order
\be
s( h ;\Delta|\Df)\sim\frac{1}{\Delta ^2}-\frac{h^2}{\Delta ^4}+\frac{h^4}{\Delta ^6}-\frac{h^6}{\Delta ^8}+\ldots = \frac{1}{\Delta^2+h^2}\,.
\ee

\small
\parskip=-10pt
\bibliography{Functionals}

\providecommand{\href}[2]{#2}\begingroup\raggedright\begin{thebibliography}{10}

\bibitem{Komargodski:2012ek}
Z.~Komargodski and A.~Zhiboedov, {\it {Convexity and Liberation at Large
  Spin}},  {\em JHEP} {\bf 11} (2013) 140,
  [\href{http://arxiv.org/abs/1212.4103}{{\tt arXiv:1212.4103}}].

\bibitem{Fitzpatrick:2012yx}
A.~L. Fitzpatrick, J.~Kaplan, D.~Poland, and D.~Simmons-Duffin, {\it {The
  Analytic Bootstrap and AdS Superhorizon Locality}},  {\em JHEP} {\bf 12}
  (2013) 004, [\href{http://arxiv.org/abs/1212.3616}{{\tt arXiv:1212.3616}}].

\bibitem{Alday:2015eya}
L.~F. Alday, A.~Bissi, and T.~Lukowski, {\it {Large spin systematics in CFT}},
  {\em JHEP} {\bf 11} (2015) 101, [\href{http://arxiv.org/abs/1502.07707}{{\tt
  arXiv:1502.07707}}].

\bibitem{Alday:2015ewa}
L.~F. Alday and A.~Zhiboedov, {\it {An Algebraic Approach to the Analytic
  Bootstrap}},  {\em JHEP} {\bf 04} (2017) 157,
  [\href{http://arxiv.org/abs/1510.08091}{{\tt arXiv:1510.08091}}].

\bibitem{Alday:2016njk}
L.~F. Alday, {\it {Large Spin Perturbation Theory for Conformal Field
  Theories}},  {\em Phys. Rev. Lett.} {\bf 119} (2017), no.~11 111601,
  [\href{http://arxiv.org/abs/1611.01500}{{\tt arXiv:1611.01500}}].

\bibitem{Caron-Huot2017b}
S.~Caron-Huot, {\it {Analyticity in Spin in Conformal Theories}},  {\em JHEP}
  {\bf 09} (2017) 078, [\href{http://arxiv.org/abs/1703.00278}{{\tt
  arXiv:1703.00278}}].

\bibitem{Li:2017lmh}
D.~Li, D.~Meltzer, and D.~Poland, {\it {Conformal Bootstrap in the Regge
  Limit}},  {\em JHEP} {\bf 12} (2017) 013,
  [\href{http://arxiv.org/abs/1705.03453}{{\tt arXiv:1705.03453}}].

\bibitem{Costa:2017twz}
M.~S. Costa, T.~Hansen, and J.~Penedones, {\it {Bounds for OPE coefficients on
  the Regge trajectory}},  {\em JHEP} {\bf 10} (2017) 197,
  [\href{http://arxiv.org/abs/1707.07689}{{\tt arXiv:1707.07689}}].

\bibitem{Kravchuk:2018htv}
P.~Kravchuk and D.~Simmons-Duffin, {\it {Light-ray operators in conformal field
  theory}},  \href{http://arxiv.org/abs/1805.00098}{{\tt arXiv:1805.00098}}.

\bibitem{Qiao:2017xif}
J.~Qiao and S.~Rychkov, {\it {A tauberian theorem for the conformal
  bootstrap}},  {\em JHEP} {\bf 12} (2017) 119,
  [\href{http://arxiv.org/abs/1709.00008}{{\tt arXiv:1709.00008}}].

\bibitem{Mukhametzhanov:2018zja}
B.~Mukhametzhanov and A.~Zhiboedov, {\it {Analytic Euclidean Bootstrap}},
  \href{http://arxiv.org/abs/1808.03212}{{\tt arXiv:1808.03212}}.

\bibitem{Mazac:2016qev}
D.~Mazac, {\it {Analytic bounds and emergence of AdS$_{2}$ physics from the
  conformal bootstrap}},  {\em JHEP} {\bf 04} (2017) 146,
  [\href{http://arxiv.org/abs/1611.10060}{{\tt arXiv:1611.10060}}].

\bibitem{Mazac:2018mdx}
D.~Mazac and M.~F. Paulos, {\it {The Analytic Functional Bootstrap I: 1D CFTs
  and 2D S-Matrices}},  \href{http://arxiv.org/abs/1803.10233}{{\tt
  arXiv:1803.10233}}.

\bibitem{Rattazzi:2008pe}
R.~Rattazzi, V.~S. Rychkov, E.~Tonni, and A.~Vichi, {\it {Bounding scalar
  operator dimensions in 4D CFT}},  {\em JHEP} {\bf 12} (2008) 031,
  [\href{http://arxiv.org/abs/0807.0004}{{\tt arXiv:0807.0004}}].

\bibitem{Poland:2018epd}
D.~Poland, S.~Rychkov, and A.~Vichi, {\it {The Conformal Bootstrap: Theory,
  Numerical Techniques, and Applications}},
  \href{http://arxiv.org/abs/1805.04405}{{\tt arXiv:1805.04405}}.

\bibitem{ElShowk:2012hu}
S.~El-Showk and M.~F. Paulos, {\it {Bootstrapping Conformal Field Theories with
  the Extremal Functional Method}},  {\em Phys. Rev. Lett.} {\bf 111} (2013),
  no.~24 241601, [\href{http://arxiv.org/abs/1211.2810}{{\tt
  arXiv:1211.2810}}].

\bibitem{El-Showk:2016mxr}
S.~El-Showk and M.~F. Paulos, {\it {Extremal bootstrapping: go with the flow}},
   {\em JHEP} {\bf 03} (2018) 148, [\href{http://arxiv.org/abs/1605.08087}{{\tt
  arXiv:1605.08087}}].

\bibitem{Simmons-Duffin:2016wlq}
D.~Simmons-Duffin, {\it {The Lightcone Bootstrap and the Spectrum of the 3d
  Ising CFT}},  {\em JHEP} {\bf 03} (2017) 086,
  [\href{http://arxiv.org/abs/1612.08471}{{\tt arXiv:1612.08471}}].

\bibitem{Rychkov:2017tpc}
J.~Qiao and S.~Rychkov, {\it {Cut-touching linear functionals in the conformal
  bootstrap}},  {\em JHEP} {\bf 06} (2017) 076,
  [\href{http://arxiv.org/abs/1705.01357}{{\tt arXiv:1705.01357}}].

\bibitem{Polyakov:1974gs}
A.~M. Polyakov, {\it {Nonhamiltonian approach to conformal quantum field
  theory}},  {\em Zh. Eksp. Teor. Fiz.} {\bf 66} (1974) 23--42. [Sov. Phys.
  JETP39,9(1974)].

\bibitem{Sen:2015doa}
K.~Sen and A.~Sinha, {\it {On critical exponents without Feynman diagrams}},
  {\em J. Phys.} {\bf A49} (2016), no.~44 445401,
  [\href{http://arxiv.org/abs/1510.07770}{{\tt arXiv:1510.07770}}].

\bibitem{Gopakumar2017}
R.~Gopakumar, A.~Kaviraj, K.~Sen, and A.~Sinha, {\it {Conformal Bootstrap in
  Mellin Space}},  {\em Phys. Rev. Lett.} {\bf 118} (2017), no.~8 081601,
  [\href{http://arxiv.org/abs/1609.00572}{{\tt arXiv:1609.00572}}].

\bibitem{Gopakumar2017a}
R.~Gopakumar, A.~Kaviraj, K.~Sen, and A.~Sinha, {\it {A Mellin space approach
  to the conformal bootstrap}},  {\em JHEP} {\bf 05} (2017) 027,
  [\href{http://arxiv.org/abs/1611.08407}{{\tt arXiv:1611.08407}}].

\bibitem{Pappadopulo:2012jk}
D.~Pappadopulo, S.~Rychkov, J.~Espin, and R.~Rattazzi, {\it {OPE Convergence in
  Conformal Field Theory}},  {\em Phys. Rev.} {\bf D86} (2012) 105043,
  [\href{http://arxiv.org/abs/1208.6449}{{\tt arXiv:1208.6449}}].

\bibitem{Hogervorst:2013sma}
M.~Hogervorst and S.~Rychkov, {\it {Radial Coordinates for Conformal Blocks}},
  {\em Phys. Rev.} {\bf D87} (2013) 106004,
  [\href{http://arxiv.org/abs/1303.1111}{{\tt arXiv:1303.1111}}].

\bibitem{Hartman:2015lfa}
T.~Hartman, S.~Jain, and S.~Kundu, {\it {Causality Constraints in Conformal
  Field Theory}},  {\em JHEP} {\bf 05} (2016) 099,
  [\href{http://arxiv.org/abs/1509.00014}{{\tt arXiv:1509.00014}}].

\bibitem{Maldacena:2015waa}
J.~Maldacena, S.~H. Shenker, and D.~Stanford, {\it {A bound on chaos}},  {\em
  JHEP} {\bf 08} (2016) 106, [\href{http://arxiv.org/abs/1503.01409}{{\tt
  arXiv:1503.01409}}].

\bibitem{Perlmutter:2016pkf}
E.~Perlmutter, {\it {Bounding the Space of Holographic CFTs with Chaos}},  {\em
  JHEP} {\bf 10} (2016) 069, [\href{http://arxiv.org/abs/1602.08272}{{\tt
  arXiv:1602.08272}}].

\bibitem{Roberts:2014ifa}
D.~A. Roberts and D.~Stanford, {\it {Two-dimensional conformal field theory and
  the butterfly effect}},  {\em Phys. Rev. Lett.} {\bf 115} (2015), no.~13
  131603, [\href{http://arxiv.org/abs/1412.5123}{{\tt arXiv:1412.5123}}].

\bibitem{Dey:2016mcs}
P.~Dey, A.~Kaviraj, and A.~Sinha, {\it {Mellin space bootstrap for global
  symmetry}},  {\em JHEP} {\bf 07} (2017) 019,
  [\href{http://arxiv.org/abs/1612.05032}{{\tt arXiv:1612.05032}}].

\bibitem{Dey:2017fab}
P.~Dey, K.~Ghosh, and A.~Sinha, {\it {Simplifying large spin bootstrap in
  Mellin space}},  {\em JHEP} {\bf 01} (2018) 152,
  [\href{http://arxiv.org/abs/1709.06110}{{\tt arXiv:1709.06110}}].

\bibitem{Gopakumar:2018xqi}
R.~Gopakumar and A.~Sinha, {\it {On the Polyakov-Mellin bootstrap}},
  \href{http://arxiv.org/abs/1809.10975}{{\tt arXiv:1809.10975}}.

\bibitem{Alday:2017gde}
L.~F. Alday, A.~Bissi, and E.~Perlmutter, {\it {Holographic Reconstruction of
  AdS Exchanges from Crossing Symmetry}},  {\em JHEP} {\bf 08} (2017) 147,
  [\href{http://arxiv.org/abs/1705.02318}{{\tt arXiv:1705.02318}}].

\bibitem{Sleight:2018ryu}
C.~Sleight and M.~Taronna, {\it {Anomalous Dimensions from Crossing Kernels}},
  {\em JHEP} {\bf 11} (2018) 089, [\href{http://arxiv.org/abs/1807.05941}{{\tt
  arXiv:1807.05941}}].

\bibitem{Liu:2018jhs}
J.~Liu, E.~Perlmutter, V.~Rosenhaus, and D.~Simmons-Duffin, {\it
  {$d$-dimensional SYK, AdS Loops, and $6j$ Symbols}},
  \href{http://arxiv.org/abs/1808.00612}{{\tt arXiv:1808.00612}}.

\bibitem{Mazac:2018qmi}
D.~Mazac, {\it {A Crossing-Symmetric OPE Inversion Formula}},
  \href{http://arxiv.org/abs/1812.02254}{{\tt arXiv:1812.02254}}.

\bibitem{Zhou:2018sfz}
X.~Zhou, {\it {Recursion Relations in Witten Diagrams and Conformal Partial
  Waves}},  \href{http://arxiv.org/abs/1812.01006}{{\tt arXiv:1812.01006}}.

\bibitem{pauloszan}
B.~Zan and M.~F. Paulos, {\it {A new approach to the numerical bootstrap}},
  {\em (to appear)}.

\bibitem{Paulos:2015jfa}
M.~F. Paulos, S.~Rychkov, B.~C. van Rees, and B.~Zan, {\it {Conformal
  Invariance in the Long-Range Ising Model}},  {\em Nucl. Phys.} {\bf B902}
  (2016) 246--291, [\href{http://arxiv.org/abs/1509.00008}{{\tt
  arXiv:1509.00008}}].

\bibitem{Behan2017}
C.~Behan, L.~Rastelli, S.~Rychkov, and B.~Zan, {\it {Long-range critical
  exponents near the short-range crossover}},  {\em Phys. Rev. Lett.} {\bf 118}
  (2017), no.~24 241601, [\href{http://arxiv.org/abs/1703.03430}{{\tt
  arXiv:1703.03430}}].

\bibitem{Behan:2017emf}
C.~Behan, L.~Rastelli, S.~Rychkov, and B.~Zan, {\it {A scaling theory for the
  long-range to short-range crossover and an infrared duality}},  {\em J.
  Phys.} {\bf A50} (2017), no.~35 354002,
  [\href{http://arxiv.org/abs/1703.05325}{{\tt arXiv:1703.05325}}].

\bibitem{Behan:2018hfx}
C.~Behan, {\it {Bootstrapping the long-range Ising model in three dimensions}},
   \href{http://arxiv.org/abs/1810.07199}{{\tt arXiv:1810.07199}}.

\bibitem{Paulos:2016fap}
M.~F. Paulos, J.~Penedones, J.~Toledo, B.~C. van Rees, and P.~Vieira, {\it {The
  S-matrix bootstrap. Part I: QFT in AdS}},  {\em JHEP} {\bf 11} (2017) 133,
  [\href{http://arxiv.org/abs/1607.06109}{{\tt arXiv:1607.06109}}].

\bibitem{Carmi:2018qzm}
D.~Carmi, L.~Di~Pietro, and S.~Komatsu, {\it {A Study of Quantum Field Theories
  in AdS at Finite Coupling}},  \href{http://arxiv.org/abs/1810.04185}{{\tt
  arXiv:1810.04185}}.

\bibitem{Heemskerk:2009pn}
I.~Heemskerk, J.~Penedones, J.~Polchinski, and J.~Sully, {\it {Holography from
  Conformal Field Theory}},  {\em JHEP} {\bf 10} (2009) 079,
  [\href{http://arxiv.org/abs/0907.0151}{{\tt arXiv:0907.0151}}].

\bibitem{Heemskerk:2010ty}
I.~Heemskerk and J.~Sully, {\it {More Holography from Conformal Field Theory}},
   {\em JHEP} {\bf 09} (2010) 099, [\href{http://arxiv.org/abs/1006.0976}{{\tt
  arXiv:1006.0976}}].

\bibitem{Cutkosky:1960sp}
R.~E. Cutkosky, {\it {Singularities and discontinuities of Feynman
  amplitudes}},  {\em J. Math. Phys.} {\bf 1} (1960) 429--433.

\bibitem{Aharony:2016dwx}
O.~Aharony, L.~F. Alday, A.~Bissi, and E.~Perlmutter, {\it {Loops in AdS from
  Conformal Field Theory}},  {\em JHEP} {\bf 07} (2017) 036,
  [\href{http://arxiv.org/abs/1612.03891}{{\tt arXiv:1612.03891}}].

\bibitem{2016arXiv160304246V}
M.~{Viazovska}, {\it {The sphere packing problem in dimension 8}},  {\em ArXiv
  e-prints} (Mar., 2016) [\href{http://arxiv.org/abs/1603.04246}{{\tt
  arXiv:1603.04246}}].

\bibitem{2016arXiv160306518C}
H.~{Cohn}, A.~{Kumar}, S.~D. {Miller}, D.~{Radchenko}, and M.~{Viazovska}, {\it
  {The sphere packing problem in dimension 24}},  {\em ArXiv e-prints} (Mar.,
  2016) [\href{http://arxiv.org/abs/1603.06518}{{\tt arXiv:1603.06518}}].

\bibitem{ModularSphere}
D.~Maz\'{a}\v{c} and L.~Rastelli, {\it {Modular Bootstrap and Sphere-Packing}},
   {\em (in preparation)}.

\bibitem{Hellerman:2009bu}
S.~Hellerman, {\it {A Universal Inequality for CFT and Quantum Gravity}},  {\em
  JHEP} {\bf 08} (2011) 130, [\href{http://arxiv.org/abs/0902.2790}{{\tt
  arXiv:0902.2790}}].

\bibitem{Friedan:2013cba}
D.~Friedan and C.~A. Keller, {\it {Constraints on 2d CFT partition functions}},
   {\em JHEP} {\bf 10} (2013) 180, [\href{http://arxiv.org/abs/1307.6562}{{\tt
  arXiv:1307.6562}}].

\bibitem{Collier:2016cls}
S.~Collier, Y.-H. Lin, and X.~Yin, {\it {Modular Bootstrap Revisited}},  {\em
  JHEP} {\bf 09} (2018) 061, [\href{http://arxiv.org/abs/1608.06241}{{\tt
  arXiv:1608.06241}}].

\bibitem{Hyp}
M.~Milgram, {\it {Variations on a hypergeometric theme}},  {\em Journal of
  Classical Analysis} {\bf 13} (2018), no.~1 1--43,
  [\href{http://arxiv.org/abs/1803.03135}{{\tt arXiv:1803.03135}}].

\end{thebibliography}\endgroup
\bibliographystyle{jhep}

\end{document}